\documentclass[prb,aps,epsf,twocolumn,showpacs,longbibliography ]{revtex4}
\usepackage{times}
\usepackage{graphicx}
\usepackage{float}
\usepackage{latexsym,amsmath,amssymb,bm,euscript}
\usepackage{color}
\usepackage{subfigure}
\usepackage{epstopdf}
\usepackage[colorlinks=true,linkcolor=blue,citecolor=blue]{hyperref}
\usepackage{hyperref}
\usepackage{soul}
\usepackage{ulem}
\usepackage{appendix}

\begin{document}

\title{Possible Nematic Spin Liquid in Spin-$1$ Antiferromagnetic System on
the Square Lattice: Implication for the Nematic Paramagnetic State of FeSe}
\author{Shou-Shu Gong$^{1}$, W. Zhu$^2$, D. N. Sheng$^2$, and Kun Yang$^3$}
\affiliation{$^{1}$National High Magnetic Field Laboratory, Florida State University, Tallahassee, FL 32310\\
$^{2}$Department of Physics and Astronomy, California State University, Northridge, CA 91330\\
$^{3}$National High Magnetic Field Laboratory and Department of Physics, Florida State University, Tallahassee, FL 32306}

\begin{abstract}
The exotic normal state of iron chalcogenide superconductor FeSe, which exhibits vanishing 
magnetic order and possesses an electronic nematic order, triggered extensive
explorations of its magnetic ground state. To understand its novel properties,
we study the ground state of a highly frustrated spin-$1$ system with bilinear-biquadratic
interactions using unbiased large-scale density matrix renormalization group.
Remarkably, with increasing biquadratic interactions, we find a paramagnetic phase between
N\'eel and stripe magnetic ordered phases. We identify this phase as a candidate
of nematic quantum spin liquid by the compelling evidences, including vanished spin and 
quadrupolar orders, absence of lattice translational symmetry breaking, and a persistent non-zero
lattice nematic order in the thermodynamic limit. The established quantum phase diagram natually 
explains the observations of enhanced spin fluctuations of FeSe in neutron scattering measurement
and the phase transition with increasing pressure. This identified paramagnetic phase
provides a new possibility to understand the novel properties of FeSe.
\end{abstract}

\pacs{74.25.-q, 74.70.Xa, 75.10.Kt}
\maketitle

\section{Introduction}

In spin-$1/2$ antiferromagnets, the interplay between quantum fluctuations and
geometric frustration may generate exotic paramagnetic states such as quantum spin liquid \cite{balents2010, savary2016}. 
With rapidly suppressed quantum fluctuations, it is usually believed that the higher spin
system such as spin-$1$ would favor magnetic order. Interestingly, some spin-$1$ systems
may have additional biquadratic interaction, and the competing interactions
can also lead to unusual paramagnetic states such as the Affleck-Kennedy-Lieb-Tasaki (AKLT)
state \cite{affleck1987, affleck1988} and quadrupolar state \cite{blume1969, lauchli2006}. 
While these states have been found in both theoretical models and realistic systems, the studies
on spin liquid are limited in contrived models \cite{yao2007} and effective field 
theories \cite{grover2011, xu2012, bieri2012, lai2013}.
The exotic spin liquid has not been found in any realistic microscopic model.
Recent exploration of this question \cite{grover2011, xu2012, bieri2012, lai2013} 
is further stimulated by spin-$1$ triangular antiferromagnets NiGa$_2$S$_4$ \cite{nakatsuji2005} 
and Ba$_3$NiSb$_2$O$_9$ \cite{cheng2011}, which behave like gapless spin liquids in experiments.

In recent studies on iron-based superconductors \cite{johnston2010, stewart2011, dai2015},
the iron chalgogenide FeSe \cite{hsu2008} is attracting much attention because of its paramagnetic
normal state, which differs from the conventional magnetic ordered normal states of
cuprates \cite{lee2006} and iron pnictides \cite{johnston2010, stewart2011, dai2015}.
Besides, FeSe possesses an electronic nematic order after a tetragonal-to-orthorhombic
structural transition at $T_s \simeq 90$K \cite{cava2009, medvedev2009, shimojima2014, nakayama2014}.
Although the primary origin of this nematic order is still unclear \cite{rahn2015, wang2015, zhao2015, 
bohmer2015, baek2015, watson20151, watson20152, massat2016, taichi2016, goldman2016, yu2016, fanfarillo2016}, 
neutron scattering measurements indicate the important role of spin degree of freedom \cite{wang2015, zhao2015}.
These novel properties have triggered wide interests in the magnetic ground state of FeSe \cite{chubukov2015, valenti2015, wangshuai2015, wangfa2015, yu2015, lai2016, wang2016, yao2016, wagner2016, zhu2016, zhuo2016}.
Neutron experiment finds a large effective spin of $S \simeq 0.74$ \cite{zhao2015}, 
which strongly supports the relevance of the spin-$1$ model as a
starting point for understanding the magnetism of FeSe. Along this line, first principles
calculations \cite{cao2015, valenti2015, wangshuai2015, zhu2016} find that in FeSe
the magnetic interactions are highly frustrated and biquadratic interaction
plays an important role \cite{valenti2015, wangshuai2015, zhu2016}. This naturally leads us to the spin model
\begin{equation} \label{eq:ham}
H = J_{i,j} \sum_{(i,j)}\vec{S}_i\cdot \vec{S}_j + K_{i,j} \sum_{(i,j)}(\vec{S}_i\cdot \vec{S}_j)^2,
\end{equation}
which contains further-neighbor interactions and is also considered to be relevant
to other iron superconductors \cite{wysocki2011, hu2012, yu2012, mazin2014}.
Semiclassical calculations for this model find various magnetic ordered phases to
interpret the observed magnetic orders in iron pnictides and FeTe \cite{wysocki2011, stanek2011,
hu2012, yu2012, mazin2014, bilbao2015, valenti2015, yao2016}. Recent mean-field studies
propose an antiferroquadrupolar (AFQ) state for FeSe \cite{yu2015, lai2016}, which exhibits a nematic order
accompanied by the quadrupolar fluctuations at wave vector $\vec{q} = (0,\pi)/(\pi, 0)$.
While mean-field approach can efficiently detect magnetic and quadrupolar ordered phases, it may
not accurately predict the paramagnetic states generated from the frustrated
competing interactions in Hamiltonian~\eqref{eq:ham}. Such possibilities for FeSe may include the
paramagnetic state that might be continuously connected to decoupled spin-$1$ chains \cite{wangfa2015, jiang2009} and nematic spin liquid. To accurately determine the phase diagram of such
a strongly frustrated system and uncover new quantum phases, unbiased studies are highly desired.

\begin{figure}
\includegraphics[width = 0.9\linewidth]{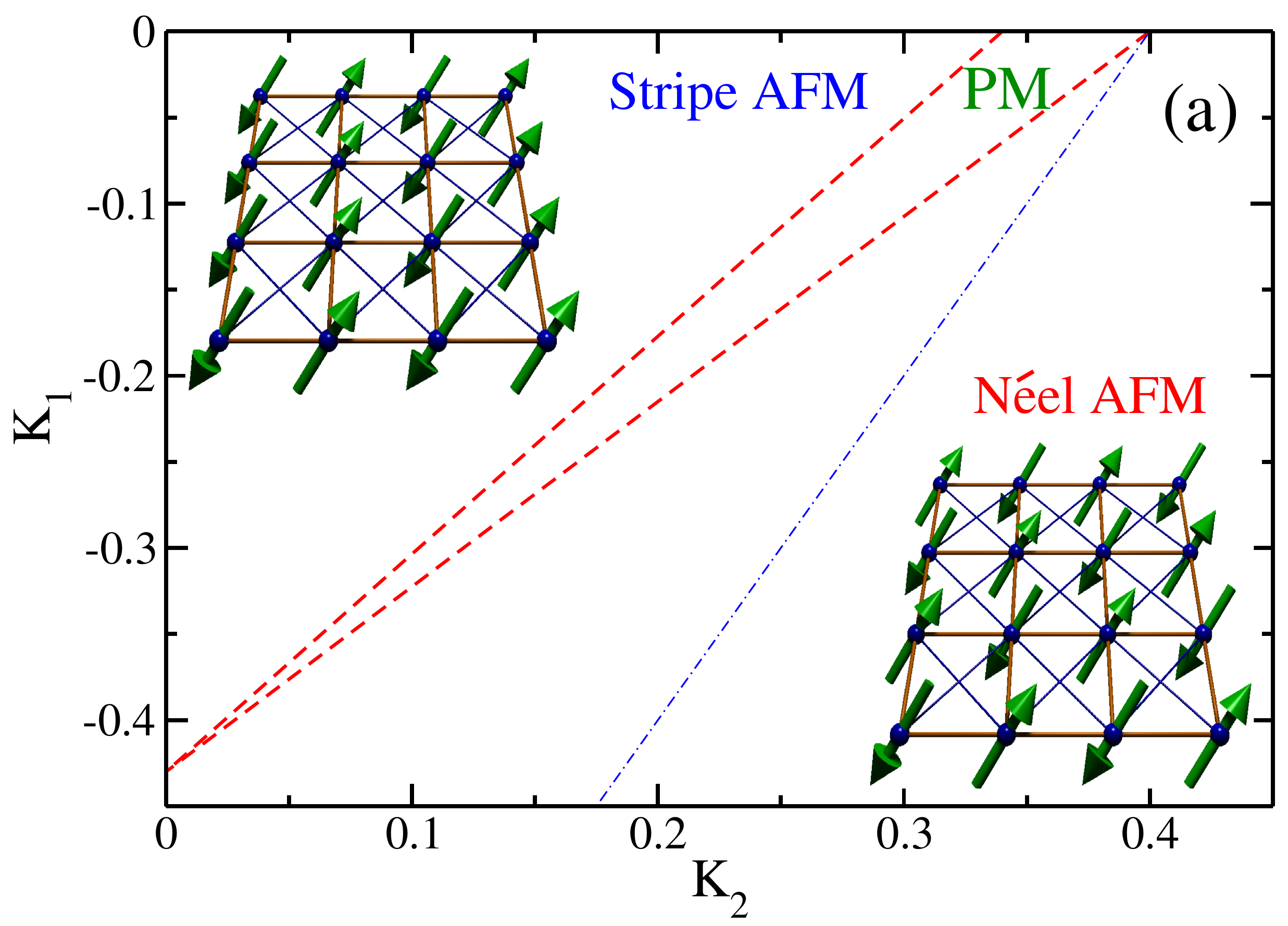}
\includegraphics[width = 1.0\linewidth]{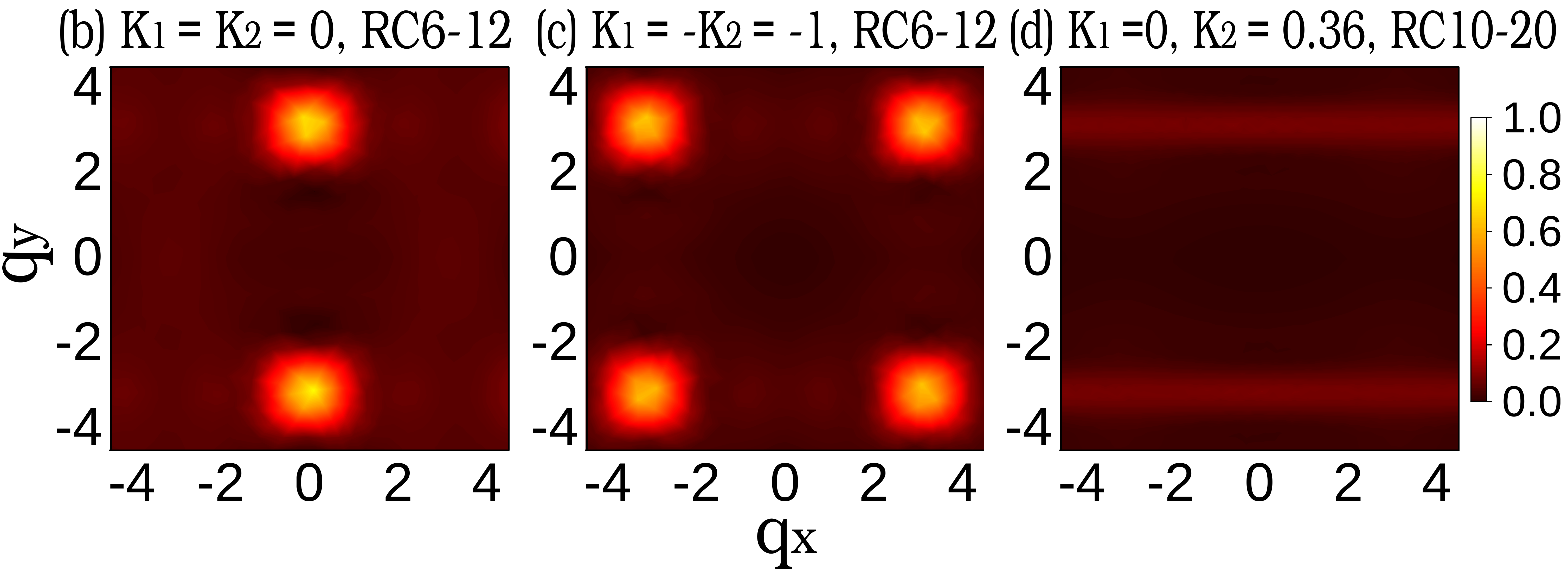}
\caption{(Color online) Different quantum phases in the spin-$1$ $J_1$-$J_2$-$K_1$-$K_2$
model on the square lattice. (a) Quantum phase diagram for $J_2 = 0.7$ in the $K_1$-$K_2$ plane.
With varying $K_1$ and  $K_2$, the system has a stripe and a N\'eel AFM phase. Between these two phases,
we find a paramagnetic (PM) phase with lattice rotational symmetry breaking, which is between the red
dash lines. The blue dash-dot line is the semiclassical phase boundary between the stripe and N\'eel AFM phase.
(b)-(d) are the magnetic order parameter $m^2(\vec{q})$ in
momentum space for the different phases. In the stripe (b) and N\'eel phase (c), $m^2$ has a peak at
$\vec{q}=(0,\pi)$ and $(\pi,\pi)$, respectively. In the paramagnetic phase, $m^2$ is featureless as shown in (d).}
\label{fig:phase}
\end{figure}

In this article, we study the ground state of the frustrated spin-$1$ model \eqref{eq:ham} on the square
lattice with first- ($J_1, K_1$) and second-neighbor ($J_2, K_2$) interactions using unbiased
density matrix renormalization group (DMRG) \cite{white1992}.
We set $J_1 = 1.0$ as energy scale. Considering stripe spin
fluctuations in FeSe \cite{rahn2015, wang2015, zhao2015} and the first principles
simulation results \cite{valenti2015, wangshuai2015}, we fix $J_2 = 0.7$ and set
$K_1 <0$. For such a parameter setup, $K_2 < 0$ only enhances ferroquadrupolar (FQ) order~\cite{wang2016};
thus, we consider $K_2 > 0$.
In the semiclassical phase diagram obtained from the site-factorized
wavefunction calculation \cite{semi}, this system possesses a stripe antiferromagnetic (AFM) and
a N\'eel AFM phase separated by the dash-dot line in Fig.~\ref{fig:phase}(a). In DMRG
calculations, through finite-size scaling of magnetic order parameters, we
find a paramagnetic regime sandwiched by the magnetic ordered phases
as shown in Fig.~\ref{fig:phase}(a). We identify this phase as a candidate of nematic
quantum spin liquid by observing vanished spin and quadrupolar orders,
no lattice translational symmetry breaking, and non-zero lattice nematic order
in the thermodynamic limit. The neighboring stripe phase can naturally explain
the enhanced stripe spin fluctuations in neutron scattering measurement of
FeSe \cite{wang2015, zhao2015}. This identified paramagnetic phase not only 
provides a new possibility to understand the exotic normal state of FeSe, 
but also sheds more light on quantum spin liquid in spin-$1$ magnetic systems.

In our  DMRG calculations, we study the rectangular cylinder (RC) system
with periodic boundary in the $y$ direction and open boundaries in the $x$ direction.
We denote the cylinder as RC$L_y$-$L_x$, where $L_y$ and $L_x$ are the number of sites in
the $y$ and $x$ directions; the width of the cylinder is $L = L_y$ (see the inset of the RC4-4
cylinder in Fig.~\ref{fig:phase}(a)). By implementing spin rotational $SU(2)$ symmetry \cite{mcculloch2002},
we study cylinder system with $L$ up to $10$ by keeping  up to $20000$ $U(1)$-equivalent states
with truncation error below $1\times 10^{-5}$ in most calculations. Our simulations allow us to obtain
accurate quantum phase diagram based on different measurements.


\begin{figure}
  \includegraphics[width = 0.49\linewidth]{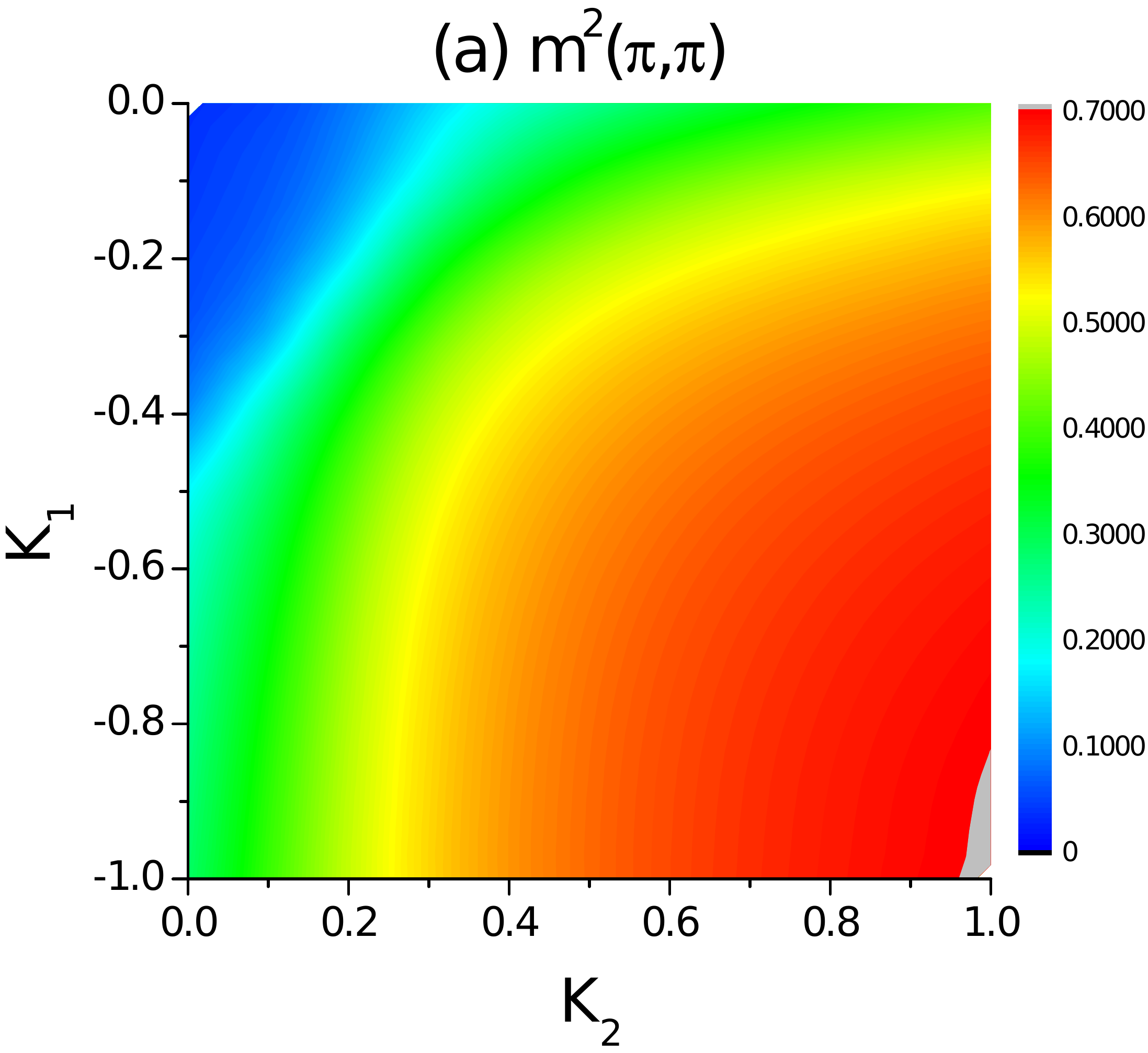}
  \includegraphics[width = 0.49\linewidth]{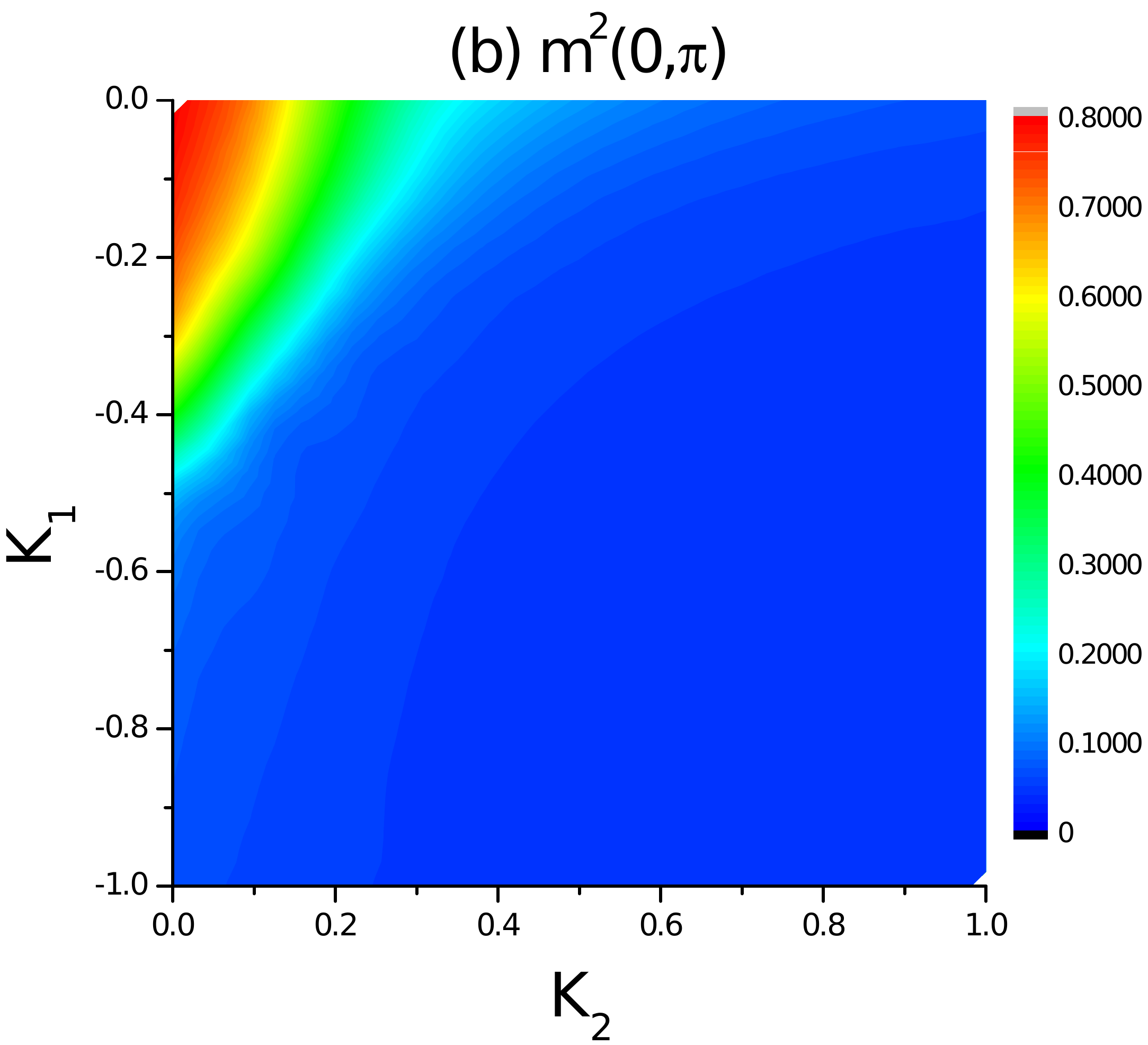}
  \caption{(Color online) $K_1$ and $K_2$ dependence of magnetic order parameters
    for the $J_1$-$J_2$-$K_1$-$K_2$ square model with $J_2 = 0.7$ on the RC$6$-$12$ cylinder.
    (a) and (b) are N\'eel order parameter $m^2(\pi,\pi)$ and stripe order parameter $m^2(0,\pi)$, respectively.}
\label{supfig:m_k1k2}
\end{figure}

\begin{figure}
  \includegraphics[width = 1.0\linewidth]{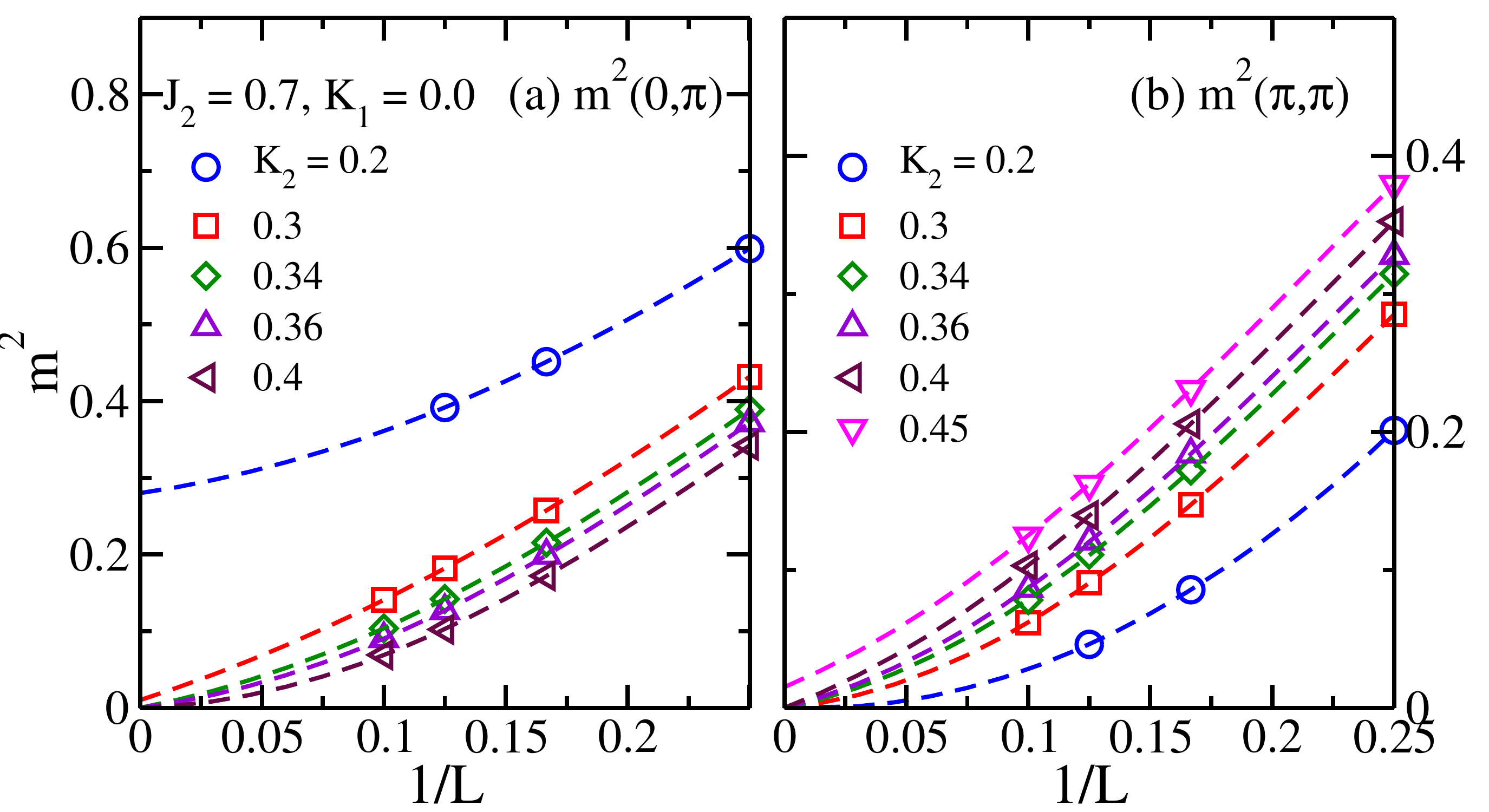}
  \includegraphics[width = 1.0\linewidth]{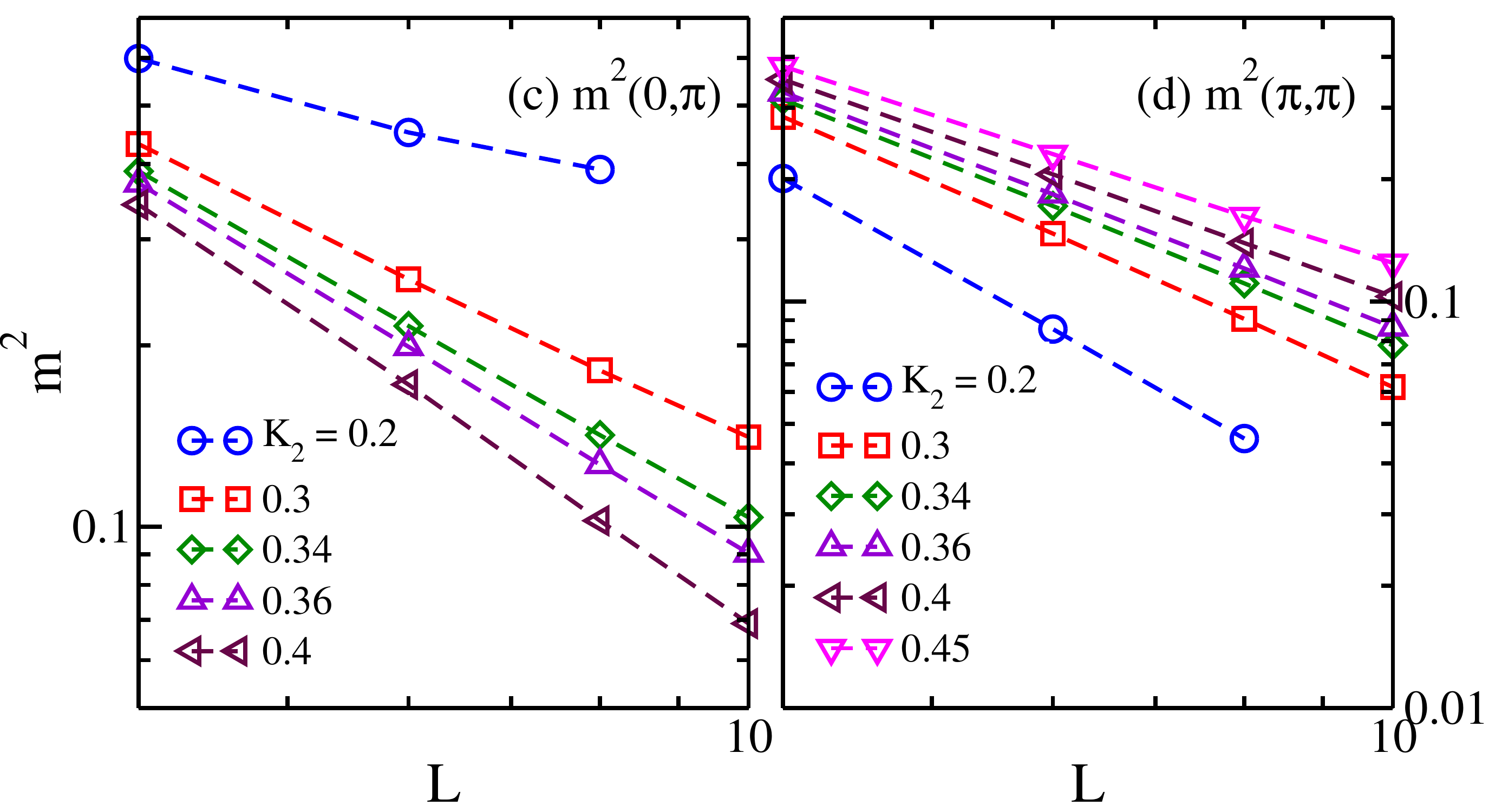}
  \caption{(Color online) Finite-size scaling of magnetic order parameters.
    (a) and (b) are the size extrapolations of stripe order $m^2(0,\pi)$
    and N\'eel order $m^2(\pi,\pi)$ versus $1/L$, respectively. We have the system
    with $J_2 = 0.7, K_1 = 0.0$ on the RC$L$-$2L$ cylinders with $L = 4 -10$.
    Dashed lines are polynomial fits up to fourth order.
    (c) and (d) are log-log plots of the two magnetic
    orders versus width $L$.}
\label{fig:m}
\end{figure}

\section{Magnetic and quadrupolar orders} 

First of all, we show the biquadratic coupling dependence of magnetic order parameters on the RC6-12 cylinder in Fig.~\ref{supfig:m_k1k2}. For this system, we have $J_2 = 0.7$. With growing $K_2$,
the stripe AFM order at small $|K_1|$ side is suppressed and N\'eel order develops.
In the large $|K_1|$ regime, the N\'eel order persists with increased $K_2$.
The global picture of Fig.~\ref{supfig:m_k1k2} is consistent with the quantum phase diagram Fig.~\ref{fig:phase}(a).

\begin{figure}
  \includegraphics[width = 1\linewidth]{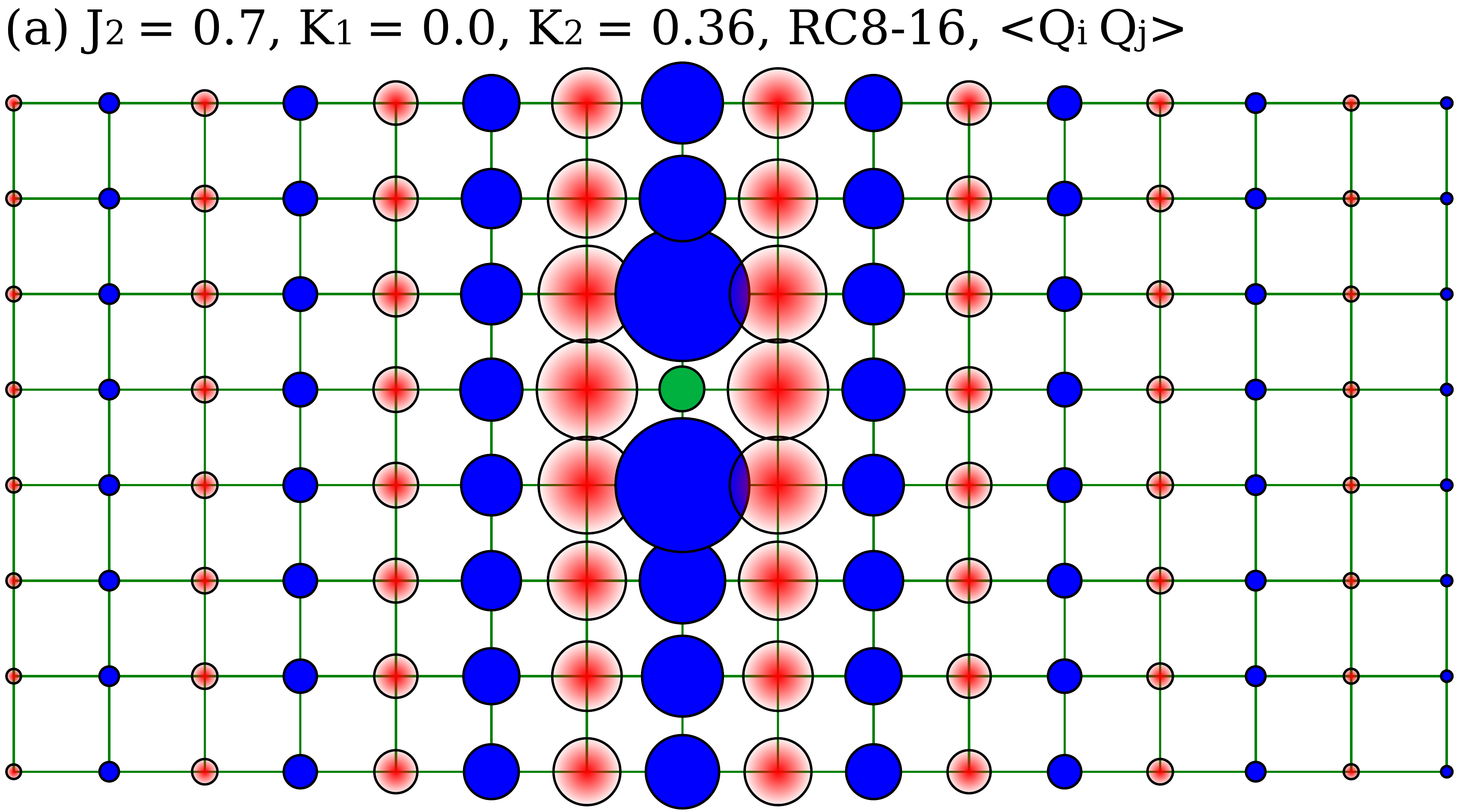}
  \includegraphics[width = 0.49\linewidth]{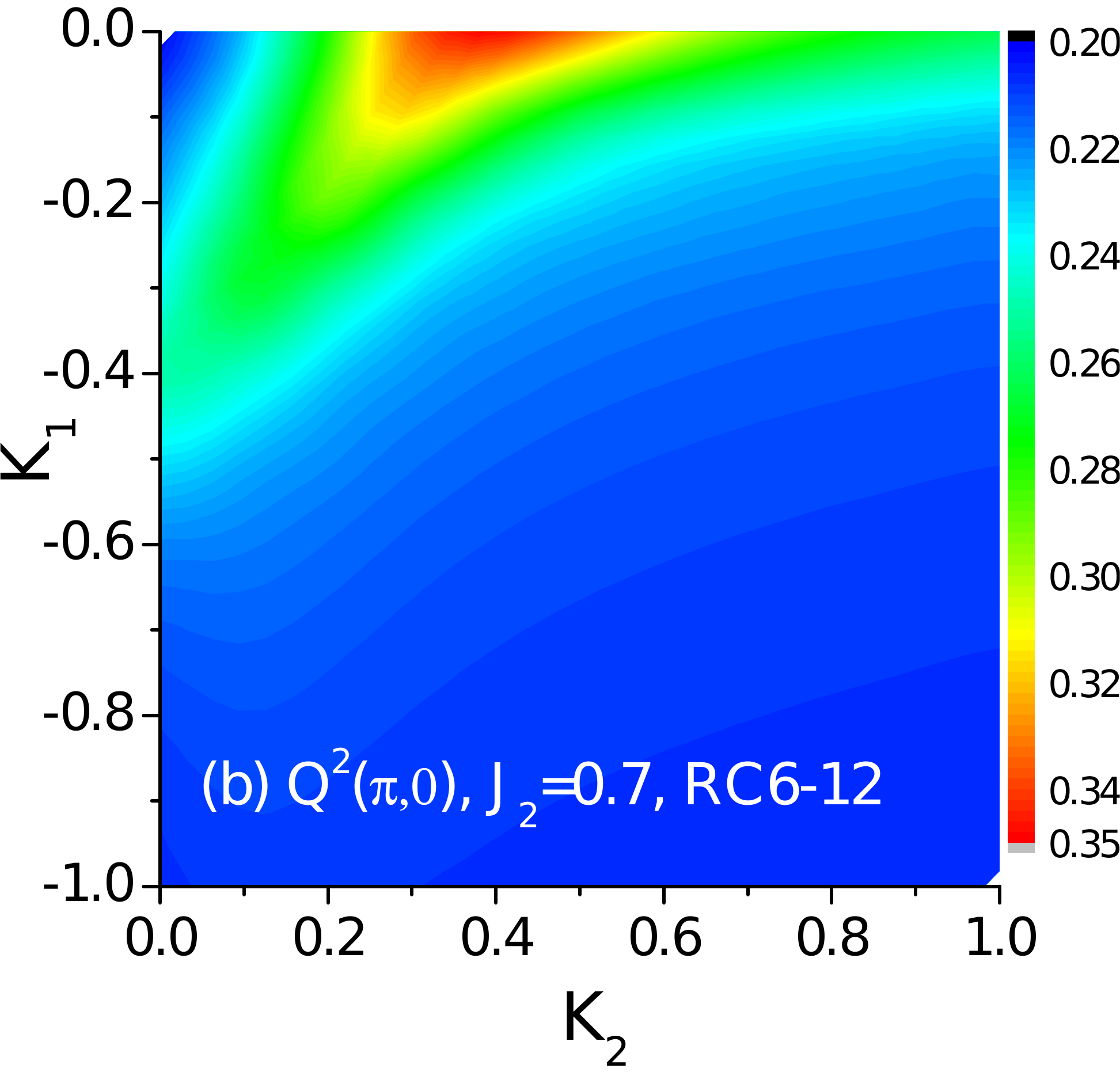}
  \includegraphics[width = 0.49\linewidth]{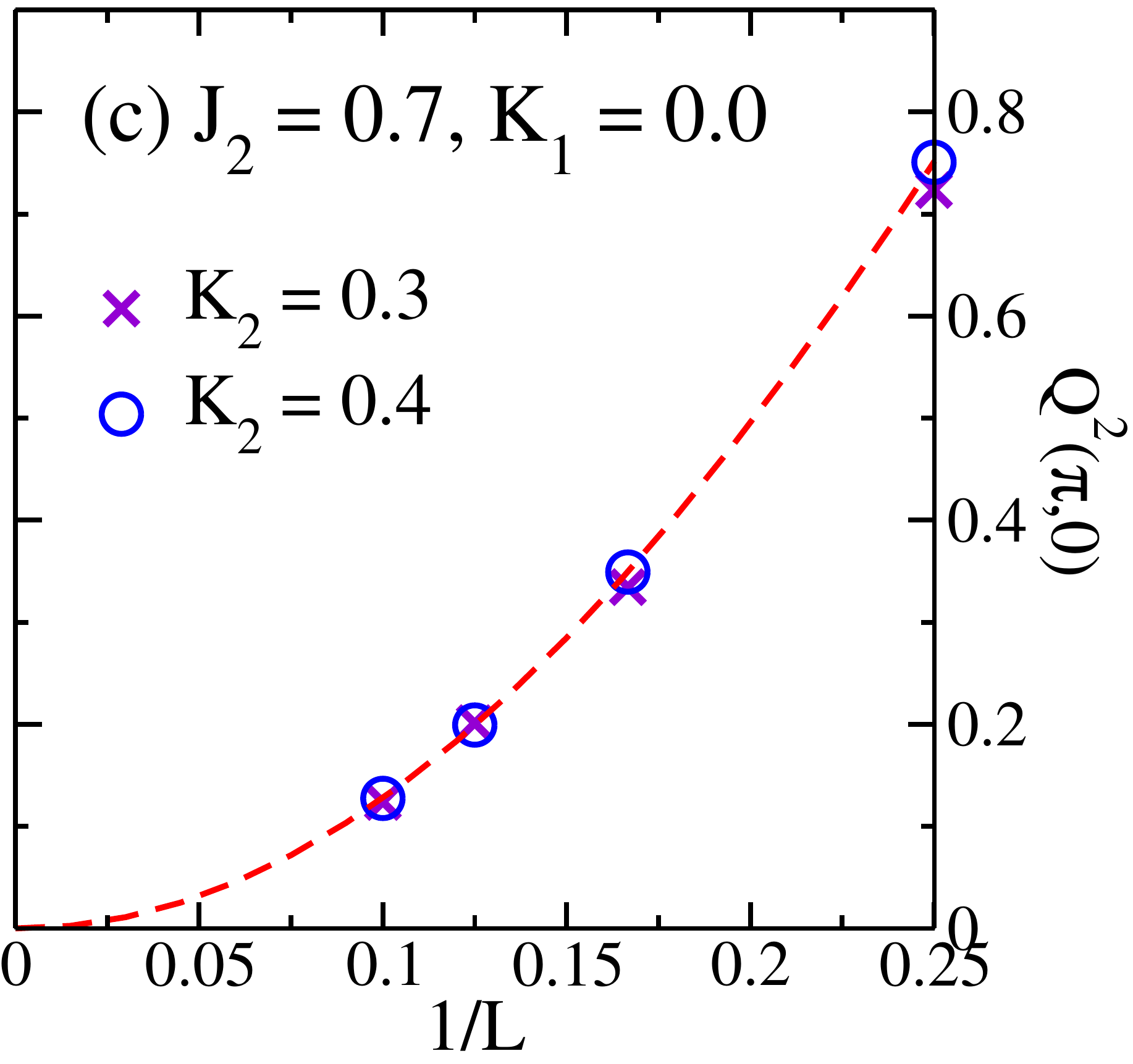}
  \caption{(Color online) The absence of stripe AFQ order.
    (a) Stripe $(\pi,0)$ AFQ correlation $\langle {\bf Q}_i \cdot {\bf Q}_j \rangle$ for
    $J_2 = 0.7, K_1 = 0.0, K_2 = 0.36$ on the RC8-16 cylinder. The solid green circle in the
    middle denotes the reference site. The solid blue and shaded red circles denote
    the positive and negative AFQ correlations, respectively.
    (b) $K_1, K_2$ dependence of stripe AFQ order parameter $Q^{2}(\pi,0)$ on the RC6-12 cylinder.
    (c) Finite-size scaling of $Q^{2}(\pi,0)$ up to width $L = 10$.}
\label{fig:q}
\end{figure}

To further study magnetic order, we calculate spin structure factor 
$m^{2}(\vec{q}) = \frac{1}{N^2}\sum_{i,j}\langle \vec{S}_{i}\cdot \vec{S}_{j}\rangle e^{i\vec{q}\cdot(\vec{r}_i-\vec{r}_j)}$
($N$ is the total numer of sites) from the spin correlations $\langle \vec{S}_i \cdot \vec{S}_j \rangle$ of the $L \times L$ sites in the middle of the RC$L$-$2L$ cylinder, which efficiently reduces edge effects of open
cylinder \cite{white2007, gong2013, gong2014}. In the stripe and N\'eel AFM states, $m^2(\vec{q})$
has the characteristic peak at $\vec{q} = (0,\pi)/(\pi,0)$ and $(\pi,\pi)$, respectively; these are
shown in Figs.~\ref{fig:phase}(b) (the stripe state selects the peak at $(0,\pi)$ because of
the cylinder geometry) and \ref{fig:phase}(c) . In the intermediate regime, $m^2(\vec{q})$
is featureless as shown in Fig.~\ref{fig:phase}(d). Compared with the semiclassical phase boundary,
one finds that our DMRG  phase boundaries shift dramatically to the small $K_2$ side, where
the semiclassical calculations may overestimate the stripe order.
In Figs.~\ref{fig:m}(a-b), we show $m^2(0,\pi)$ and $m^2(\pi,\pi)$ for $K_1 = 0.0$ with growing $K_2$ and
$L = 4 - 10$. The appropriate finite-size scaling suggests that the stripe
order vanishes at $K_2 \simeq 0.34$, and the N\'eel order develops at $K_2 \simeq 0.4$,
leaving an intermediate regime with no magnetic order. The log-log plots of magnetic orders versus system width are shown in Figs.~\ref{fig:m}(c-d), where both orders appear to  vanish
in a power-law manner in the intermediate regime. Thus, we establish a paramagnetic
phase in this regime, possibly with critical magnetic fluctuations. 
To demonstrate the stability of the intermediate phase, we examine the extended parameter regime 
with $J_2 = 0.75, 0.8$ and we also identify the intermediate phase by tuning biquadratic coupling (see Appendix), which supports a stable non-magnetic phase.  Next, we will demonstrate various measurement 
results to characterize the physics in the intermediate phase.

Since biquadratic interaction is present in the system,
we investigate the quadrupolar order $\mathbf{Q}_i$ \cite{blume1969, lauchli2006},
where $\mathbf{Q}_i = (Q^{3z^2-r^2}_i, Q^{x^2-y^2}_i, Q^{xy}_i, Q^{yz}_i, Q^{zx}_i)$ is
a rank-two tensor operator with five components
$Q^{3z^2-r^2}_i=[2(S^z_i)^2 - (S^x_i)^2 - (S^y_i)^2]/\sqrt{3}$, $Q^{x^2-y^2}_i=(S^x_i)^2 - (S^y_i)^2$,
$Q^{xy}_i=S^x_i S^y_i+ S^y_i S^x_i$, $Q^{yz}_i= S^y_i S^z_i+ S^z_i S^y_i$, $Q^{zx}_i= S^z_i S^x_i+ S^x_i S^z_i$.
In Fig.~\ref{fig:q}(a), we show that the quadrupolar correlation in the intermediate regime
exhibits a stripe AFQ pattern. To detect stripe AFQ order, we calculate quadrupolar structure factor
$Q^2(\vec q)=\frac{1}{N^2}\sum_{i,j}\langle \mathbf{Q}_{i}\cdot \mathbf{Q}_{j}\rangle e^{i\vec{q}\cdot(\vec{r}_i-\vec{r}_j)}$
defined in a way similar to $m^2(\vec q)$. In Fig.~\ref{fig:q}(b), we show
the stripe AFQ order parameter $Q^2(\pi,0)$ on the RC6-12 cylinder in the $K_1$-$K_2$ plane,
where the finite-size $Q^2(\pi,0)$ is enhanced in the intermediate regime.
However, the size extrapolation in Fig.~\ref{fig:q}(c) shows that $Q^2(\pi,0)$ approaches zero for $L \rightarrow \infty$,
indicating the vanishing AFQ order in the thermodynamic limit.

\begin{figure}
  \includegraphics[width = 0.7\linewidth]{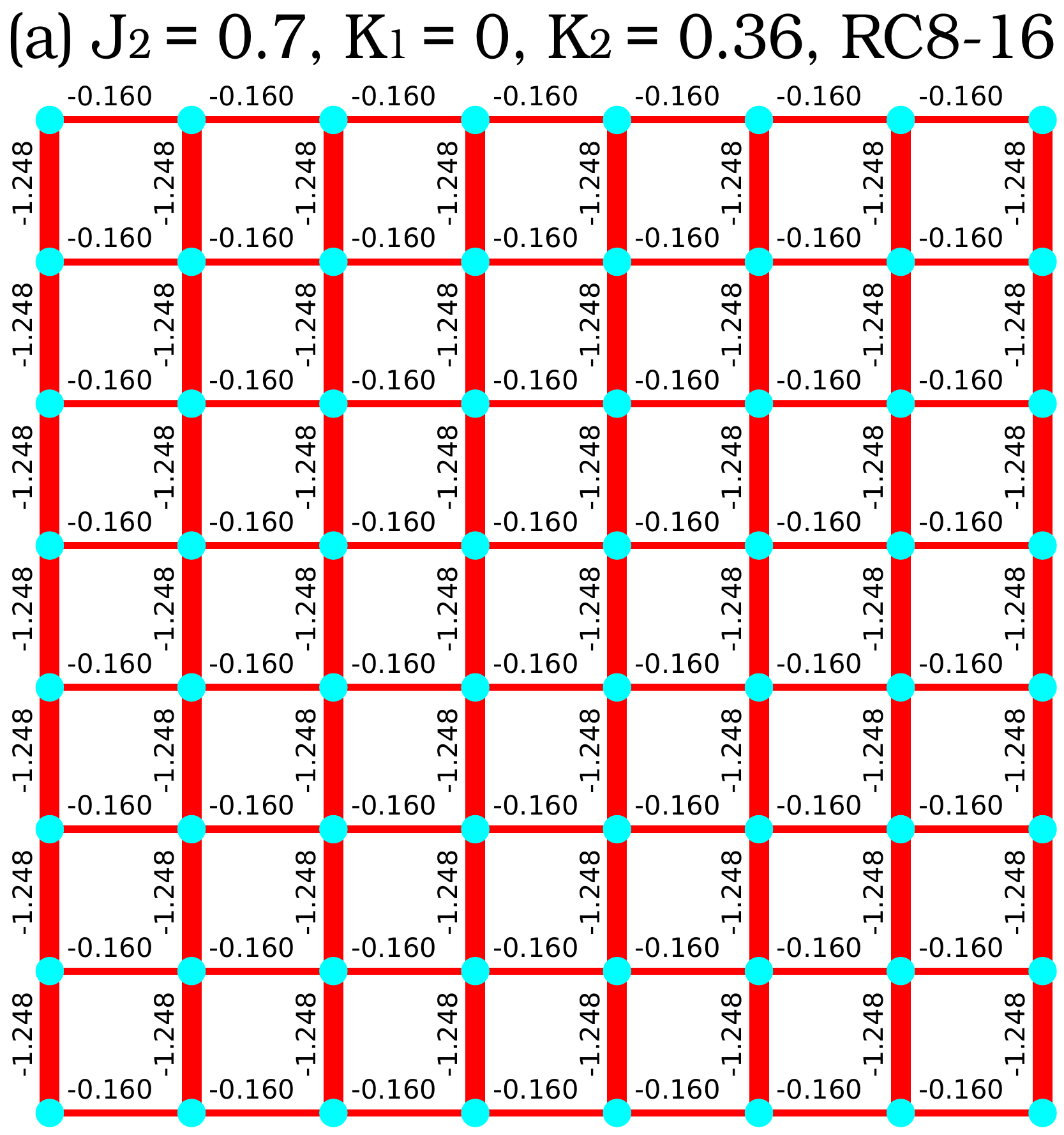}
  \includegraphics[width = 0.8\linewidth]{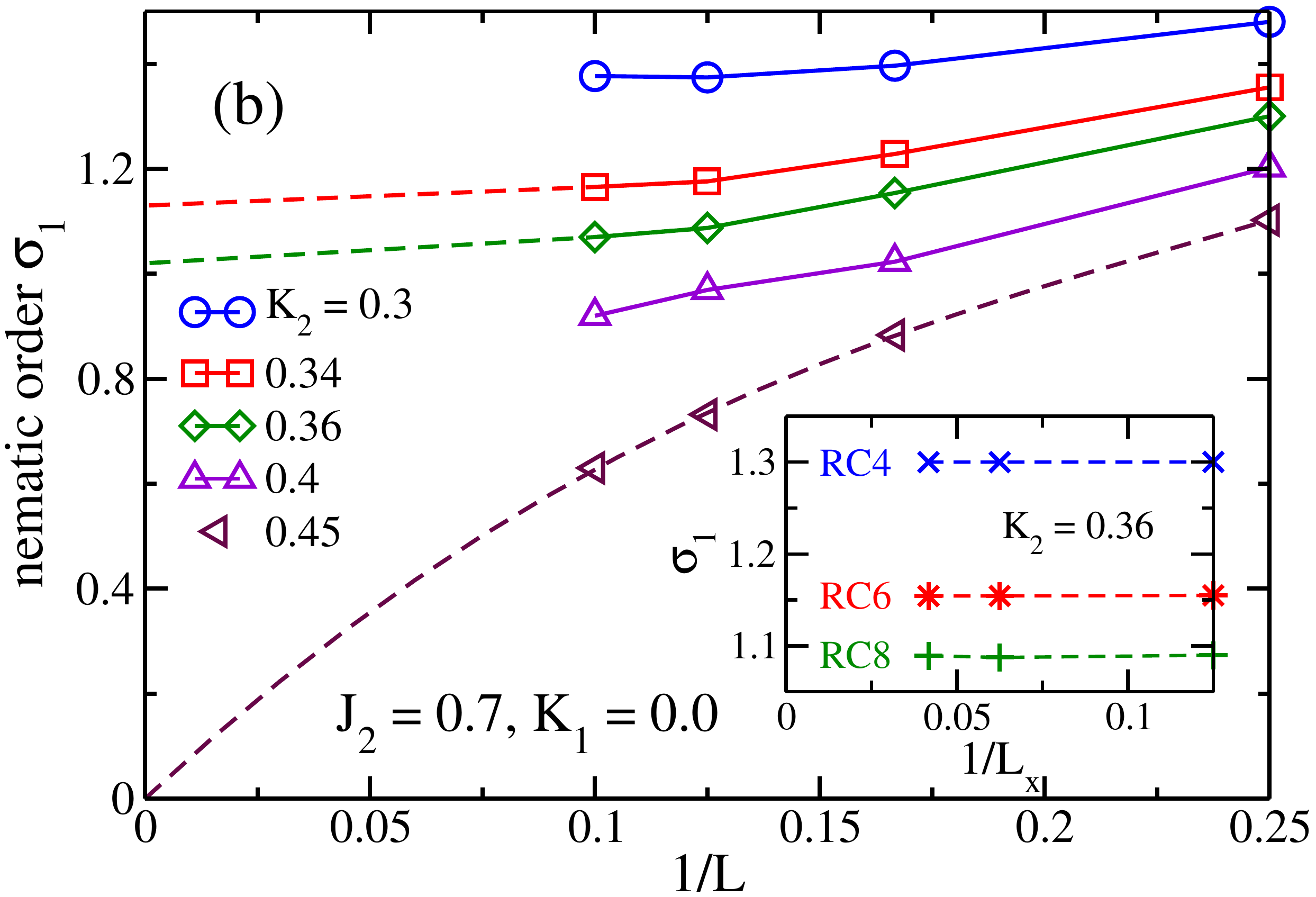}
  \caption{(Color online) Lattice symmetry breaking in the intermediate phase.
    (a) $J_1$ bond energy $\langle \vec{S}_i \cdot \vec{S}_j \rangle$ for
    $K_1 = 0.0, K_2 = 0.36$ on the RC8-16 cylinder. Here, we only show the middle $8 \times 8$ sites.
    (b) Finite-size scaling of bond nematic order $\sigma_1$.
    The inset shows the cylinder length dependence of $\sigma_1$ for $K_2 = 0.36$ and different $L_y$.}
\label{fig:nematic}
\end{figure}

\section{Nematic order} 

Next, we study lattice symmetry breaking by measuring the
nearest-neighbor $J_1$ bond energy $\langle \vec{S}_i \cdot \vec{S}_j \rangle$.
In Fig.~\ref{fig:nematic}(a), we show the bond energy for $K_1 = 0.0, K_2 = 0.36$ on the RC8-16
cylinder, which is quite tranlationally uniform in the bulk of cylinder. Note that
the open boundary conditions in the $x$ direction of cylinder system usually induce 
a bond translational symmetry breaking, and the corresponding dimer 
order (the bond energy difference along the $x$ direction $\langle \vec{S}_i \cdot \vec{S}_{i+1}\rangle 
- \langle \vec{S}_{i+1} \cdot \vec{S}_{i+2}\rangle$) decays from the edge to the bulk. 
For a valence-bond crystal (VBC) phase, the dimer order decay length would increase 
fast while in a non-VBC phase the decay length is finite in the thermodynamic limit~\cite{sandvik2012}. 
In our DMRG calculations, we find that the bond dimer order always decay quite fast with a 
very short decay length on our studied system size, indicating the preserved lattice 
translational symmetry.

Importantly, one can see a strong nematicity between horizontal and vertical bond energy. 
We define a bond nematic order as $\sigma_1 \equiv \langle \vec{S}_i \cdot \vec{S}_{i+\hat{x}} \rangle 
- \langle \vec{S}_i \cdot \vec{S}_{i+\hat{y}} \rangle$ with the bond energy in the bulk of cylinder.
Note that here the bond energy is not translationally invariant only for few columns on the edge. 
$\sigma_1$ versus $1/L$ is presented in Fig.~\ref{fig:nematic}(b) for different $K_2$.
We show the cylinder length dependence of $\sigma_1$ in the inset of Fig.~\ref{fig:nematic}(b),
which indicates the extremely small finite-size effects of $\sigma_1$ versus $L_x$. In the stripe 
AFM phase for $K_2 \lesssim 0.34$, $\sigma_1$ scales to finite value with $1/L$, supporting the 
rotational symmetry breaking of stripe magnetic ordered phase. For $K_2 > 0.4$, $\sigma_1$ decreases 
fast and tends to vanish, which strongly indicates a transition to a phase without lattice rotational 
symmetry breaking. This transition is compatible with the developing N\'eel order at $K_2 \simeq 0.4$ 
found in Fig.~\ref{fig:m}(b). Interestingly, in the intermediate phase, we find that the nematic 
order also decreases slowly and approaches finite value for $L \rightarrow \infty$, indicating lattice
rotational symmetry breaking in this intermediate phase.

We remark that the finite nematic order observed in the intermediate phase is not induced by 
cylinder geometry but intrinsic. For the geometry induced nematic order such as the order in the
neighboring N\'eel phase without a $C_4$ symmetry breaking, one can see that the order decays 
very fast to vanish with growing cylinder width, in contrast to the scaling behavior in the intermediate
phase. As a numerical method, we would like to point out that for detecting lattice symmetry 
breaking, edge bond pinning has been shown effective in quantum Monte Carlo~\cite{sandvik2012} 
and DMRG simulations~\cite{zhu2013, gong2013, gong2014}. In the recent DMRG calculations for 
the spin-$1/2$ $J_1-J_2$ triangular Heisenberg model~\cite{zhu2015, hu2015, ian2016},
a strong nematic order is also found, which is considered as an evidence of a spontaneous
rotational symmetry breaking of the identified spin liquid phase.

\begin{figure}
  \includegraphics[width = 1.0\linewidth]{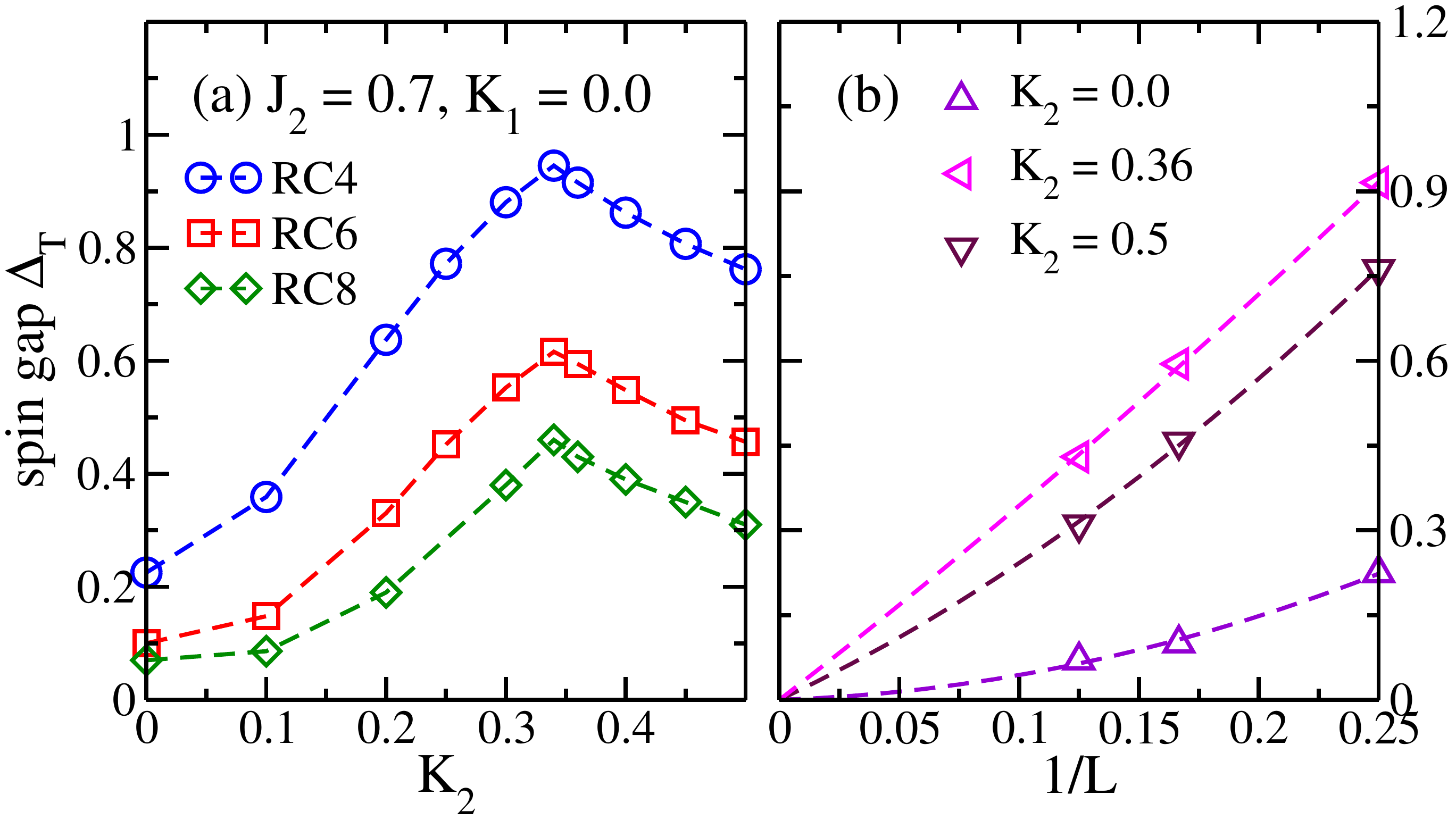}
  \caption{(Color online) Spin gap $\Delta_T$ in the different phases. (a) $\Delta_T$ versus 
    $K_2$ for $J_1 = 0.7, K_1 = 0.0$ on different cylinders. (b) Finite-size scaling of spin gap $\Delta_T$ in different phases. 
    To avoid edge excitations, spin gap is obtained by sweeping the middle $L \times L$ sites with total spin
   $S = 1$ based on the ground state of the long RC$L$-$L_x$ cylinders with $L_x = 24$ and $L = 4, 6, 8$.}
\label{fig:gap}
\end{figure}

\section{Spin gap} 

The vanishing magnetic order and spontaneous
lattice rotational symmetry breaking suggest the intermediate phase as a possible AKLT 
state \cite{wangfa2015} or a nematic spin liquid. To futher characterize this phase, we 
calculate the finite-size spin-$1$ excitation gap, defined as the energy difference between the lowest
energy states in total spin-$1$ and spin-$0$ sectors for a given system size \cite{yan2011, gong2014, gong2015}. 
We demonstrate spin gap with increasing $K_2$ in Fig.~\ref{fig:gap}(a), where it exhibits
a kink at $K_2 = 0.34$. While the ground-state energy varies smoothly with growing $K_2$ (see Appendix),
the kink of spin gap indicates an energy level crossing in spin-$1$ sector, which could be compatible
with the phase transition found in Fig.~\ref{fig:m}(a). At $K_2 = 0.4$, both ground-state 
energy and spin gap exhibit no singularity on our studied system size, which suggest a possible 
continuous phase transition. The vanishing nematic order for $K_2 \gtrsim 0.4$ and the spin
gap singularity at $K_2 = 0.34$ support the intermediate phase found in the finite-size scaling
of magnetic orders. 

In Fig.~\ref{fig:gap}(b), we show finite-size scaling of the spin gap in different phases. 
In both stripe and N\'eel phases, spin gap is smoothly scaled to zero, which agrees with
the gapless spin excitations from continuous spin rotational symmetry breaking. In the 
paramagnetic phase, the spin gap also approaches zero appropriately, which seems to be
inconsistent with a spin gapped AKLT-like state \cite{wangfa2015} but leaves a possibility of
a gapless nematic spin liquid.

\section{DMRG results on the tilt cylinder} 

\begin{figure}
\includegraphics[width = 0.7\linewidth]{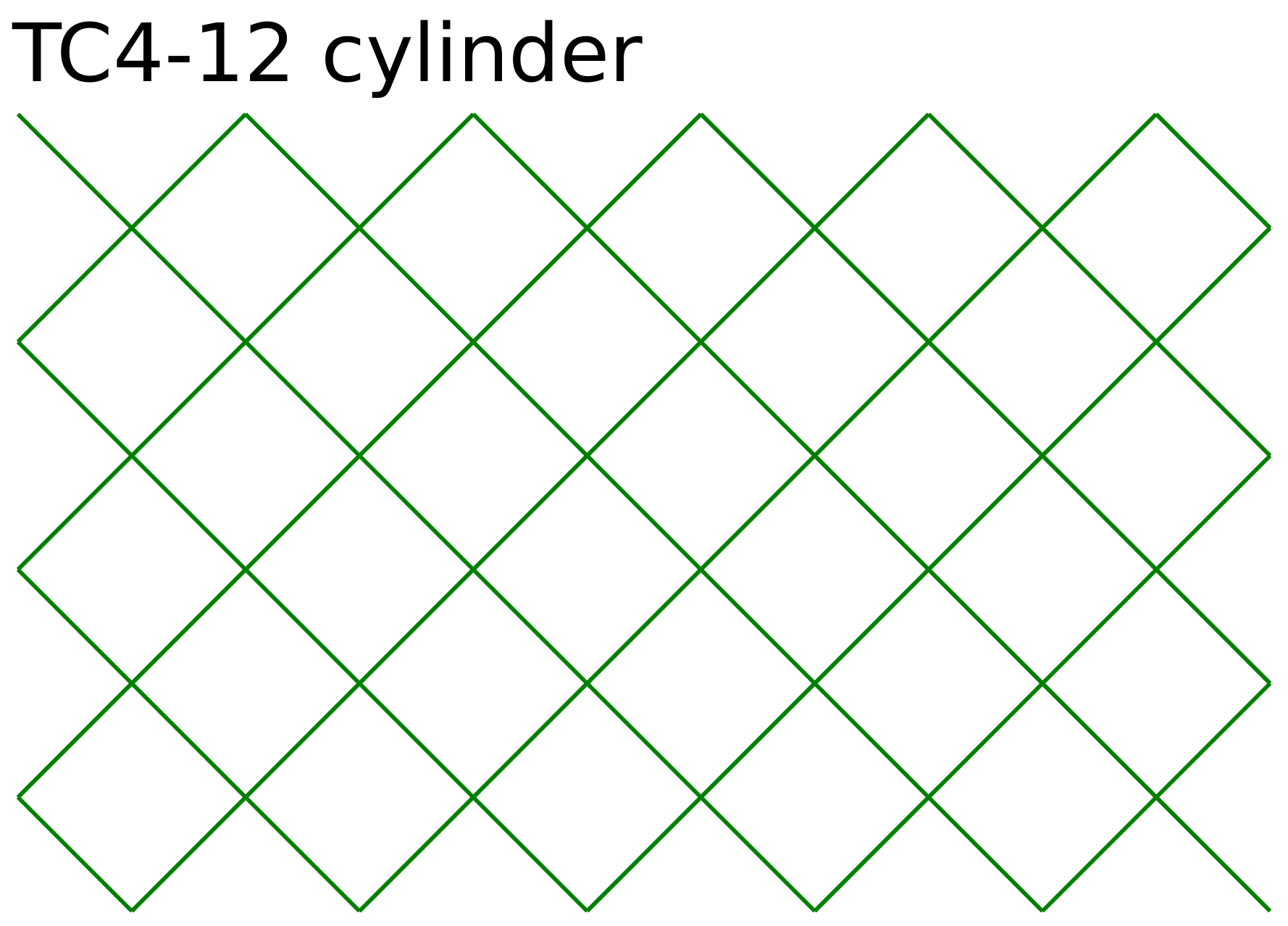}
\caption{(Color online) A schematic figure for the $45$-degree tilted cylinder (TC) on the square lattice.
Here, the cylinder width is $L_y = 4$ and the length is $L_x = 12$, which is denoted as TC4-12. For TC cylinder,
the cylinder width is $\sqrt{2}L_y$.
}
\label{supfig:tc}
\end{figure}

As a supplementary of our finite-size calculations, we also test the tilted cylinder (TC)
that is obtained by a $\pi/4$ rotation of the rectangular lattice. A schematic figure of the TC cylinder
is shown in Fig.~\ref{supfig:tc}. The cylinder width for TC cylinder is $W_y = \sqrt{2} L_y$. 
It should be noticed that different from RC, the bond $\pi/2$ rotational symmetry is not broken
by geometry on TC cylinder.

\begin{figure}
\includegraphics[width = 0.9\linewidth]{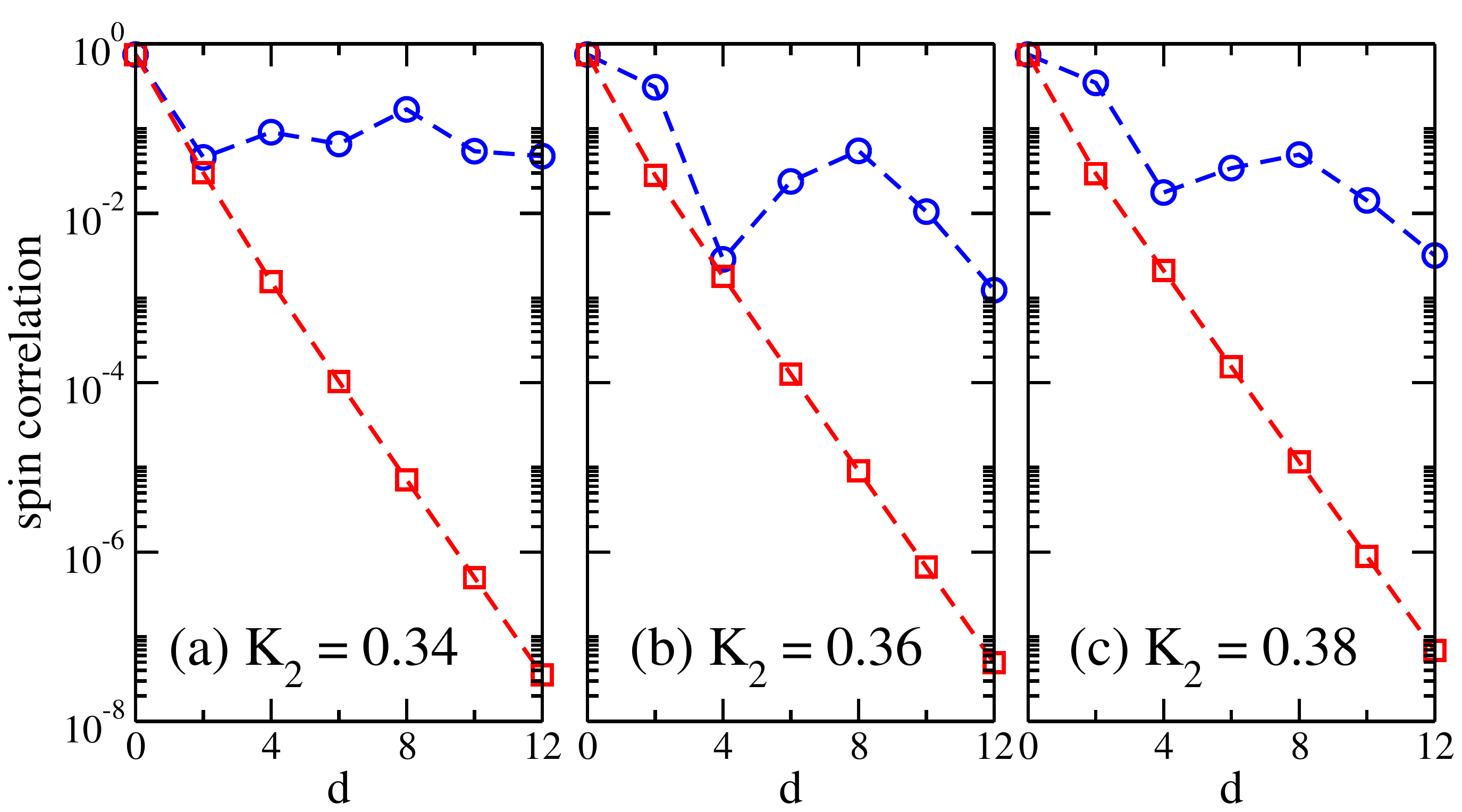}
\caption{(Color online) Log-linear plots of the spin correlations in the intermediate $K_2$ regime 
for $J_2 = 0.7, K_1 = 0.0$ on the RC6 and TC4 cylinders. The red squares denote the RC6 cylinder, and the blue
circles denote the TC4 cylinder. 
}
\label{supfig:spin}
\end{figure}

First of all, we calculate the spin order on the TC cylinder. We find the consistent 
$(0,\pi)$ and $(\pi,\pi)$ magnetic orders in the small $K_2$ and large $K_2$ regimes, respectively. 
However, in the intermediate $K_2$ regime where we find a non-magnetic state on the RC cylinder, 
DMRG calculations obtain a state with strong spin correlations on TC cylinder. As shown in 
Fig.~\ref{supfig:spin}, while the spin correlations on the RC cylinder decay exponentially 
to vanish, those on the TC cylinder decay quite slowly, which does not support a non-magnetic state.

\begin{figure}
\includegraphics[width = 1.0\linewidth]{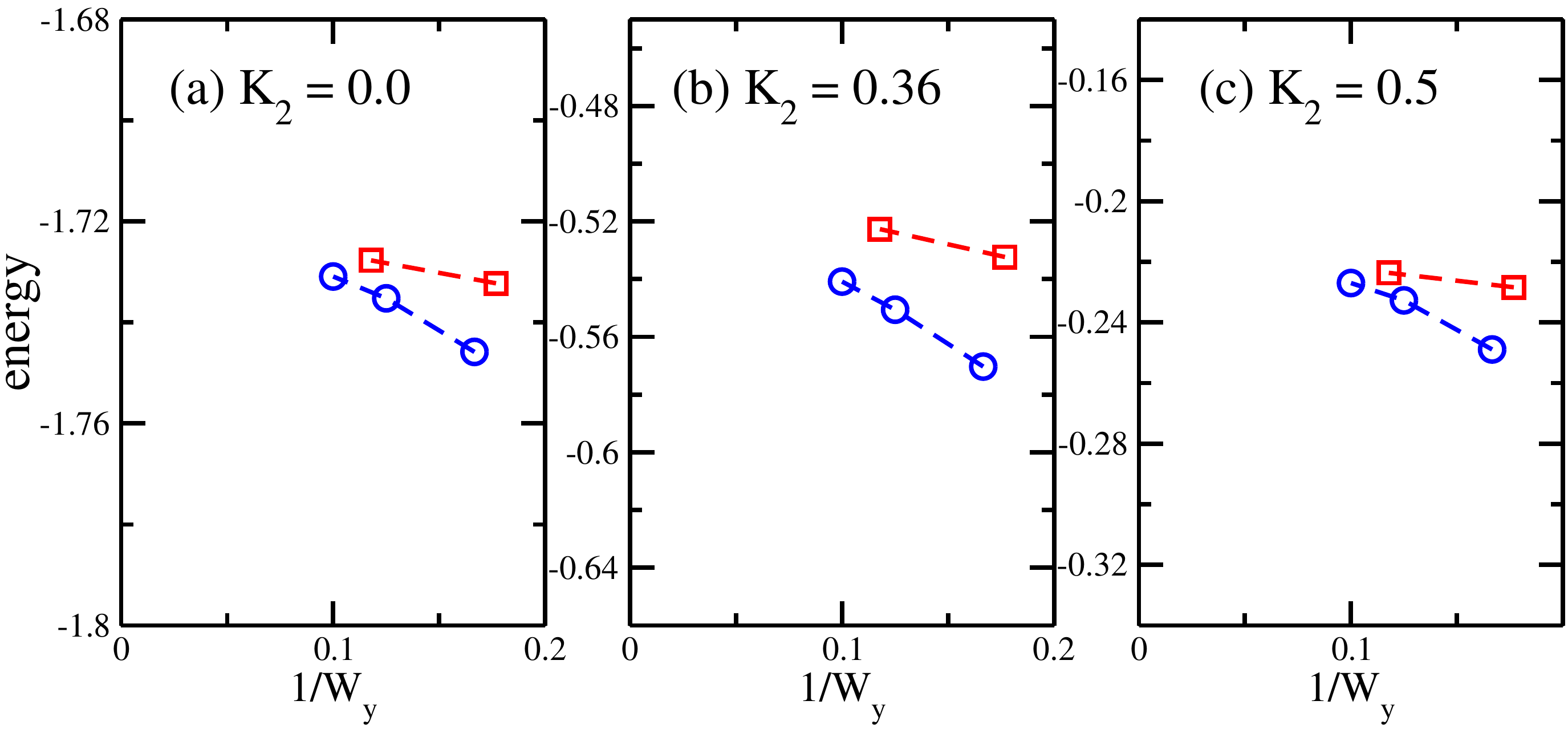}
\caption{(Color online) Bulk energy versus cylinder width $W_y$ on the RC and TC cylinders. The system has $J_2 = 0.7, K_1 = 0.0$ and different $K_2$. For (a) $K_2 = 0.0$, the system is in the $(0,\pi)$ magnetic order phase. For (b) $K_2 = 0.36$, the system is in the intermediate regime. For (c) $K_2 = 0.5$, the system is in the $(\pi,\pi)$ magnetic order
phase. The blue circles are the bulk energy for the RC6, RC8, and RC10 cylinders. The red squares denote the energy
for the TC4 and TC6 cylinders. For RC cylinder, cylinder width $W_y = L_y$; for TC cylinder, $W_y = \sqrt{2}L_y$. In the
two magnetic order phases, the energies on the two geometries approach each other. However, in the intermeidate
regime, the TC cylinder appears to have the higher energy than the RC cylinder on our studied system size.}
\label{supfig:energy}
\end{figure}

To understand the different results on the two geometries, we compare the bulk energy on
both systems. As shown in Fig.~\ref{supfig:energy}, in the two magnetic order phases,
the bulk energies on both geometries approach to each other with increasing cylinder width, indicating
the consistent energy in large size limit. However, in the intermediate regime, the TC cylinder 
appears to have the higher energy than the RC cylinder. The close energies of the two states may
imply the gapless nature of the low-lying excitations, which is consistent with the vanishing gap
in the intermediate phase. The lower energy of the non-magnetic state supports it as the stronger 
candidate of the true ground state. We also remark that in our DMRG calculations on TC cylinder, 
convergence is very challenging and the DMRG truncation error is much bigger than the RC cylinder with the
similar $W_y$, which suggests that TC cylinder may not be a proper geometry for studying the intermediate
phase.

\section{Discussion and Summary} 

Motivated by the exotic nematic paramagnetic normal state of iron 
chalcogenide superconductor FeSe, we study a spin-$1$ $J_1$-$J_2$-$K_1$-$K_2$ system 
on the square lattice using density matrix renormalization group. By implementing spin 
rotational $SU(2)$ symmetry, we study cylinder geometry with system width up to $10$ legs, 
which significantly reduces finite-size effects of order parameter scaling. With increased 
biquadratic interactions $K_1, K_2$, we find a paramagnetic phase between stripe and N\'eel 
magnetic ordered phases, which preserves all spin rotational and lattice translational 
symmetries but breaks lattice rotational symmetry. 

The nematic paramagnetic state in this $J_1$-$J_2$-$K_1$-$K_2$ system provides a new 
possibility to understand the magnetic ground state of FeSe. The current findings naturally 
match the observations of FeSe in neutron scattering \cite{wang2015, zhao2015} and high 
pressure experiments \cite{taichi2016, goldman2016, yu2016}, where the paramagnetic state 
of FeSe with substantial stripe spin fluctuations is identified to sit close to the stripe 
magnetic phase and may undergo a phase transition to the stripe magnetic ordered phase at 
high pressure. As FeSe is a bad metal that is in proximity of a Mott insulator, it would
be interesting to consider the effects of itinerant electrons on the nematicity of the localized
moments in further study. Our DMRG results suggest this paramagnetic state may be a nematic 
quantum spin liquid. Spin liquid states in spin-$1$ system have been discussed for the 
triangular antiferromagnets \cite{grover2011, xu2012, bieri2012, lai2013} related with materials 
NiGa$_2$S$_4$ \cite{nakatsuji2005} and Ba$_3$NiSb$_2$O$_9$ \cite{cheng2011}, but have not been 
found in unbiased calculations besides our work. Our work provides insight for the interplay 
between spin Heisenberg and biquadratic interactions, and sheds more light on interesting phases 
in spin-$1$ system.

\acknowledgments

We acknowledge the discussions with Z.-F.~Wang, F.~Wang, W.-J.~Hu, H.-H.~Lai,
and Q.-M.~Si. This research is supported by the state of Florida (S.S.G.),
National Science Foundation Grants DMR-1157490 (S.S.G. and K.Y.), DMR-1442366 (K.Y.),
PREM DMR-1205734 (W.Z.), and DMR-1408560 (D.N.S.). S.S.G. acknowledges the
computation support of project DMR-160004 from the Extreme
Science and Engineering Discovery Environment (XSEDE) \cite{xsede}, which is supported by
National Science Foundation grant number ACI-1053575.

\appendix

\section{$J_1$-$J_2$-$K_1$-$K_2$ square model}

\begin{figure}
  \includegraphics[width = 1.0\linewidth]{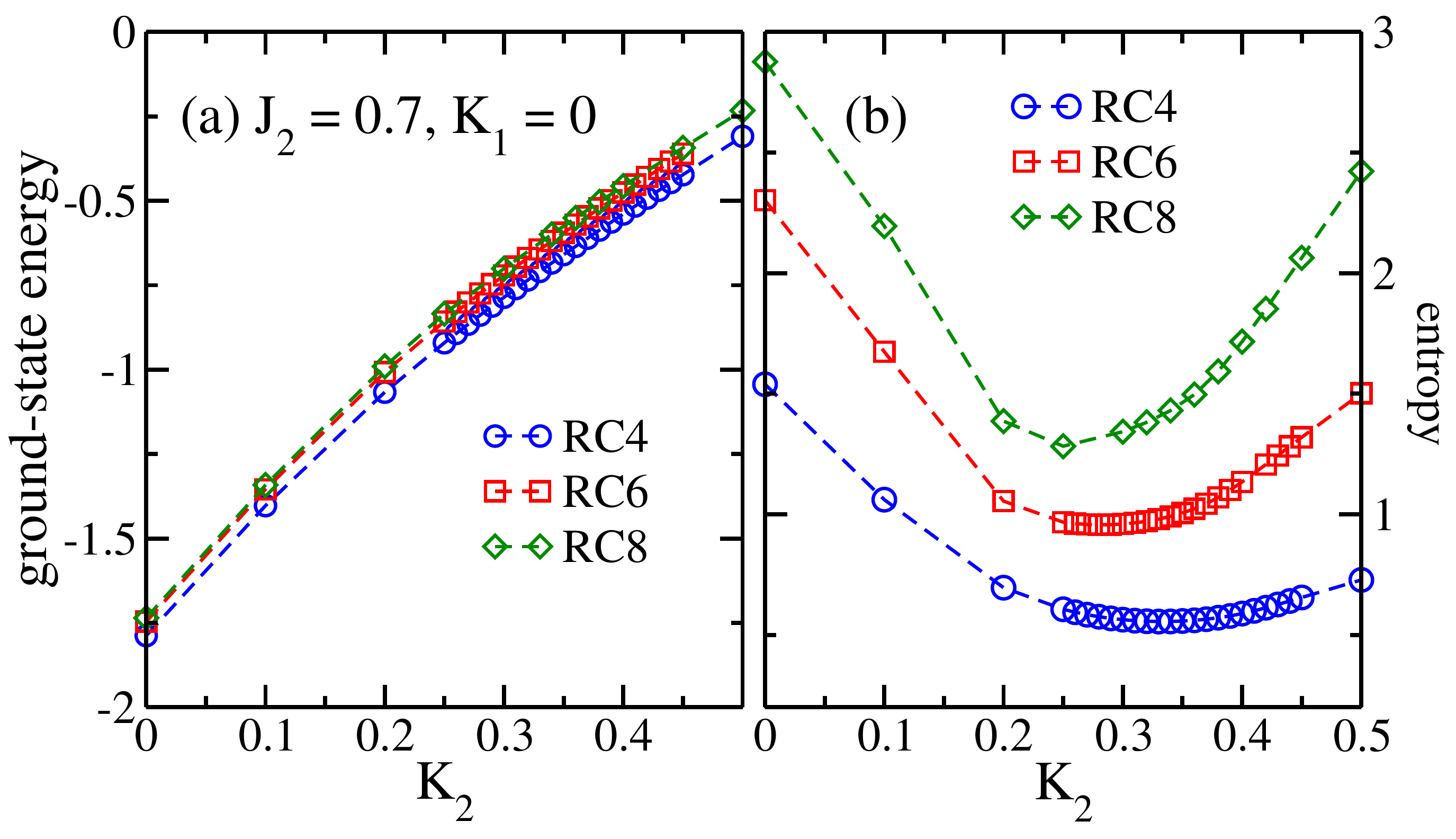}
    \caption{(Color online) $K_2$ coupling dependence of (a) ground-state energy and (b) entanglement entropy
     for $J_2 = 0.7, K_1 = 0.0$ in different long cylinders with $L_x = 24$.}
\label{supfig:transition_k1k2}
\end{figure}

In Fig.~\ref{supfig:transition_k1k2}, we show the $K_2$ coupling dependence of the ground-state energy
and entanglement entropy in the bulk of cylinder for $J_2 = 0.7, K_1 = 0.0$ on different cylinders. In the main
text, we show that the system has an intermediate phase for $0.34 \lesssim K_2 \lesssim 0.4$. Here, we find
that both the ground-state energy and the entropy appear smooth near the phase boundaries, which
indicates possible continuous transitions. Generally, a direct phase transition from N\'eel to
stripe AFM phase would be first order in Landau's paradigm. The smooth transition behaviors could be
compatible with an intermediate paramagnetic phase between the two magnetic ordered phases.

\begin{figure}
\includegraphics[width = 0.49\linewidth]{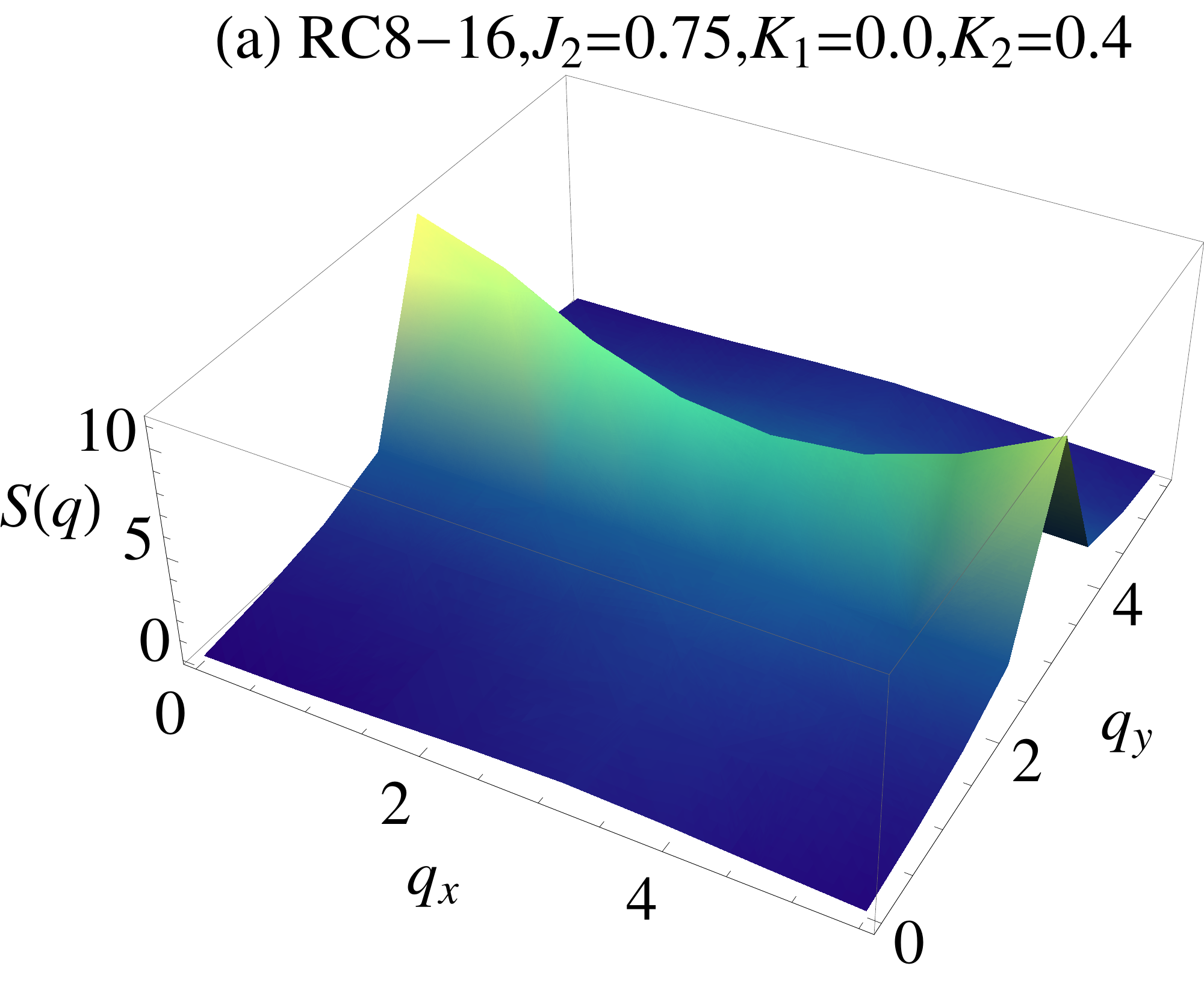}
\includegraphics[width = 0.49\linewidth]{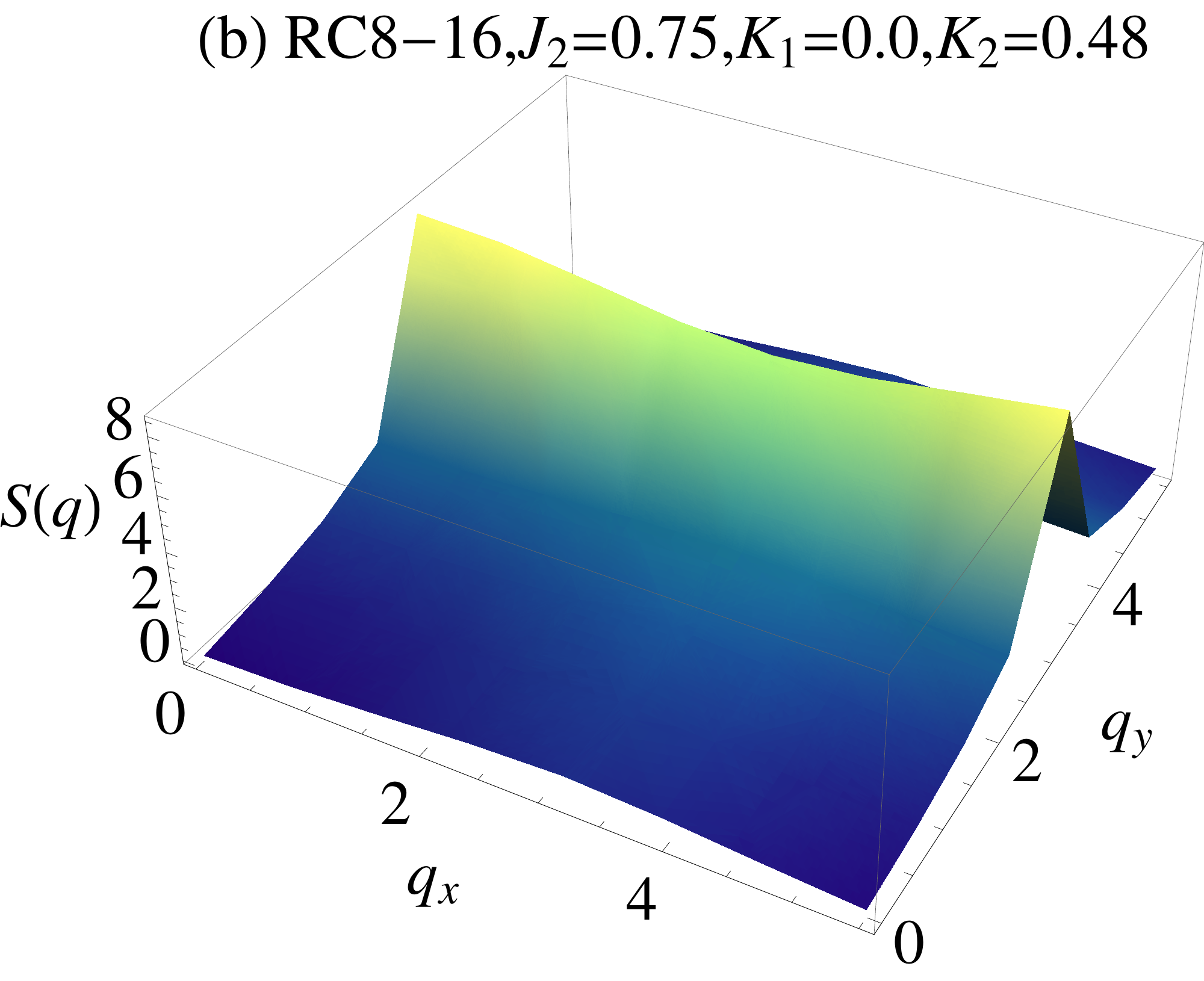}
\includegraphics[width = 0.49\linewidth]{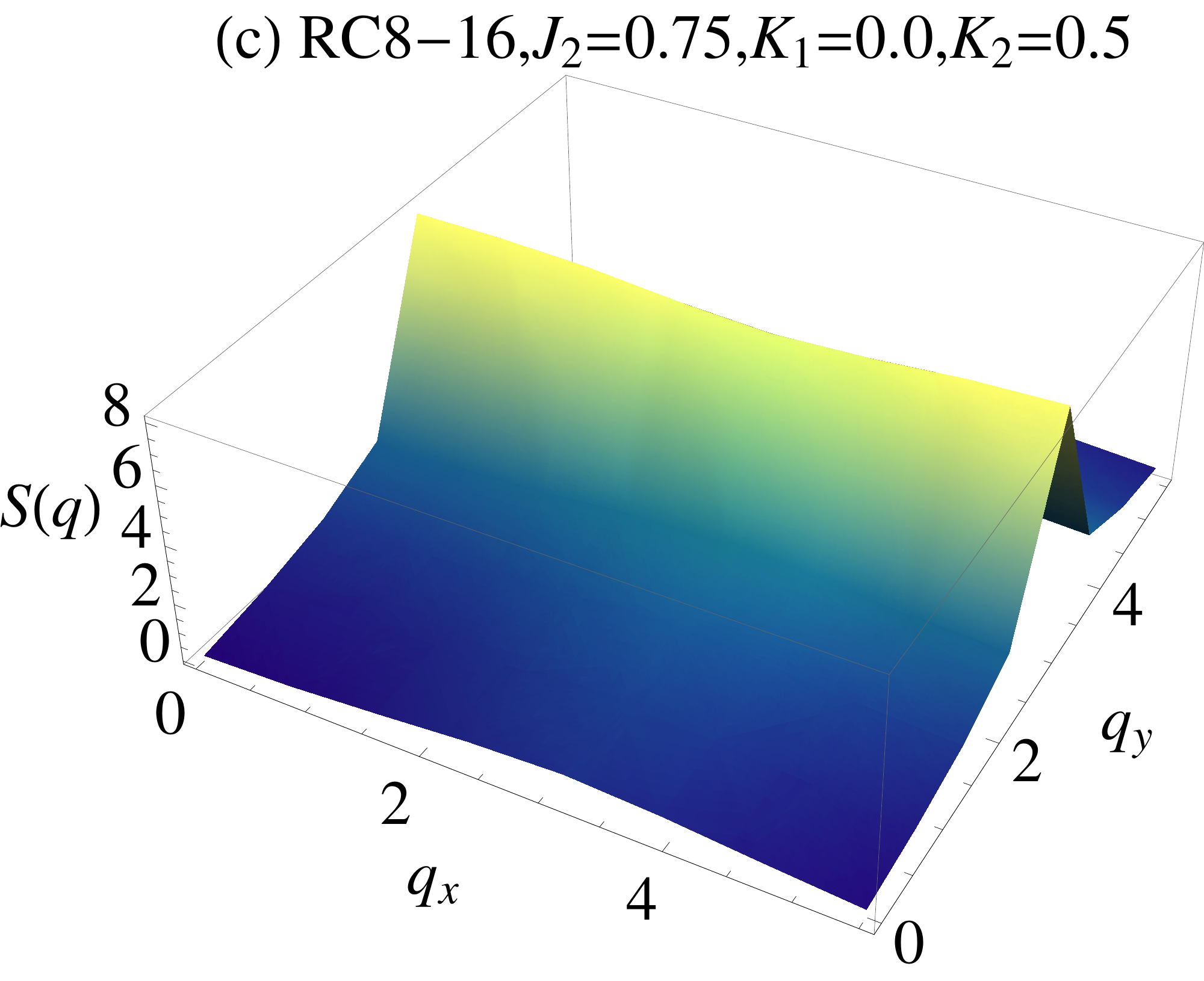}
\includegraphics[width = 0.49\linewidth]{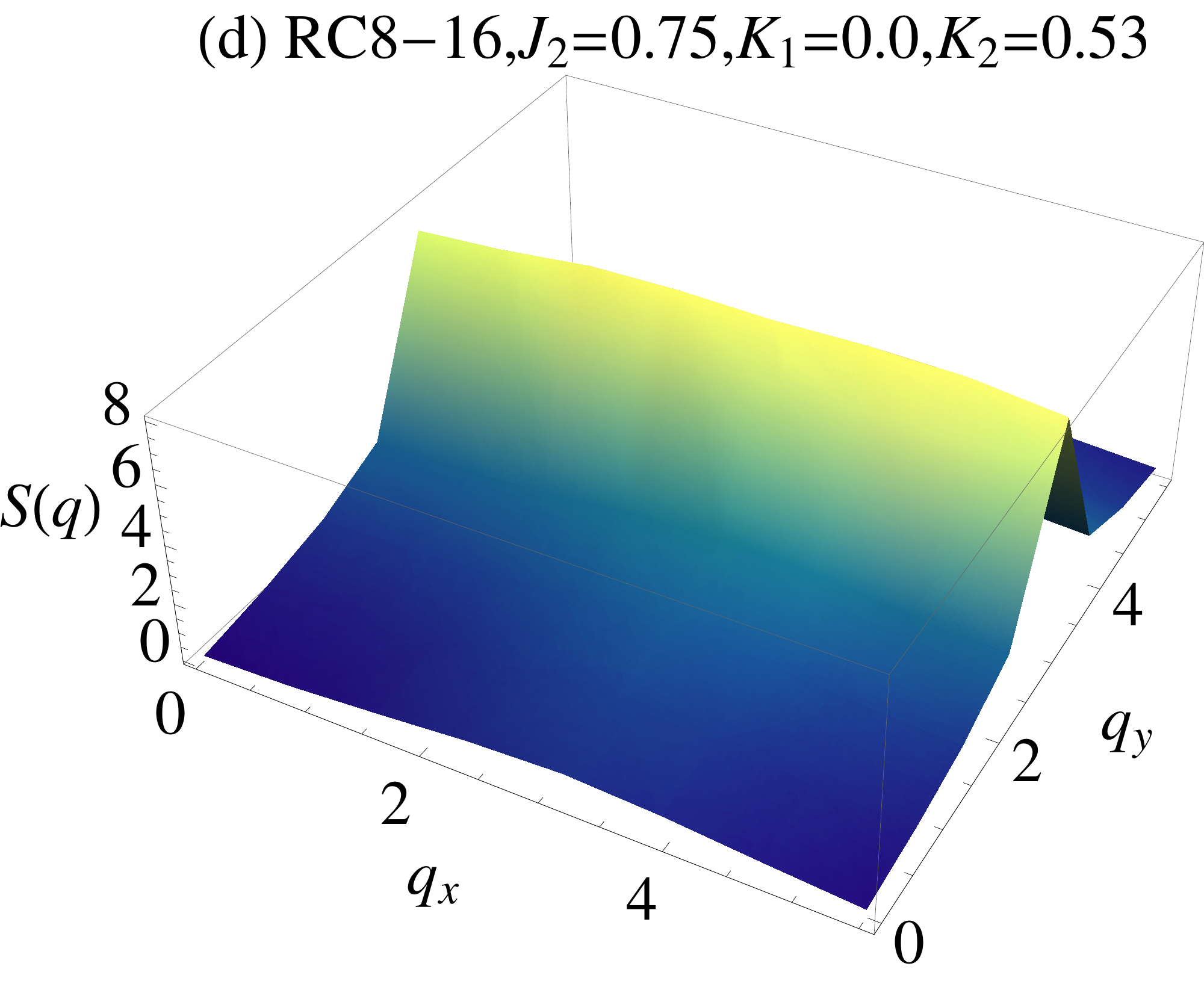}
\includegraphics[width = 0.49\linewidth]{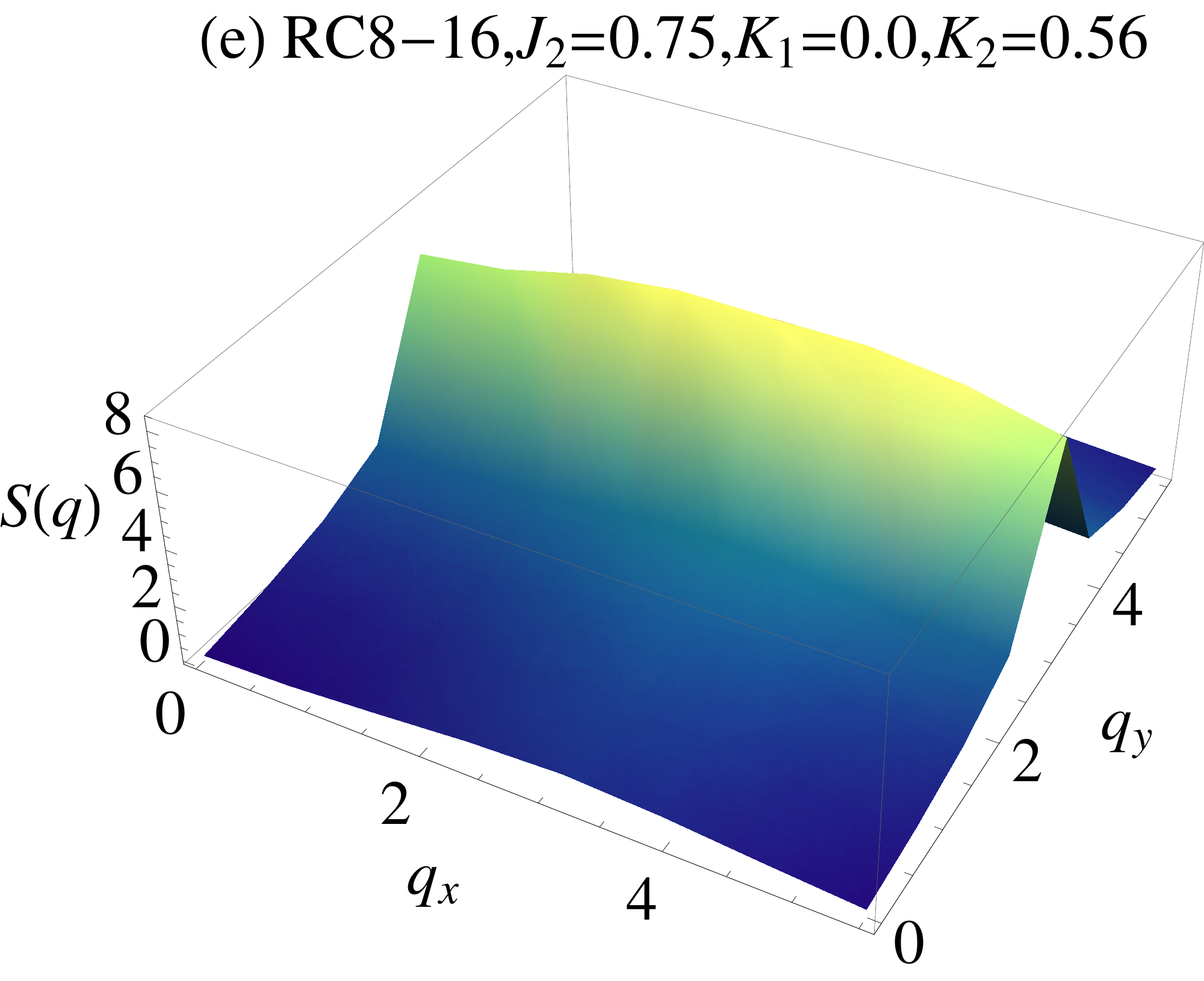}
\includegraphics[width = 0.49\linewidth]{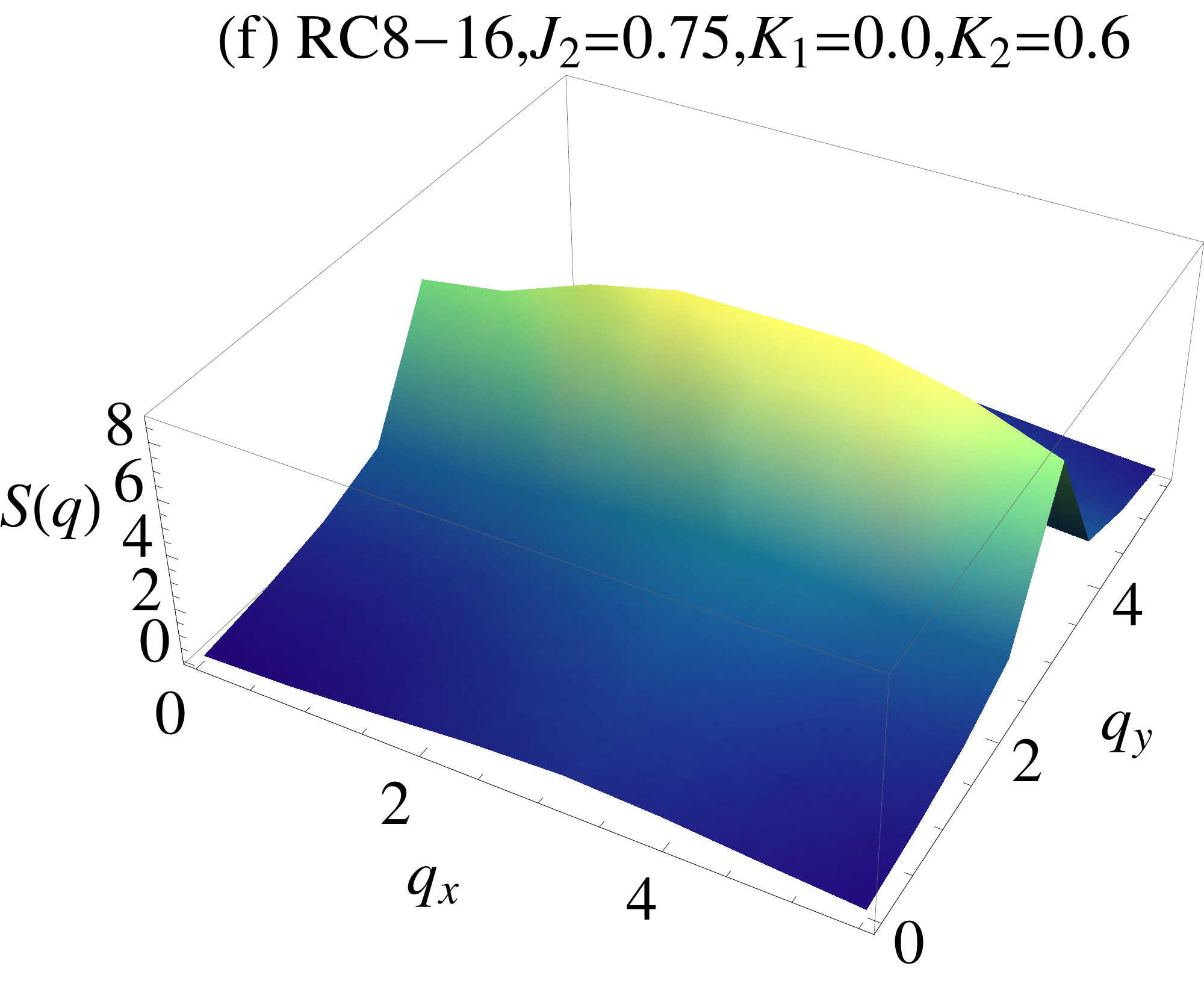}
\caption{(Color online) Spin structure factor $S(\vec{q})$ for $J_2 = 0.75, K_1 = 0.0$ and different $K_2$ on
the RC8-16 cylinder. The structure factor is obtained by the Fourier transform from the spin correlations of the
middle $8\times 8$ sites. For $K_2 = 0.4$, $S(\vec{q})$ has the stripe characteristic peak at $\vec{q} = (0,\pi)$. For $0.48
\lesssim K_2 \lesssim 0.6$, $S(\vec{q})$ is featureless, consistent with the non-magnetic intermediate phase.
}
\label{supfig:sq_075}
\end{figure}
\begin{figure}
\includegraphics[width = 0.49\linewidth]{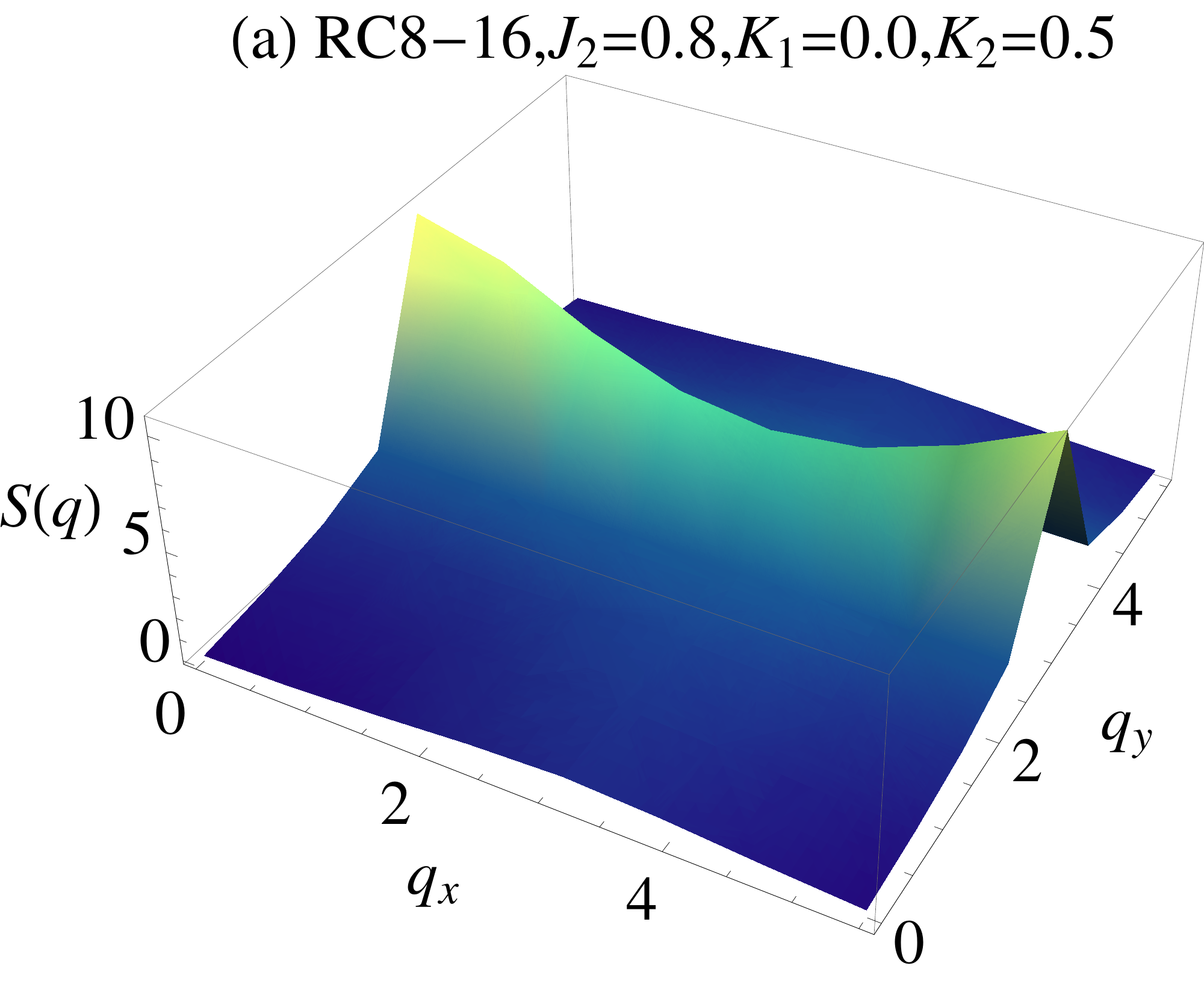}
\includegraphics[width = 0.49\linewidth]{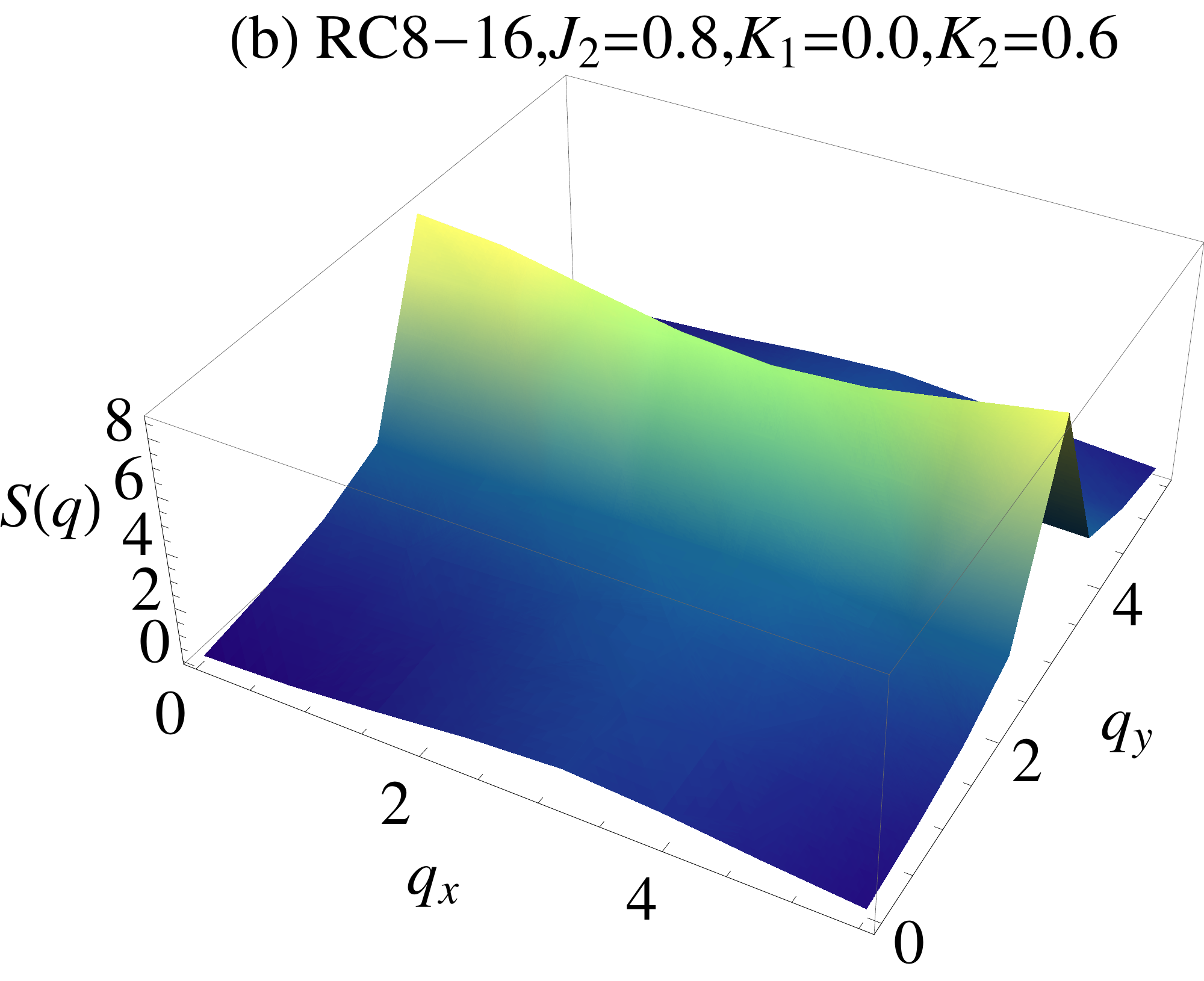}
\includegraphics[width = 0.49\linewidth]{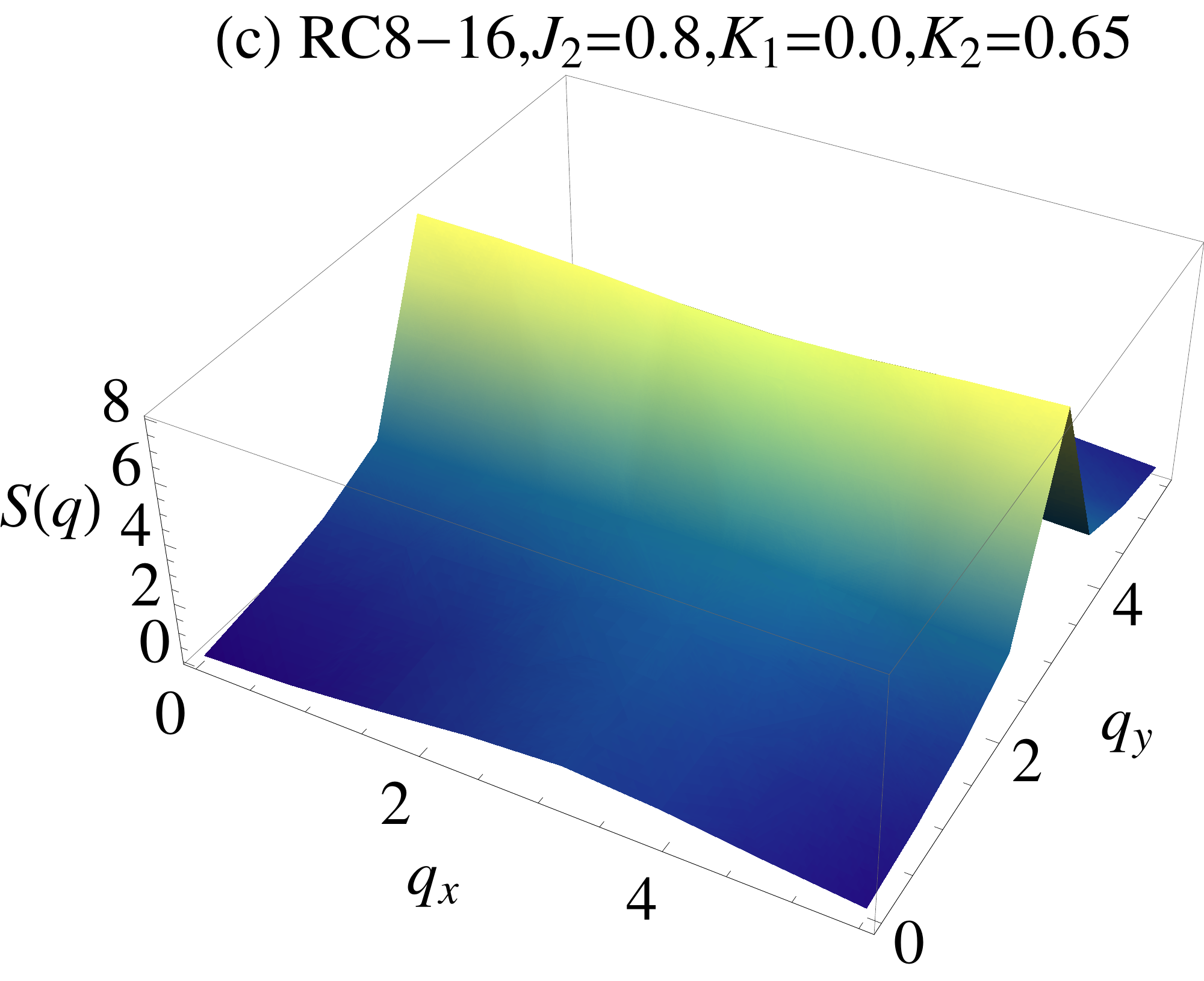}
\includegraphics[width = 0.49\linewidth]{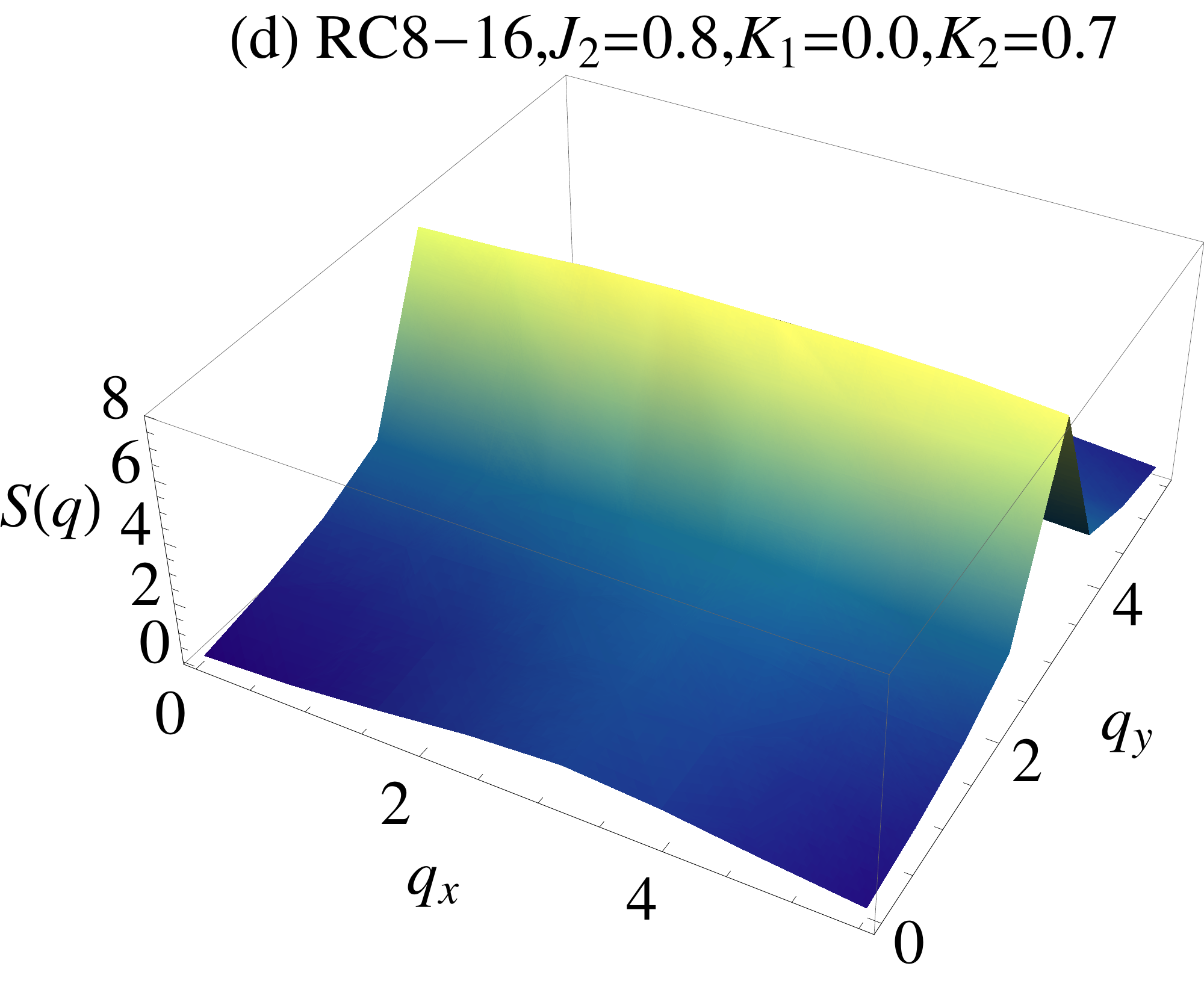}
\includegraphics[width = 0.49\linewidth]{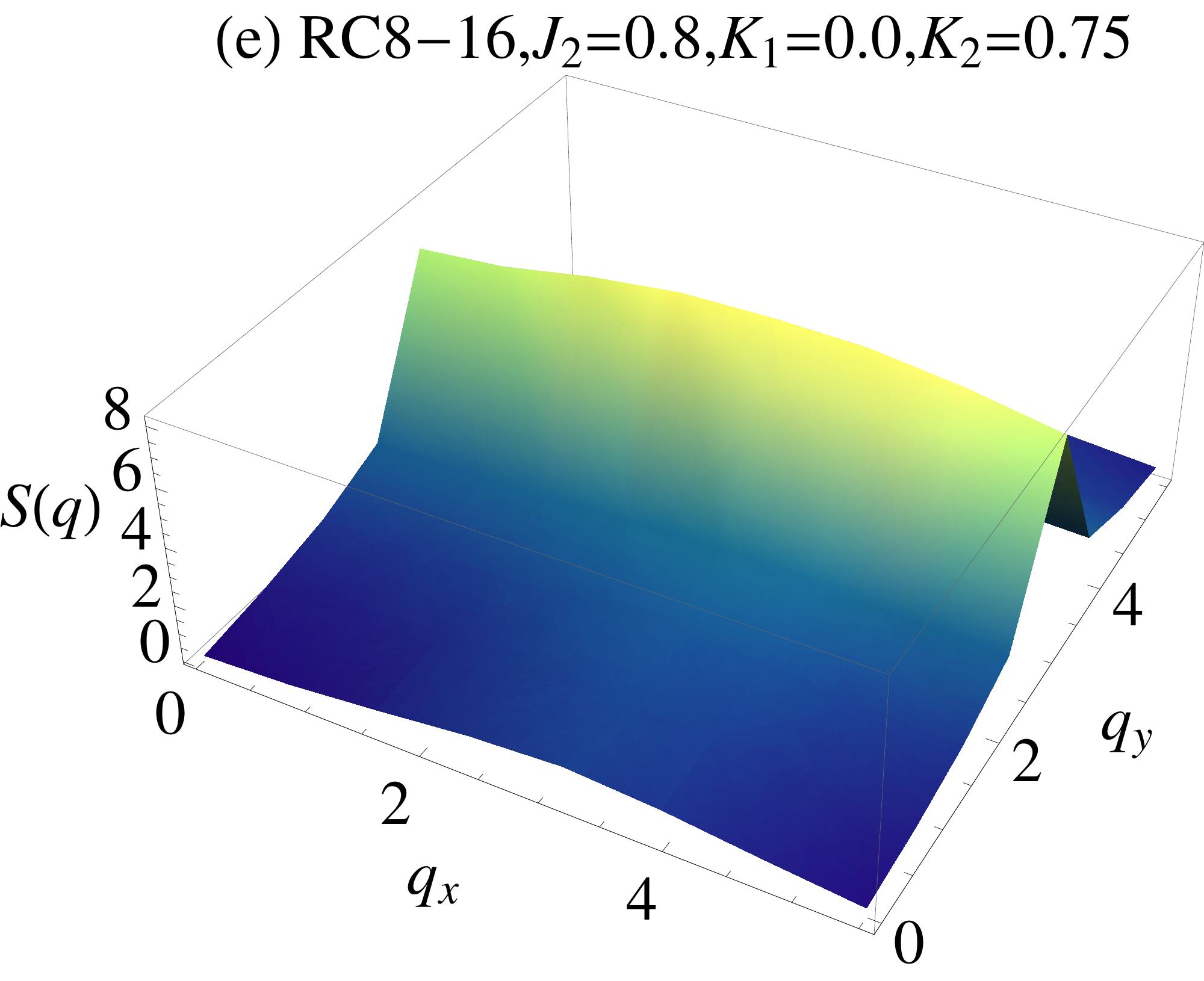}
\caption{(Color online) Spin structure factor $S(\vec{q})$ for $J_2 = 0.8, K_1 = 0.0$ and different $K_2$ on
the RC8-16 cylinder. We obtain the data following the way described in the caption of Fig.~\ref{supfig:sq_075}.
Here, for $J_2 = 0.8$, we also find the featureless $S(\vec{q})$ for $0.6 \lesssim K_2 \lesssim 0.75$.
}
\label{supfig:sq_08}
\end{figure}
\begin{figure}
\includegraphics[width = 0.8\linewidth]{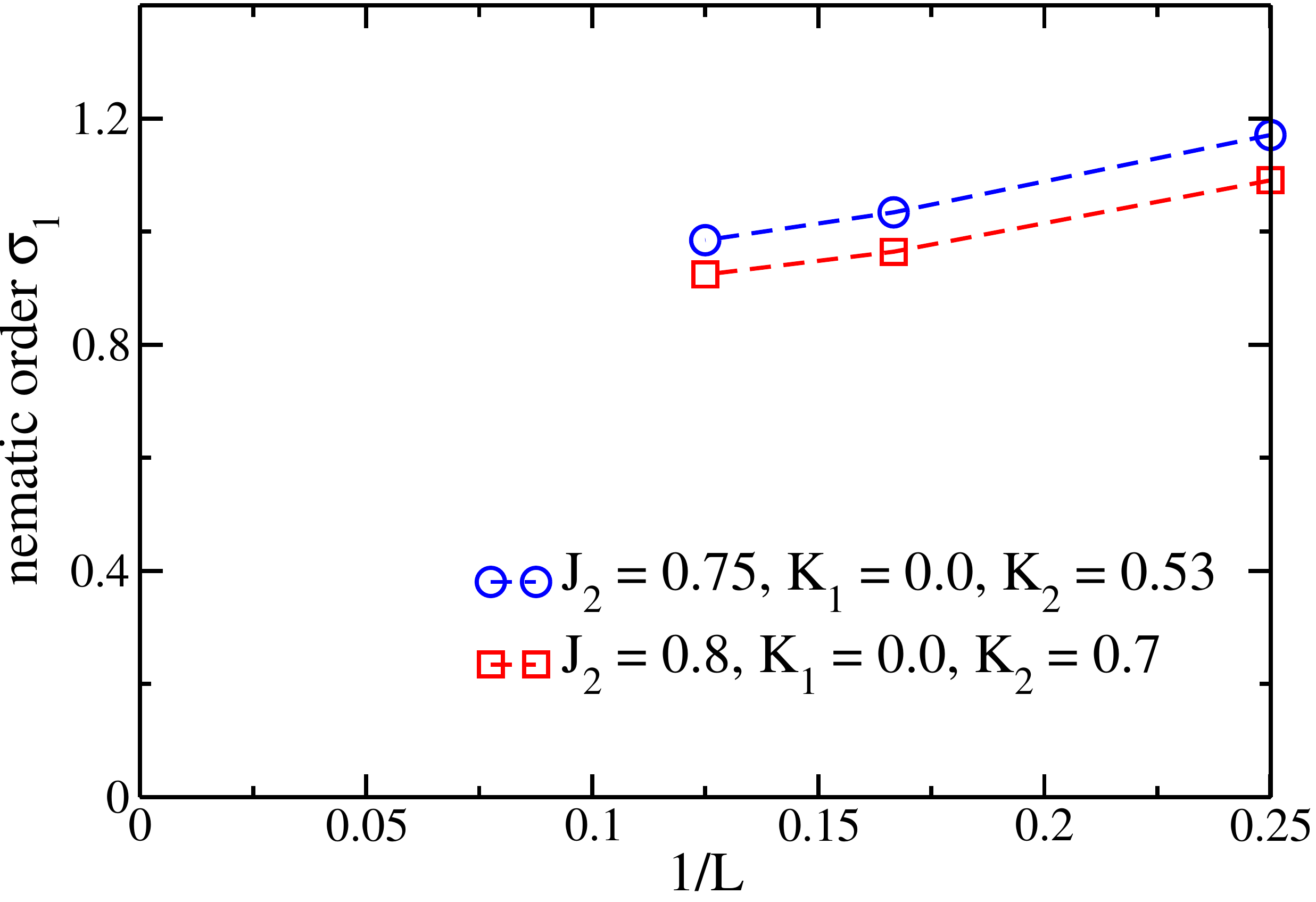}
\caption{(Color online) Size dependence of lattice nematic order $\sigma_1$ for the parameter points in the
intermediate phase regime for $J_2 = 0.75, 0.8, K_1 = 0.0$.
}
\label{supfig:nematic_j2}
\end{figure}

To demonstrate the stability of the intermediate phase, we also extend the studied
parameter regime to $J_2 = 0.75$ and $0.8$. Following the setup for $J_2 = 0.7$, we fix
$K_1 = 0.0$ and tune $K_2$. In Figs.~\ref{supfig:sq_075} and \ref{supfig:sq_08}, we show
the magnetic spin dipole structure factor 
$S(q) = \frac{1}{N}\sum_{i,j}e^{iq\cdot(r_i - r_j)}\langle S_i \cdot S_j \rangle$ on the 
RC8-16 cylinder. For $J_2 = 0.75$, one can find that the spin structure factor 
is featureless for $0.48 \lesssim K_2 \lesssim 0.6$; and for $J_2 = 0.8$, the structure factor is featureless 
for $0.6 \lesssim K_2 \lesssim 0.75$. In the non-magnetic intermediate 
regime for $J_2 = 0.75, 0.8$, we also examine the quadrupolar order (not shown here), which 
exhits the same $(\pi,0)$ AFQ fluctuations as we find for $J_2 = 0.7$ in the intermediate phase. 
In Fig.~\ref{supfig:nematic_j2}, we also show the finite-size scaling of the nematic order 
$\sigma_1 \equiv \langle \vec{S}_i \cdot \vec{S}_{i+\hat{x}} \rangle - 
\langle \vec{S}_i \cdot \vec{S}_{i+\hat{y}} \rangle$ in the intermediate regime for $J_2 = 0.75$ 
and $0.8$. Consistently, the size scaling also indicates the finite nematic order. Therefore, our results
indicate that the non-magnetic nematic intermediate phase is stable by tuning $J_2$.

\section{$J_1$-$J_2$-$K_1$ square model}

\begin{figure}
\includegraphics[width = 0.8\linewidth]{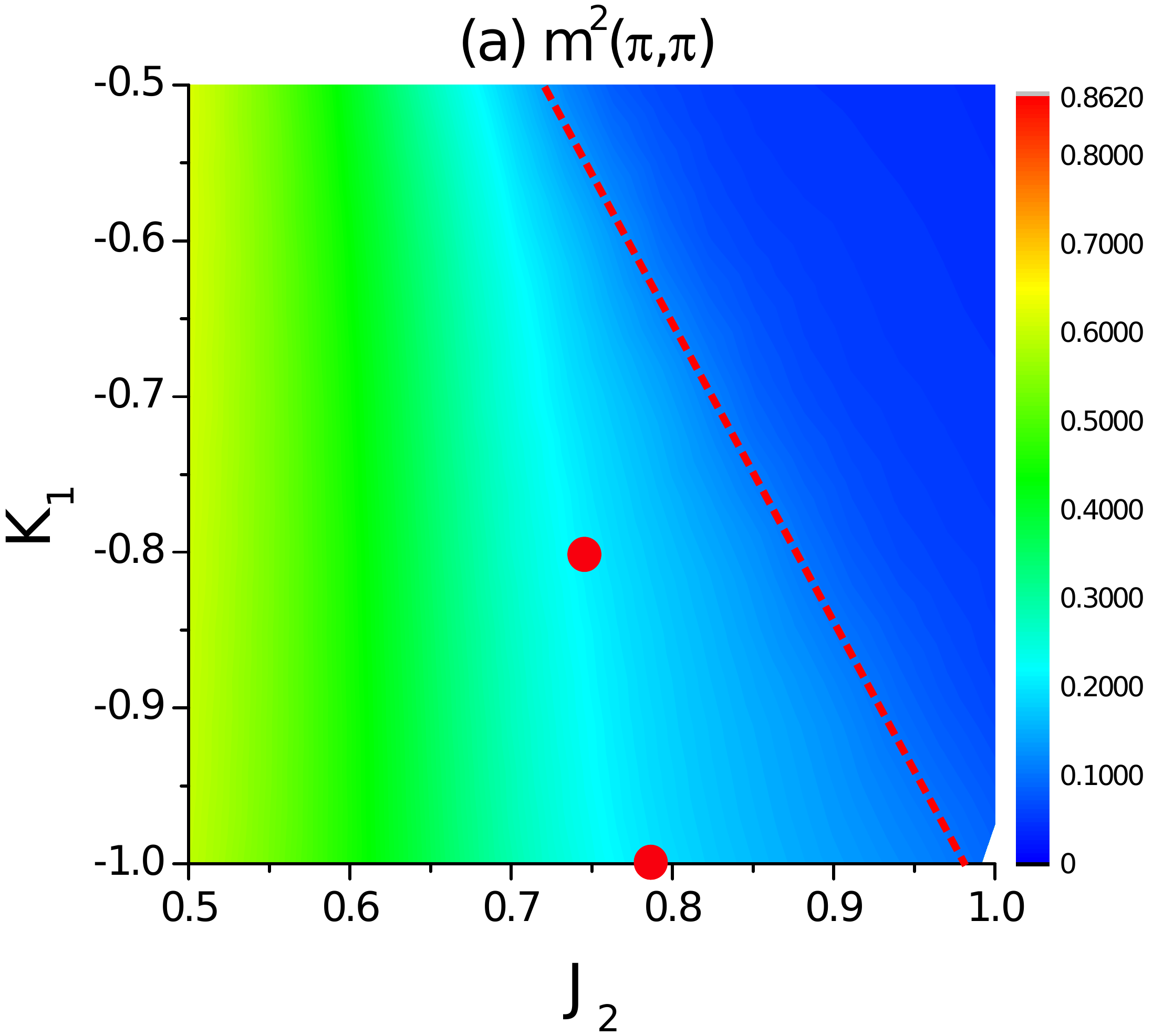}
\includegraphics[width = 0.8\linewidth]{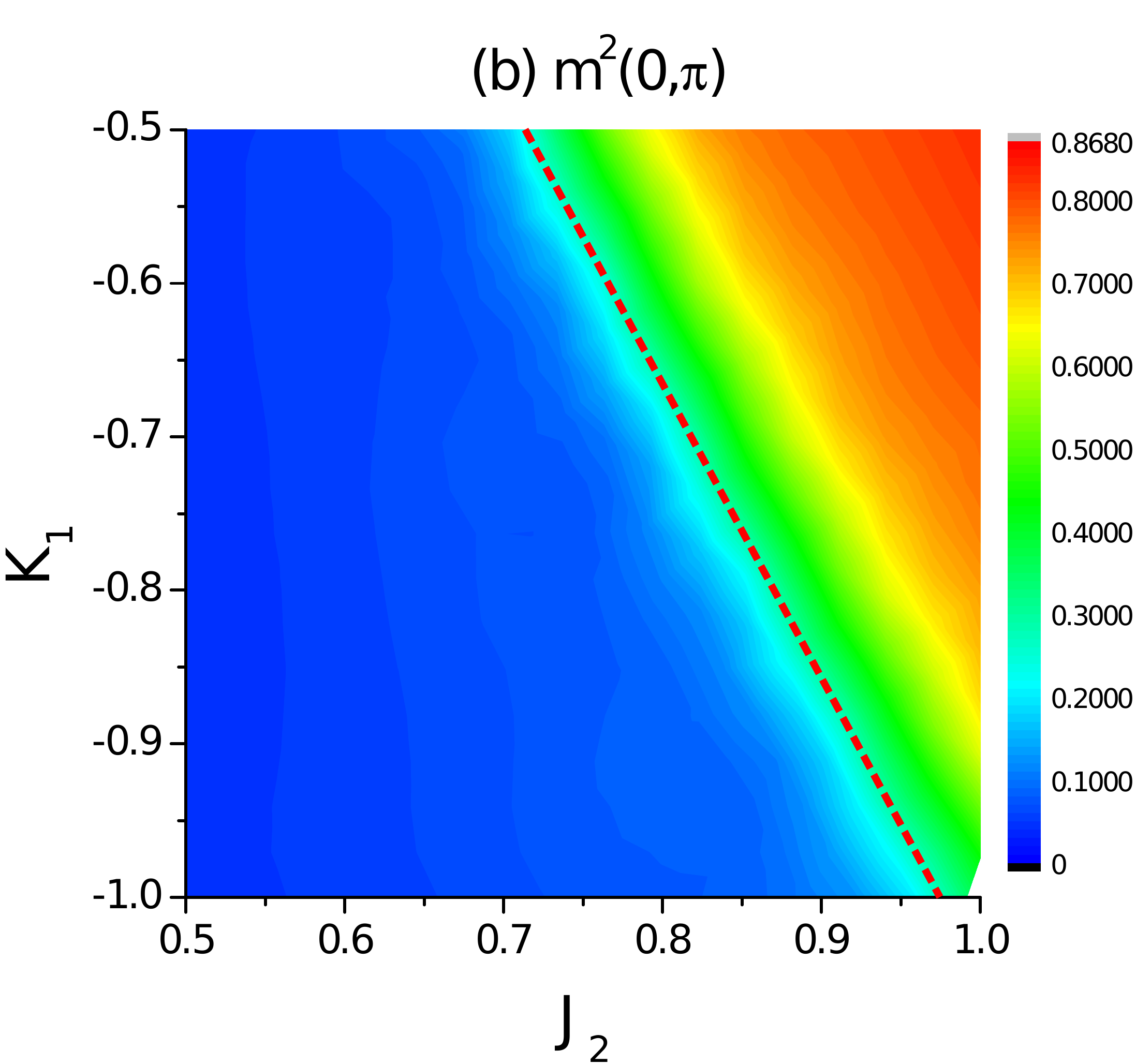}
\caption{(Color online) N\'{e}el AFM order parameter $m^2(\pi,\pi)$ (a)
and stripe AFM order parameter $m^2(0,\pi)$ (b) versus $J_2$ and $K_1$
interactions for the $J_1$-$J_2$-$K_1$ square model on the RC6-12 cylinder.
In both figures, the red dash line denotes the phase transition to the stripe AFM
order. The red dots in subfigure (a) denote the phase transition from N\'eel to the
intermediate ferroquadrupolar phase, which are determined from the finite-size scaling of
magnetic order parameters as shown in Fig.~\ref{supfig:m_scaling_j2k1}.}
\label{supfig:m_j2k1}
\end{figure}

{\it Magnetic orders.---} We show the magnetic order parameters on the RC6-12
cylinder for $0.5 \leq J_2 \leq 1.0, 0.5 \leq |K_1| \leq 1.0$ in Fig.~\ref{supfig:m_j2k1}.
We find that the stripe AFM order develops very fast above a critical $J_2$.
This phase transition is denoted by the red dash line in Fig.~\ref{supfig:m_j2k1}.
For the N\'eel phase, we can find that the blue regime with weak N\'eel order before
the transition to stripe phase in Fig.~\ref{supfig:m_j2k1}(a) is enlarged with increasing
$|K_1|$, which may indicate an intermediate regime.

\begin{figure}
\includegraphics[width = 1.0\linewidth]{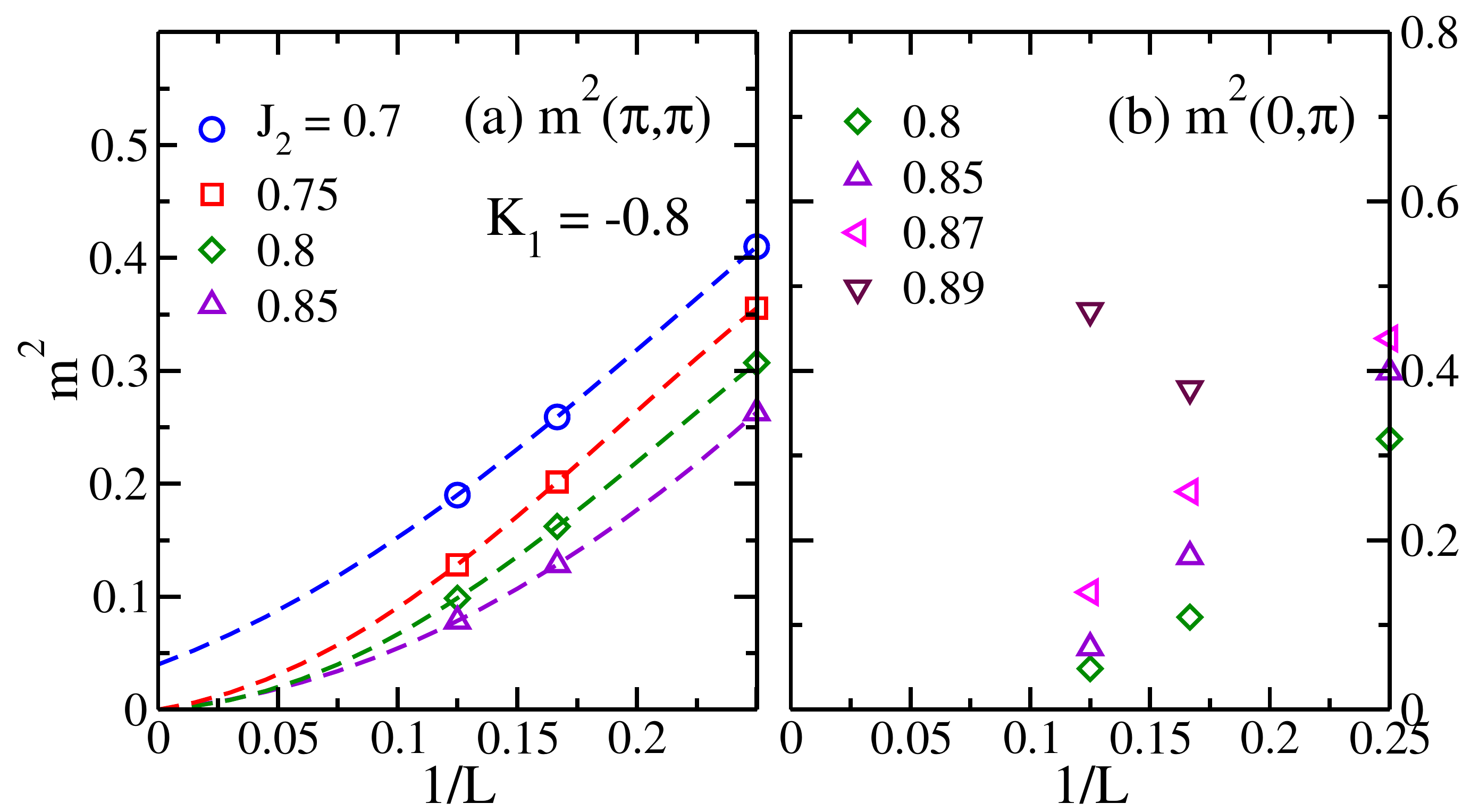}
\caption{(Color online) Finite-size scaling of magnetic order parameters for the
$J_1$-$J_2$-$K_1$ square model on the RC$L$-$2L$ cylinders with $L = 4, 6, 8$.
(a) and (b) are the N\'{e}el and stripe magnetic order parameters $m^2(\pi,\pi)$ and $m^2(0,\pi)$
versus $1/L$, respectively. Lines are polynomial fits.}
\label{supfig:m_scaling_j2k1}
\end{figure}

To determine whether there is an intermediate phase, we make finite-size scaling of
magnetic order parameters. In Fig.~\ref{supfig:m_scaling_j2k1}, we show the size scaling
of N\'eel and stripe order parameters for $K_1 = -0.8$ with increased $J_2$. Here, as the
convergence challenge in DMRG calculations in the intermediate regime, we only show
the data up to $L = 8$. Through the appropriate extrapolation, we find that the N\'eel
order vanishes at $J_2 \simeq 0.75$ and the stripe order develops at $J_2 \simeq 0.88$,
which give us the transition points shown in Fig.~\ref{supfig:m_j2k1}(a) and identify an
intermediate paramagnetic phase.

\begin{figure}
  \includegraphics[width = 0.52\linewidth]{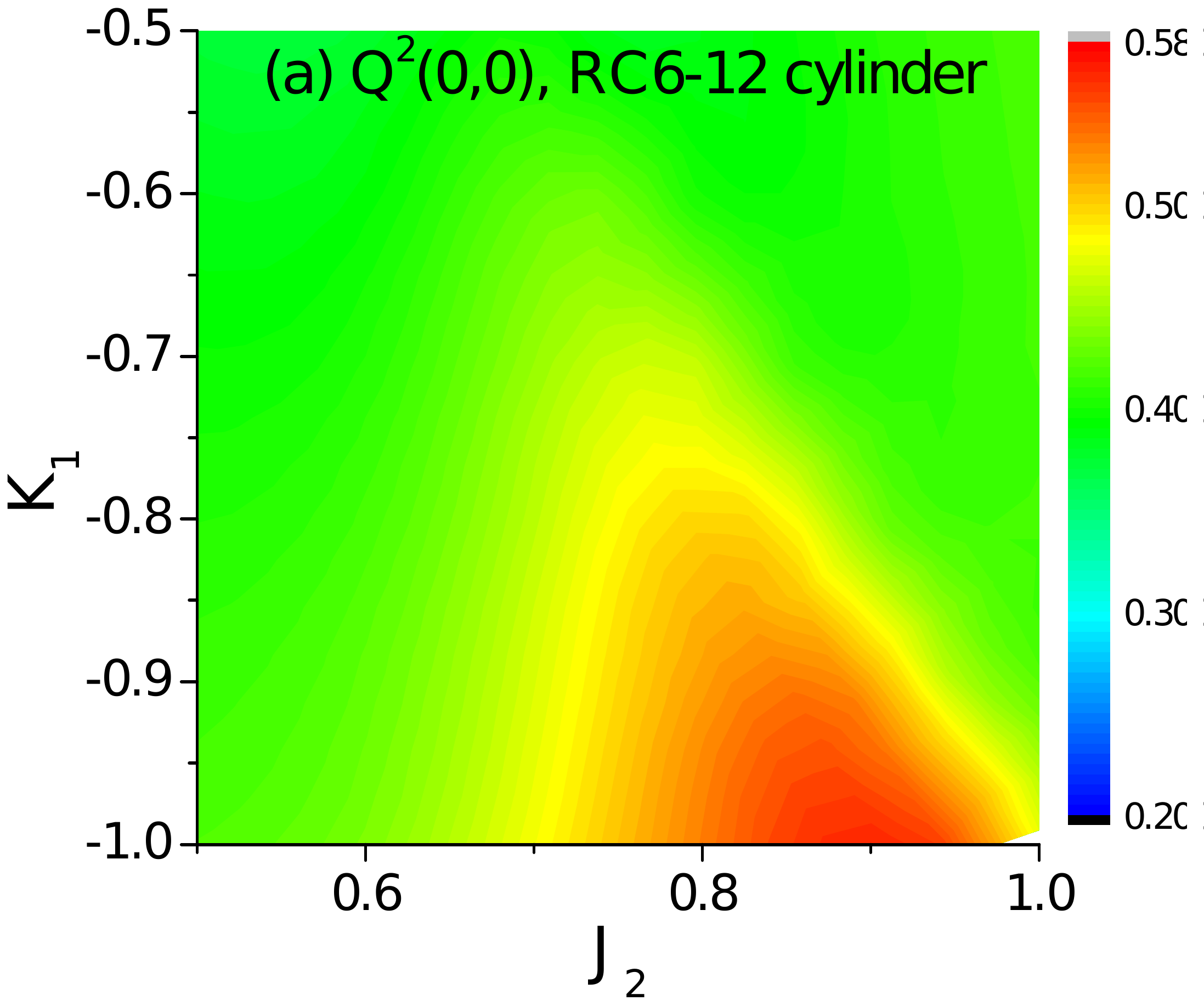}
  \includegraphics[width = 0.47\linewidth]{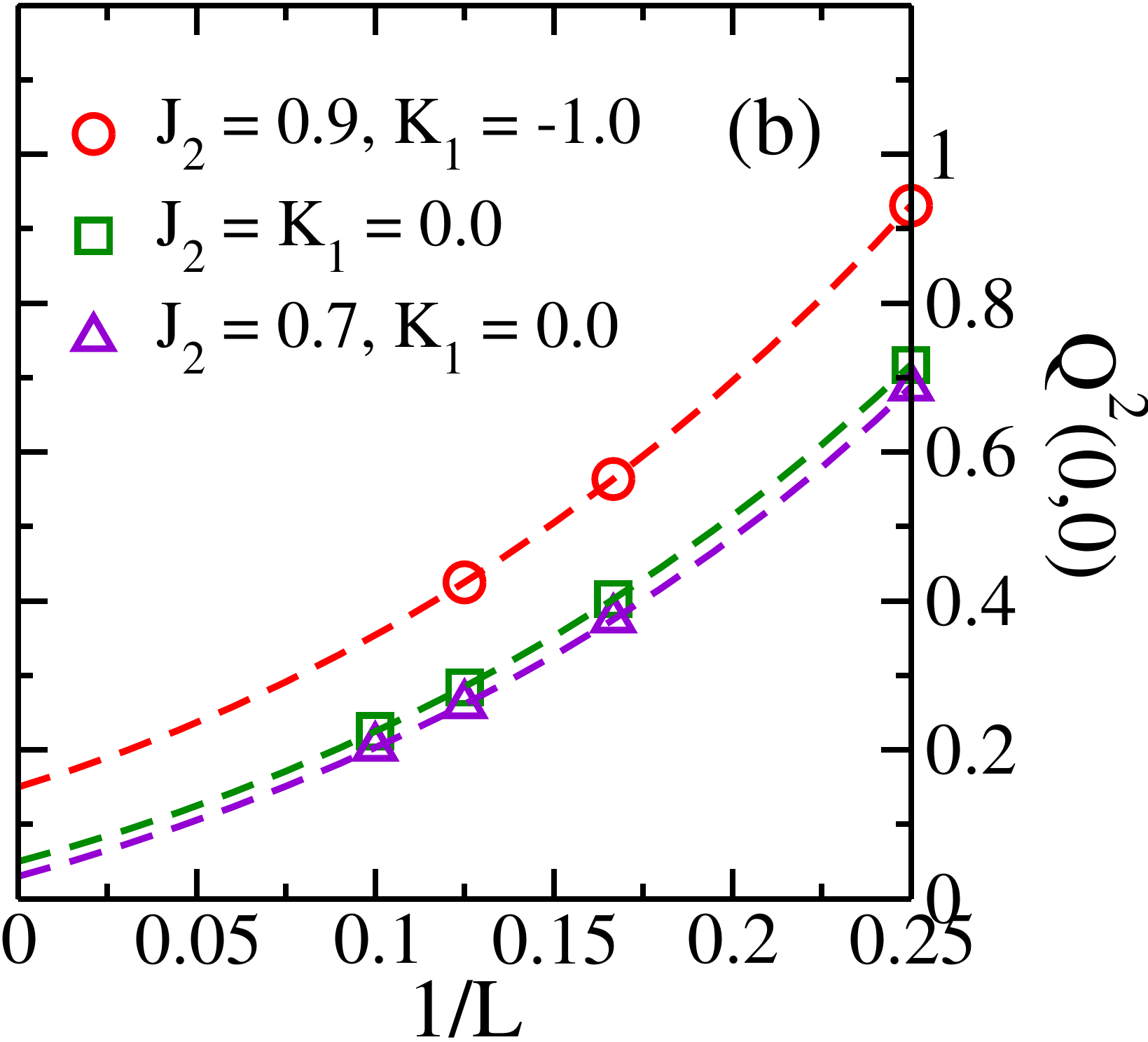}
  \includegraphics[width = 0.8\linewidth]{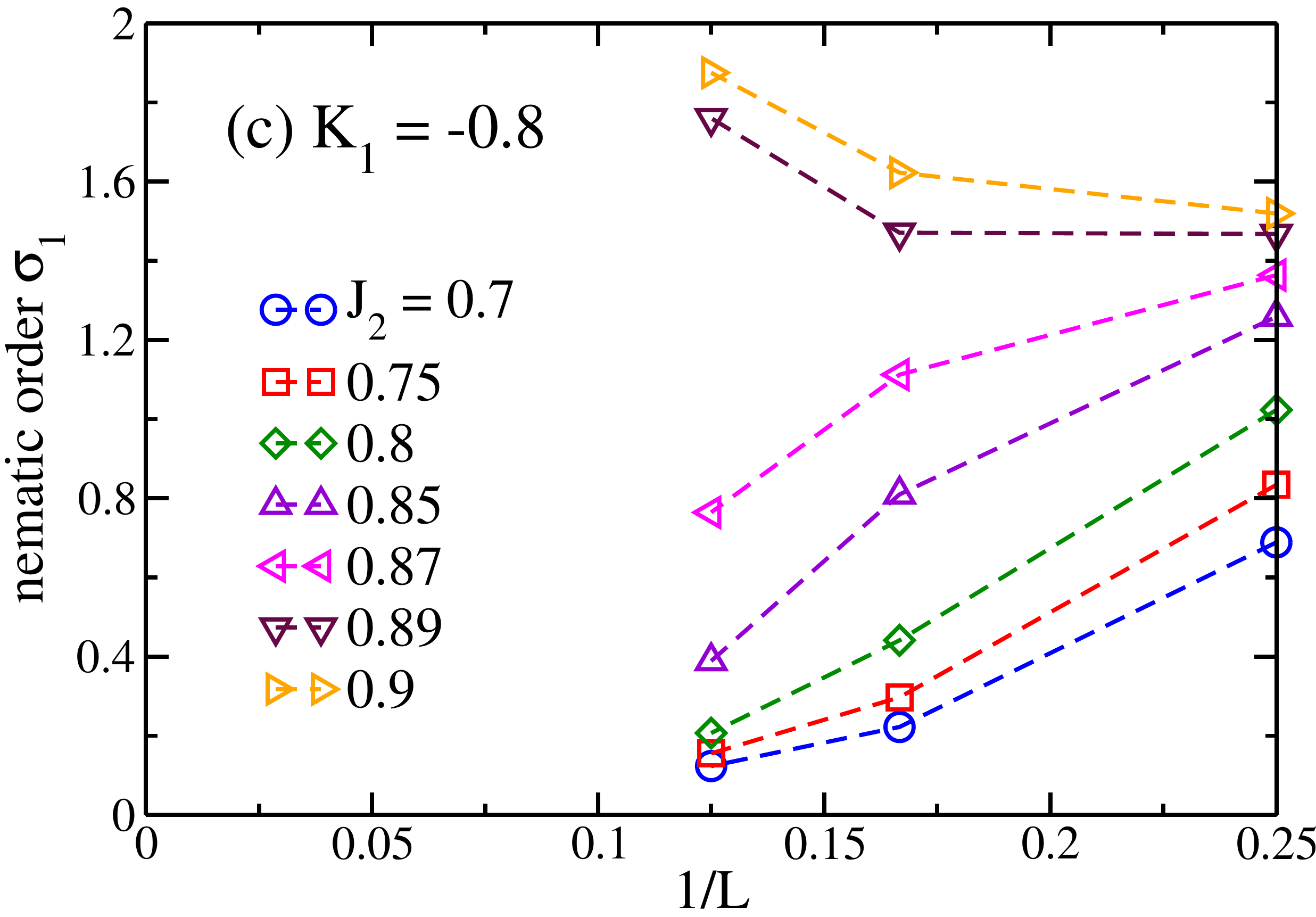}
  \caption{(Color online) FQ phase in the $J_1$-$J_2$-$K_1$ model.
    (a) $J_2, K_1$ dependence of FQ order parameter $Q^2(0,0)$ on RC6-12 cylinder.
    (b) Finite-size scaling of $Q^2(0,0)$ in different phases.
    (c) Finite-size scaling of lattice nematic order $\sigma_1$ for $K_1 = -0.8$ and different $J_2$.}
\label{supfig:fq}
\end{figure}

{\it Ferroquadrupolar phase.---} Next, we study ferroquadrupolar (FQ) order in the intermediate
phase. In Fig.~\ref{supfig:fq}(a), we show the $J_2, K_1$ coupling dependence of the FQ order
parameter $Q^2(0,0)$ on the RC6-12 cylinder. We can find the strong enhancement of $Q^2(0,0)$
in the large $J_2, |K_1|$ regime, which is consistent with the intermediate regime identified by studying
magnetic orders in Fig.~\ref{supfig:m_j2k1}. In Fig.~\ref{supfig:fq}(b), we show the finite-size scaling
of the FQ order, which unambiguously shows the finite FQ order in the thermodynamic limit. Thus,
the vanished magnetic order and finite FQ order identify this intermediate regime as a FQ phase.

{\it Preserved lattice symmetry.---} We also calcuate the nearest-neighbor bond energy $\langle \vec{S}_i \cdot \vec{S}_j \rangle$
to detect lattice symmetry breaking. We find that the bond energy is quite uniform in the bulk of cylinder, indicating the
translational invariance. In Fig.~\ref{supfig:fq}(c), we demonstrate the
size scaling of the bond nematic order $\sigma_1$ for $K_1 = -0.8$. Similar to the main
text, the nematic order $\sigma_1$ is defined as the difference between the horizontal and vertical
bond energy as $\sigma_1 = \langle \vec{S}_i \cdot \vec{S}_{i+\hat{x}} \rangle - \langle \vec{S}_i \cdot \vec{S}_{i+\hat{y}} \rangle$.
While $\sigma_1$ is strong and scales to finite value in the stripe AFM phase
for $J_2 \gtrsim 0.88$, it decays very fast to vanish in both the N\'{e}el
and FQ phases.

\section{Origin of biquadratic interaction}

We briefly discuss the origin of biquadratic coupling in our model.
Generally speaking, there are two different mechanisms to generate biquadratic interaction.
One is spin-phonon coupling or lattice distortion effect.
The other one is microscopic description of the isotropic non-Heisenberg Hamiltonian
extracted at the fourth order of perturbation from a Hubbard Hamiltonian.

\subsubsection{Phonon coupling}
As a phenomenological origin, one might think of the coupling between spin and
lattice degrees of freedom that results from the exchange
integrals on the atomic positions in a crystal. Since the exchange integrals
are linear functions of the displacement coordinates, while the elastic
energy of the deformation shows quadratic behaviour, a frustrated system
may gain energy by distorting the lattice. Alternatively, the competition between the
lattice distortion and the associated energy gain may lead to a quadratic coupling.
This effect is discussed in detail for the case of a single tetrahedral molecule
with four spins \cite{Mila}.
If we assume that the exchange integral for a pair of nearborhood spins $S_i$ and $S_j$ depends only
on the inter-atomic distance $r_{ij}$ (a reasonable assumption for direct exchange),
the elastic energy associated with a bond distortion can be
written as $\kappa \delta r_{ij}^2 /2$, where $\delta r_{ij}$ is the variation of the bond length and $\kappa$ is
the elastic constant. Thus we reach the so-called bond-phonon model:
\begin{equation}
H^{bp}=J\sum_{ij} (1-\alpha \delta r_{ij}) S_i \cdot S_j + \kappa \delta r_{ij}^2 /2
\end{equation}
where $\alpha$ is the spin-lattice coupling constant. Considering $\delta r_{ij}$
as independent parameters, we may integrate them out and find an effective spin
Hamiltonian:
\begin{equation}
H = J \sum_{\langle i,j\rangle}S_i\cdot S_j + K \sum_{\langle i,j\rangle}(S_i\cdot S_j)^2,
\end{equation}
where $K=-J\alpha/2\kappa$ is a dimensionless constant.
Here, based on the bond-phonon model, we get a quadratic interaction in addition to
the original Heisenberg spin exchange coupling, despite that this derivation is a semi-classic description \cite{Kittel}.

\subsubsection{Microscopic origin from Hubbard model}
We will derive an effective Hamiltonian for iron-based superconductor  based on simple arguments.
Since iron-based superconductors have six electrons occupying the nearly degenerate $3d$ Fe orbitals,
the system is intrinsically multi-orbital in microscopic Hamiltonian.
Band structure calculations on iron-based superconductors have shown the primary Fe orbitals are
$d_{xz}$, $d_{yz}$ and $d_{xy}$. Based the further approximation that the role of the $d_{xy}$
can be replaced by a next-near-neighbor hybridization between $d_{xz}$ and $d_{yz}$ orbtials,
we get a two-dimensional square lattice with two degenerated $d_{xz}$ and $d_{yz}$ orbitals per site,
which is proposed as minimal two-band model for iron-based superconductor \cite{Raghu2008}.
The itinerant electrons of the degenerated $d_{xz}$ and $d_{yz}$ orbitals
are described by a tight-binding Hamiltonian
\begin{equation}
H=H_t+H_{intra}+H_{inter}+H_{Hund},
\end{equation}
The itinerant electrons of the degenerate $d_{xz}$ and $d_{yz}$ orbitals are described by a tight-binding Hamiltonian,
\begin{equation}
H_t=\sum_{(ij),(\alpha\beta),\sigma} t_{ij,\alpha\beta} c^{\dagger}_{i,\alpha,\sigma} c_{j,\beta,\sigma}+ h.c.
\end{equation}
where $c^{\dagger}_{i,\alpha,\sigma}$ creates an electron with spin $\sigma$ at site $i$
on orbital $\alpha=d_{xz(yz)}$. For simplicity, we first assume $t_{ij,\alpha\alpha}=t$ for nearest neighbors.
We define the intraband Hubbard interaction $H_U$ and interband Hubbard interaction $H_V$ as
\begin{eqnarray}
H_{intra}= U \sum_{i,\alpha} n_{i,\alpha,\uparrow} n_{i,\alpha,\downarrow}, \\
H_{inter}= V \sum_{i,\sigma,\sigma'} n_{i,\alpha,\sigma} n_{i,\beta,\sigma'}
\end{eqnarray}
and the Hund's rule coupling as
\begin{equation}
H_{Hund}= - J_H \sum_{i,\alpha,\beta} \left[ c^{\dagger}_{i,\alpha,\uparrow} c_{i,\alpha,\downarrow} c^{\dagger}_{i,\beta,\downarrow} c_{i,\beta,\uparrow} +h.c. \right],
\end{equation}
where the Hund coupling ensures that two electrons forming a spin triplet if they occupying different orbitals on the same site.

To derive an effective Hamilitonan, let us first consider the limit of strong interaction defined by $U\neq 0$, $J_H\neq 0$ and $t=0$.
For one-site, the ground-state manifold is spanned by configurations with
two electrons on each site, one in each orbital, and the two electrons of
a given site forming a triplet. Thus, the spin-1 model is likely suitable to describe the iron-based superconductor,
which also matches the very recent neutral scattering measurements on FeSe samples \cite{zhao2015}.
Next we consider two-sites.
Two $S = 1$ spins can be combined into a total $S=2, 1, 0$
with the corresponding levels $5-$, and $3-$fold degenerate, and non-degenerated,
where we labeled as $|S,S^z\rangle$ and the total spin $S$ and
its a-component $S^z$ are good quantum numbers.
When a small hopping $t$ is added, the fluctuations will lift the groundstate degeneracy and favor the spin singlet state.
Here, the discussion is parallel to the case of the simple $e_g$ molecule with two orbitals in each site \cite{Fazekas_Book}.
We just quote the results, up to fourth-order perturbation $\propto t^4$:
\begin{widetext}
\begin{eqnarray}
H_{eff} = \left(\frac{2t^2}{U+J_H}-\frac{8t^4}{(U+J_H)^3} \right) S_i \cdot S_j
+ \frac{12t^4}{(U+J_H)^3} \left( \frac{1}{U+J_H}  -  \frac{2}{2(U+V)+J_H}
- \frac{2}{2(U-V)+J_H}  \right) P_{S=0},
\end{eqnarray}
\end{widetext}
where $S_i$ is spin-1 operator and $P_{S=0}$ projects to the spin single state:
\begin{equation}
P_{S=0}=\frac{1}{3}((S_i \cdot S_j)^2-1 ).
\end{equation}
Finally, we get the bilinear-biquadratic exchange Hamiltonian as
\begin{equation}
H_{eff}= J \sum_{\langle i,j\rangle}S_i\cdot S_j + K \sum_{\langle i,j\rangle}(S_i\cdot S_j)^2.
\end{equation}
For an isolated Fe atom, the intraband interaction $U$ and interband interaction $V$ are similar in magnitude,
while Hund coupling $J_H$ is an order smaller. Thus, a reasonable estimate is $J>0$ and $K<0$ for iron-based superconductor.
This model can be also extended to the next-nearest-neighbors, thus we have the $J_1$-$J_2$-$K_1$-$K_2$ model
as the start point,
\begin{eqnarray}
H_{eff} &=& J_1 \sum_{\langle i,j\rangle} S_i\cdot S_j  +J_2 \sum_{\langle\langle i,j\rangle\rangle}S_i\cdot S_j \nonumber \\
&+& K_1 \sum_{\langle i,j\rangle}(S_i\cdot S_j)^2 + K_2 \sum_{\langle\langle i,j\rangle\rangle}(S_i\cdot S_j)^2.
\end{eqnarray}

\bibliography{square}

\begin{thebibliography}{71}
\expandafter\ifx\csname natexlab\endcsname\relax\def\natexlab#1{#1}\fi
\expandafter\ifx\csname bibnamefont\endcsname\relax
  \def\bibnamefont#1{#1}\fi
\expandafter\ifx\csname bibfnamefont\endcsname\relax
  \def\bibfnamefont#1{#1}\fi
\expandafter\ifx\csname citenamefont\endcsname\relax
  \def\citenamefont#1{#1}\fi
\expandafter\ifx\csname url\endcsname\relax
  \def\url#1{\texttt{#1}}\fi
\expandafter\ifx\csname urlprefix\endcsname\relax\def\urlprefix{URL }\fi
\providecommand{\bibinfo}[2]{#2}
\providecommand{\eprint}[2][]{\url{#2}}

\bibitem[{\citenamefont{{Balents}}(2010)}]{balents2010}
\bibinfo{author}{\bibfnamefont{L.}~\bibnamefont{{Balents}}},
  \bibinfo{journal}{\nat} \textbf{\bibinfo{volume}{464}}, \bibinfo{pages}{199}
  (\bibinfo{year}{2010}),
  \urlprefix\url{http://www.nature.com/nature/journal/v464/n7286/full/nature08%
917.html}.

\bibitem[{\citenamefont{{Savary} and {Balents}}(2016)}]{savary2016}
\bibinfo{author}{\bibfnamefont{L.}~\bibnamefont{{Savary}}} \bibnamefont{and}
  \bibinfo{author}{\bibfnamefont{L.}~\bibnamefont{{Balents}}},
  \bibinfo{journal}{ArXiv e-prints}  (\bibinfo{year}{2016}),
  \eprint{1601.03742}, \urlprefix\url{http://arxiv.org/abs/1601.03742}.

\bibitem[{\citenamefont{Affleck et~al.}(1987)\citenamefont{Affleck, Kennedy,
  Lieb, and Tasaki}}]{affleck1987}
\bibinfo{author}{\bibfnamefont{I.}~\bibnamefont{Affleck}},
  \bibinfo{author}{\bibfnamefont{T.}~\bibnamefont{Kennedy}},
  \bibinfo{author}{\bibfnamefont{E.~H.} \bibnamefont{Lieb}}, \bibnamefont{and}
  \bibinfo{author}{\bibfnamefont{H.}~\bibnamefont{Tasaki}},
  \bibinfo{journal}{Phys. Rev. Lett.} \textbf{\bibinfo{volume}{59}},
  \bibinfo{pages}{799} (\bibinfo{year}{1987}),
  \urlprefix\url{http://link.aps.org/doi/10.1103/PhysRevLett.59.799}.

\bibitem[{\citenamefont{Affleck et~al.}(1988)\citenamefont{Affleck, Kennedy,
  Lieb, and Tasaki}}]{affleck1988}
\bibinfo{author}{\bibfnamefont{I.}~\bibnamefont{Affleck}},
  \bibinfo{author}{\bibfnamefont{T.}~\bibnamefont{Kennedy}},
  \bibinfo{author}{\bibfnamefont{E.~H.} \bibnamefont{Lieb}}, \bibnamefont{and}
  \bibinfo{author}{\bibfnamefont{H.}~\bibnamefont{Tasaki}},
  \bibinfo{journal}{Communications in Mathematical Physics}
  \textbf{\bibinfo{volume}{115}}, \bibinfo{pages}{477} (\bibinfo{year}{1988}),
  \urlprefix\url{http://link.springer.com/article/10.1007%2FBF01218021?LI=true%
}.

\bibitem[{\citenamefont{Blume and Hsieh}(1969)}]{blume1969}
\bibinfo{author}{\bibfnamefont{M.}~\bibnamefont{Blume}} \bibnamefont{and}
  \bibinfo{author}{\bibfnamefont{Y.~Y.} \bibnamefont{Hsieh}},
  \bibinfo{journal}{Journal of Applied Physics} \textbf{\bibinfo{volume}{40}},
  \bibinfo{pages}{1249} (\bibinfo{year}{1969}),
  \urlprefix\url{http://scitation.aip.org/content/aip/journal/jap/40/3/10.1063%
/1.1657616}.

\bibitem[{\citenamefont{L\"auchli et~al.}(2006)\citenamefont{L\"auchli, Mila,
  and Penc}}]{lauchli2006}
\bibinfo{author}{\bibfnamefont{A.}~\bibnamefont{L\"auchli}},
  \bibinfo{author}{\bibfnamefont{F.}~\bibnamefont{Mila}}, \bibnamefont{and}
  \bibinfo{author}{\bibfnamefont{K.}~\bibnamefont{Penc}},
  \bibinfo{journal}{Phys. Rev. Lett.} \textbf{\bibinfo{volume}{97}},
  \bibinfo{pages}{087205} (\bibinfo{year}{2006}),
  \urlprefix\url{http://link.aps.org/doi/10.1103/PhysRevLett.97.087205}.

\bibitem[{\citenamefont{Yao and Kivelson}(2007)}]{yao2007}
\bibinfo{author}{\bibfnamefont{H.}~\bibnamefont{Yao}} \bibnamefont{and}
  \bibinfo{author}{\bibfnamefont{S.~A.} \bibnamefont{Kivelson}},
  \bibinfo{journal}{Phys. Rev. Lett.} \textbf{\bibinfo{volume}{99}},
  \bibinfo{pages}{247203} (\bibinfo{year}{2007}),
  \urlprefix\url{http://link.aps.org/doi/10.1103/PhysRevLett.99.247203}.

\bibitem[{\citenamefont{Grover and Senthil}(2011)}]{grover2011}
\bibinfo{author}{\bibfnamefont{T.}~\bibnamefont{Grover}} \bibnamefont{and}
  \bibinfo{author}{\bibfnamefont{T.}~\bibnamefont{Senthil}},
  \bibinfo{journal}{Phys. Rev. Lett.} \textbf{\bibinfo{volume}{107}},
  \bibinfo{pages}{077203} (\bibinfo{year}{2011}),
  \urlprefix\url{http://link.aps.org/doi/10.1103/PhysRevLett.107.077203}.

\bibitem[{\citenamefont{Xu et~al.}(2012)\citenamefont{Xu, Wang, Qi, Balents,
  and Fisher}}]{xu2012}
\bibinfo{author}{\bibfnamefont{C.}~\bibnamefont{Xu}},
  \bibinfo{author}{\bibfnamefont{F.}~\bibnamefont{Wang}},
  \bibinfo{author}{\bibfnamefont{Y.}~\bibnamefont{Qi}},
  \bibinfo{author}{\bibfnamefont{L.}~\bibnamefont{Balents}}, \bibnamefont{and}
  \bibinfo{author}{\bibfnamefont{M.~P.~A.} \bibnamefont{Fisher}},
  \bibinfo{journal}{Phys. Rev. Lett.} \textbf{\bibinfo{volume}{108}},
  \bibinfo{pages}{087204} (\bibinfo{year}{2012}),
  \urlprefix\url{http://link.aps.org/doi/10.1103/PhysRevLett.108.087204}.

\bibitem[{\citenamefont{Bieri et~al.}(2012)\citenamefont{Bieri, Serbyn,
  Senthil, and Lee}}]{bieri2012}
\bibinfo{author}{\bibfnamefont{S.}~\bibnamefont{Bieri}},
  \bibinfo{author}{\bibfnamefont{M.}~\bibnamefont{Serbyn}},
  \bibinfo{author}{\bibfnamefont{T.}~\bibnamefont{Senthil}}, \bibnamefont{and}
  \bibinfo{author}{\bibfnamefont{P.~A.} \bibnamefont{Lee}},
  \bibinfo{journal}{Phys. Rev. B} \textbf{\bibinfo{volume}{86}},
  \bibinfo{pages}{224409} (\bibinfo{year}{2012}),
  \urlprefix\url{http://link.aps.org/doi/10.1103/PhysRevB.86.224409}.

\bibitem[{\citenamefont{Lai}(2013)}]{lai2013}
\bibinfo{author}{\bibfnamefont{H.-H.} \bibnamefont{Lai}},
  \bibinfo{journal}{Phys. Rev. B} \textbf{\bibinfo{volume}{87}},
  \bibinfo{pages}{205131} (\bibinfo{year}{2013}),
  \urlprefix\url{http://link.aps.org/doi/10.1103/PhysRevB.87.205131}.

\bibitem[{\citenamefont{Nakatsuji et~al.}(2005)\citenamefont{Nakatsuji, Nambu,
  Tonomura, Sakai, Jonas, Broholm, Tsunetsugu, Qiu, and Maeno}}]{nakatsuji2005}
\bibinfo{author}{\bibfnamefont{S.}~\bibnamefont{Nakatsuji}},
  \bibinfo{author}{\bibfnamefont{Y.}~\bibnamefont{Nambu}},
  \bibinfo{author}{\bibfnamefont{H.}~\bibnamefont{Tonomura}},
  \bibinfo{author}{\bibfnamefont{O.}~\bibnamefont{Sakai}},
  \bibinfo{author}{\bibfnamefont{S.}~\bibnamefont{Jonas}},
  \bibinfo{author}{\bibfnamefont{C.}~\bibnamefont{Broholm}},
  \bibinfo{author}{\bibfnamefont{H.}~\bibnamefont{Tsunetsugu}},
  \bibinfo{author}{\bibfnamefont{Y.}~\bibnamefont{Qiu}}, \bibnamefont{and}
  \bibinfo{author}{\bibfnamefont{Y.}~\bibnamefont{Maeno}},
  \bibinfo{journal}{Science} \textbf{\bibinfo{volume}{309}},
  \bibinfo{pages}{1697} (\bibinfo{year}{2005}),
  \urlprefix\url{http://science.sciencemag.org/content/309/5741/1697}.

\bibitem[{\citenamefont{Cheng et~al.}(2011)\citenamefont{Cheng, Li, Balicas,
  Zhou, Goodenough, Xu, and Zhou}}]{cheng2011}
\bibinfo{author}{\bibfnamefont{J.~G.} \bibnamefont{Cheng}},
  \bibinfo{author}{\bibfnamefont{G.}~\bibnamefont{Li}},
  \bibinfo{author}{\bibfnamefont{L.}~\bibnamefont{Balicas}},
  \bibinfo{author}{\bibfnamefont{J.~S.} \bibnamefont{Zhou}},
  \bibinfo{author}{\bibfnamefont{J.~B.} \bibnamefont{Goodenough}},
  \bibinfo{author}{\bibfnamefont{C.}~\bibnamefont{Xu}}, \bibnamefont{and}
  \bibinfo{author}{\bibfnamefont{H.~D.} \bibnamefont{Zhou}},
  \bibinfo{journal}{Phys. Rev. Lett.} \textbf{\bibinfo{volume}{107}},
  \bibinfo{pages}{197204} (\bibinfo{year}{2011}),
  \urlprefix\url{http://link.aps.org/doi/10.1103/PhysRevLett.107.197204}.

\bibitem[{\citenamefont{Johnston}(2010)}]{johnston2010}
\bibinfo{author}{\bibfnamefont{D.~C.} \bibnamefont{Johnston}},
  \bibinfo{journal}{Advances in Physics} \textbf{\bibinfo{volume}{59}},
  \bibinfo{pages}{803} (\bibinfo{year}{2010}),
  \urlprefix\url{http://www.tandfonline.com/doi/abs/10.1080/00018732.2010.5134%
80}.

\bibitem[{\citenamefont{Stewart}(2011)}]{stewart2011}
\bibinfo{author}{\bibfnamefont{G.~R.} \bibnamefont{Stewart}},
  \bibinfo{journal}{Rev. Mod. Phys.} \textbf{\bibinfo{volume}{83}},
  \bibinfo{pages}{1589} (\bibinfo{year}{2011}),
  \urlprefix\url{http://link.aps.org/doi/10.1103/RevModPhys.83.1589}.

\bibitem[{\citenamefont{Dai}(2015)}]{dai2015}
\bibinfo{author}{\bibfnamefont{P.}~\bibnamefont{Dai}}, \bibinfo{journal}{Rev.
  Mod. Phys.} \textbf{\bibinfo{volume}{87}}, \bibinfo{pages}{855}
  (\bibinfo{year}{2015}),
  \urlprefix\url{http://link.aps.org/doi/10.1103/RevModPhys.87.855}.

\bibitem[{\citenamefont{Hsu et~al.}(2008)\citenamefont{Hsu, Luo, Yeh, Chen,
  Huang, Wu, Lee, Huang, Chu, Yan et~al.}}]{hsu2008}
\bibinfo{author}{\bibfnamefont{F.-C.} \bibnamefont{Hsu}},
  \bibinfo{author}{\bibfnamefont{J.-Y.} \bibnamefont{Luo}},
  \bibinfo{author}{\bibfnamefont{K.-W.} \bibnamefont{Yeh}},
  \bibinfo{author}{\bibfnamefont{T.-K.} \bibnamefont{Chen}},
  \bibinfo{author}{\bibfnamefont{T.-W.} \bibnamefont{Huang}},
  \bibinfo{author}{\bibfnamefont{P.~M.} \bibnamefont{Wu}},
  \bibinfo{author}{\bibfnamefont{Y.-C.} \bibnamefont{Lee}},
  \bibinfo{author}{\bibfnamefont{Y.-L.} \bibnamefont{Huang}},
  \bibinfo{author}{\bibfnamefont{Y.-Y.} \bibnamefont{Chu}},
  \bibinfo{author}{\bibfnamefont{D.-C.} \bibnamefont{Yan}},
  \bibnamefont{et~al.}, \bibinfo{journal}{Proceedings of the National Academy
  of Sciences} \textbf{\bibinfo{volume}{105}}, \bibinfo{pages}{14262}
  (\bibinfo{year}{2008}),
  \urlprefix\url{http://www.pnas.org/content/105/38/14262.short}.

\bibitem[{\citenamefont{Lee et~al.}(2006)\citenamefont{Lee, Nagaosa, and
  Wen}}]{lee2006}
\bibinfo{author}{\bibfnamefont{P.~A.} \bibnamefont{Lee}},
  \bibinfo{author}{\bibfnamefont{N.}~\bibnamefont{Nagaosa}}, \bibnamefont{and}
  \bibinfo{author}{\bibfnamefont{X.-G.} \bibnamefont{Wen}},
  \bibinfo{journal}{Rev. Mod. Phys.} \textbf{\bibinfo{volume}{78}},
  \bibinfo{pages}{17} (\bibinfo{year}{2006}),
  \urlprefix\url{http://link.aps.org/doi/10.1103/RevModPhys.78.17}.

\bibitem[{\citenamefont{McQueen et~al.}(2009)\citenamefont{McQueen, Williams,
  Stephens, Tao, Zhu, Ksenofontov, Casper, Felser, and Cava}}]{cava2009}
\bibinfo{author}{\bibfnamefont{T.~M.} \bibnamefont{McQueen}},
  \bibinfo{author}{\bibfnamefont{A.~J.} \bibnamefont{Williams}},
  \bibinfo{author}{\bibfnamefont{P.~W.} \bibnamefont{Stephens}},
  \bibinfo{author}{\bibfnamefont{J.}~\bibnamefont{Tao}},
  \bibinfo{author}{\bibfnamefont{Y.}~\bibnamefont{Zhu}},
  \bibinfo{author}{\bibfnamefont{V.}~\bibnamefont{Ksenofontov}},
  \bibinfo{author}{\bibfnamefont{F.}~\bibnamefont{Casper}},
  \bibinfo{author}{\bibfnamefont{C.}~\bibnamefont{Felser}}, \bibnamefont{and}
  \bibinfo{author}{\bibfnamefont{R.~J.} \bibnamefont{Cava}},
  \bibinfo{journal}{Phys. Rev. Lett.} \textbf{\bibinfo{volume}{103}},
  \bibinfo{pages}{057002} (\bibinfo{year}{2009}),
  \urlprefix\url{http://link.aps.org/doi/10.1103/PhysRevLett.103.057002}.

\bibitem[{\citenamefont{Medvedev et~al.}(2009)\citenamefont{Medvedev, McQueen,
  Troyan, Palasyuk, Eremets, Cava, Naghavi, Casper, Ksenofontov, Wortmann
  et~al.}}]{medvedev2009}
\bibinfo{author}{\bibfnamefont{S.}~\bibnamefont{Medvedev}},
  \bibinfo{author}{\bibfnamefont{T.}~\bibnamefont{McQueen}},
  \bibinfo{author}{\bibfnamefont{I.}~\bibnamefont{Troyan}},
  \bibinfo{author}{\bibfnamefont{T.}~\bibnamefont{Palasyuk}},
  \bibinfo{author}{\bibfnamefont{M.}~\bibnamefont{Eremets}},
  \bibinfo{author}{\bibfnamefont{R.}~\bibnamefont{Cava}},
  \bibinfo{author}{\bibfnamefont{S.}~\bibnamefont{Naghavi}},
  \bibinfo{author}{\bibfnamefont{F.}~\bibnamefont{Casper}},
  \bibinfo{author}{\bibfnamefont{V.}~\bibnamefont{Ksenofontov}},
  \bibinfo{author}{\bibfnamefont{G.}~\bibnamefont{Wortmann}},
  \bibnamefont{et~al.}, \bibinfo{journal}{Nature materials}
  \textbf{\bibinfo{volume}{8}}, \bibinfo{pages}{630} (\bibinfo{year}{2009}),
  \urlprefix\url{http://www.nature.com/nmat/journal/v8/n8/abs/nmat2491.html}.

\bibitem[{\citenamefont{Shimojima et~al.}(2014)\citenamefont{Shimojima, Suzuki,
  Sonobe, Nakamura, Sakano, Omachi, Yoshioka, Kuwata-Gonokami, Ono, Kumigashira
  et~al.}}]{shimojima2014}
\bibinfo{author}{\bibfnamefont{T.}~\bibnamefont{Shimojima}},
  \bibinfo{author}{\bibfnamefont{Y.}~\bibnamefont{Suzuki}},
  \bibinfo{author}{\bibfnamefont{T.}~\bibnamefont{Sonobe}},
  \bibinfo{author}{\bibfnamefont{A.}~\bibnamefont{Nakamura}},
  \bibinfo{author}{\bibfnamefont{M.}~\bibnamefont{Sakano}},
  \bibinfo{author}{\bibfnamefont{J.}~\bibnamefont{Omachi}},
  \bibinfo{author}{\bibfnamefont{K.}~\bibnamefont{Yoshioka}},
  \bibinfo{author}{\bibfnamefont{M.}~\bibnamefont{Kuwata-Gonokami}},
  \bibinfo{author}{\bibfnamefont{K.}~\bibnamefont{Ono}},
  \bibinfo{author}{\bibfnamefont{H.}~\bibnamefont{Kumigashira}},
  \bibnamefont{et~al.}, \bibinfo{journal}{Phys. Rev. B}
  \textbf{\bibinfo{volume}{90}}, \bibinfo{pages}{121111}
  (\bibinfo{year}{2014}),
  \urlprefix\url{http://link.aps.org/doi/10.1103/PhysRevB.90.121111}.

\bibitem[{\citenamefont{Nakayama et~al.}(2014)\citenamefont{Nakayama, Miyata,
  Phan, Sato, Tanabe, Urata, Tanigaki, and Takahashi}}]{nakayama2014}
\bibinfo{author}{\bibfnamefont{K.}~\bibnamefont{Nakayama}},
  \bibinfo{author}{\bibfnamefont{Y.}~\bibnamefont{Miyata}},
  \bibinfo{author}{\bibfnamefont{G.~N.} \bibnamefont{Phan}},
  \bibinfo{author}{\bibfnamefont{T.}~\bibnamefont{Sato}},
  \bibinfo{author}{\bibfnamefont{Y.}~\bibnamefont{Tanabe}},
  \bibinfo{author}{\bibfnamefont{T.}~\bibnamefont{Urata}},
  \bibinfo{author}{\bibfnamefont{K.}~\bibnamefont{Tanigaki}}, \bibnamefont{and}
  \bibinfo{author}{\bibfnamefont{T.}~\bibnamefont{Takahashi}},
  \bibinfo{journal}{Phys. Rev. Lett.} \textbf{\bibinfo{volume}{113}},
  \bibinfo{pages}{237001} (\bibinfo{year}{2014}),
  \urlprefix\url{http://link.aps.org/doi/10.1103/PhysRevLett.113.237001}.

\bibitem[{\citenamefont{Rahn et~al.}(2015)\citenamefont{Rahn, Ewings,
  Sedlmaier, Clarke, and Boothroyd}}]{rahn2015}
\bibinfo{author}{\bibfnamefont{M.~C.} \bibnamefont{Rahn}},
  \bibinfo{author}{\bibfnamefont{R.~A.} \bibnamefont{Ewings}},
  \bibinfo{author}{\bibfnamefont{S.~J.} \bibnamefont{Sedlmaier}},
  \bibinfo{author}{\bibfnamefont{S.~J.} \bibnamefont{Clarke}},
  \bibnamefont{and} \bibinfo{author}{\bibfnamefont{A.~T.}
  \bibnamefont{Boothroyd}}, \bibinfo{journal}{Phys. Rev. B}
  \textbf{\bibinfo{volume}{91}}, \bibinfo{pages}{180501}
  (\bibinfo{year}{2015}),
  \urlprefix\url{http://link.aps.org/doi/10.1103/PhysRevB.91.180501}.

\bibitem[{\citenamefont{Wang et~al.}(2015)\citenamefont{Wang, Shen, Pan, Hao,
  Ma, Zhou, Steffens, Schmalzl, Forrest, Abdel-Hafiez et~al.}}]{wang2015}
\bibinfo{author}{\bibfnamefont{Q.}~\bibnamefont{Wang}},
  \bibinfo{author}{\bibfnamefont{Y.}~\bibnamefont{Shen}},
  \bibinfo{author}{\bibfnamefont{B.}~\bibnamefont{Pan}},
  \bibinfo{author}{\bibfnamefont{Y.}~\bibnamefont{Hao}},
  \bibinfo{author}{\bibfnamefont{M.}~\bibnamefont{Ma}},
  \bibinfo{author}{\bibfnamefont{F.}~\bibnamefont{Zhou}},
  \bibinfo{author}{\bibfnamefont{P.}~\bibnamefont{Steffens}},
  \bibinfo{author}{\bibfnamefont{K.}~\bibnamefont{Schmalzl}},
  \bibinfo{author}{\bibfnamefont{T.}~\bibnamefont{Forrest}},
  \bibinfo{author}{\bibfnamefont{M.}~\bibnamefont{Abdel-Hafiez}},
  \bibnamefont{et~al.}, \bibinfo{journal}{Nature materials} p.
  \bibinfo{pages}{159} (\bibinfo{year}{2015}),
  \urlprefix\url{http://www.nature.com/nmat/journal/v15/n2/full/nmat4492.html}.

\bibitem[{\citenamefont{{Wang} et~al.}(2015)\citenamefont{{Wang}, {Shen},
  {Pan}, {Zhang}, {Ikeuchi}, {Iida}, {Christianson}, {Walker}, {Adroja},
  {Abdel-Hafiez} et~al.}}]{zhao2015}
\bibinfo{author}{\bibfnamefont{Q.}~\bibnamefont{{Wang}}},
  \bibinfo{author}{\bibfnamefont{Y.}~\bibnamefont{{Shen}}},
  \bibinfo{author}{\bibfnamefont{B.}~\bibnamefont{{Pan}}},
  \bibinfo{author}{\bibfnamefont{X.}~\bibnamefont{{Zhang}}},
  \bibinfo{author}{\bibfnamefont{K.}~\bibnamefont{{Ikeuchi}}},
  \bibinfo{author}{\bibfnamefont{K.}~\bibnamefont{{Iida}}},
  \bibinfo{author}{\bibfnamefont{A.~D.} \bibnamefont{{Christianson}}},
  \bibinfo{author}{\bibfnamefont{H.~C.} \bibnamefont{{Walker}}},
  \bibinfo{author}{\bibfnamefont{D.~T.} \bibnamefont{{Adroja}}},
  \bibinfo{author}{\bibfnamefont{M.}~\bibnamefont{{Abdel-Hafiez}}},
  \bibnamefont{et~al.}, \bibinfo{journal}{ArXiv e-prints}
  (\bibinfo{year}{2015}), \eprint{1511.02485},
  \urlprefix\url{http://arxiv.org/abs/1511.02485}.

\bibitem[{\citenamefont{B\"ohmer et~al.}(2015)\citenamefont{B\"ohmer, Arai,
  Hardy, Hattori, Iye, Wolf, L\"ohneysen, Ishida, and Meingast}}]{bohmer2015}
\bibinfo{author}{\bibfnamefont{A.~E.} \bibnamefont{B\"ohmer}},
  \bibinfo{author}{\bibfnamefont{T.}~\bibnamefont{Arai}},
  \bibinfo{author}{\bibfnamefont{F.}~\bibnamefont{Hardy}},
  \bibinfo{author}{\bibfnamefont{T.}~\bibnamefont{Hattori}},
  \bibinfo{author}{\bibfnamefont{T.}~\bibnamefont{Iye}},
  \bibinfo{author}{\bibfnamefont{T.}~\bibnamefont{Wolf}},
  \bibinfo{author}{\bibfnamefont{H.~v.} \bibnamefont{L\"ohneysen}},
  \bibinfo{author}{\bibfnamefont{K.}~\bibnamefont{Ishida}}, \bibnamefont{and}
  \bibinfo{author}{\bibfnamefont{C.}~\bibnamefont{Meingast}},
  \bibinfo{journal}{Phys. Rev. Lett.} \textbf{\bibinfo{volume}{114}},
  \bibinfo{pages}{027001} (\bibinfo{year}{2015}),
  \urlprefix\url{http://link.aps.org/doi/10.1103/PhysRevLett.114.027001}.

\bibitem[{\citenamefont{Baek et~al.}(2015)\citenamefont{Baek, Efremov, Ok, Kim,
  van~den Brink, and B{\"u}chner}}]{baek2015}
\bibinfo{author}{\bibfnamefont{S.}~\bibnamefont{Baek}},
  \bibinfo{author}{\bibfnamefont{D.}~\bibnamefont{Efremov}},
  \bibinfo{author}{\bibfnamefont{J.}~\bibnamefont{Ok}},
  \bibinfo{author}{\bibfnamefont{J.}~\bibnamefont{Kim}},
  \bibinfo{author}{\bibfnamefont{J.}~\bibnamefont{van~den Brink}},
  \bibnamefont{and}
  \bibinfo{author}{\bibfnamefont{B.}~\bibnamefont{B{\"u}chner}},
  \bibinfo{journal}{Nature materials} \textbf{\bibinfo{volume}{14}},
  \bibinfo{pages}{210} (\bibinfo{year}{2015}),
  \urlprefix\url{http://www.nature.com/nmat/journal/v14/n2/abs/nmat4138.html}.

\bibitem[{\citenamefont{Watson et~al.}(2015{\natexlab{a}})\citenamefont{Watson,
  Kim, Haghighirad, Davies, McCollam, Narayanan, Blake, Chen, Ghannadzadeh,
  Schofield et~al.}}]{watson20151}
\bibinfo{author}{\bibfnamefont{M.~D.} \bibnamefont{Watson}},
  \bibinfo{author}{\bibfnamefont{T.~K.} \bibnamefont{Kim}},
  \bibinfo{author}{\bibfnamefont{A.~A.} \bibnamefont{Haghighirad}},
  \bibinfo{author}{\bibfnamefont{N.~R.} \bibnamefont{Davies}},
  \bibinfo{author}{\bibfnamefont{A.}~\bibnamefont{McCollam}},
  \bibinfo{author}{\bibfnamefont{A.}~\bibnamefont{Narayanan}},
  \bibinfo{author}{\bibfnamefont{S.~F.} \bibnamefont{Blake}},
  \bibinfo{author}{\bibfnamefont{Y.~L.} \bibnamefont{Chen}},
  \bibinfo{author}{\bibfnamefont{S.}~\bibnamefont{Ghannadzadeh}},
  \bibinfo{author}{\bibfnamefont{A.~J.} \bibnamefont{Schofield}},
  \bibnamefont{et~al.}, \bibinfo{journal}{Phys. Rev. B}
  \textbf{\bibinfo{volume}{91}}, \bibinfo{pages}{155106}
  (\bibinfo{year}{2015}{\natexlab{a}}),
  \urlprefix\url{http://link.aps.org/doi/10.1103/PhysRevB.91.155106}.

\bibitem[{\citenamefont{Watson et~al.}(2015{\natexlab{b}})\citenamefont{Watson,
  Kim, Haghighirad, Blake, Davies, Hoesch, Wolf, and Coldea}}]{watson20152}
\bibinfo{author}{\bibfnamefont{M.~D.} \bibnamefont{Watson}},
  \bibinfo{author}{\bibfnamefont{T.~K.} \bibnamefont{Kim}},
  \bibinfo{author}{\bibfnamefont{A.~A.} \bibnamefont{Haghighirad}},
  \bibinfo{author}{\bibfnamefont{S.~F.} \bibnamefont{Blake}},
  \bibinfo{author}{\bibfnamefont{N.~R.} \bibnamefont{Davies}},
  \bibinfo{author}{\bibfnamefont{M.}~\bibnamefont{Hoesch}},
  \bibinfo{author}{\bibfnamefont{T.}~\bibnamefont{Wolf}}, \bibnamefont{and}
  \bibinfo{author}{\bibfnamefont{A.~I.} \bibnamefont{Coldea}},
  \bibinfo{journal}{Phys. Rev. B} \textbf{\bibinfo{volume}{92}},
  \bibinfo{pages}{121108} (\bibinfo{year}{2015}{\natexlab{b}}),
  \urlprefix\url{http://link.aps.org/doi/10.1103/PhysRevB.92.121108}.

\bibitem[{\citenamefont{{Massat} et~al.}(2016)\citenamefont{{Massat}, {Farina},
  {Paul}, {Karlsson}, {Strobel}, {Toulemonde}, {Measson}, {Cazayous}, {Sacuto},
  {Kasahara} et~al.}}]{massat2016}
\bibinfo{author}{\bibfnamefont{P.}~\bibnamefont{{Massat}}},
  \bibinfo{author}{\bibfnamefont{D.}~\bibnamefont{{Farina}}},
  \bibinfo{author}{\bibfnamefont{I.}~\bibnamefont{{Paul}}},
  \bibinfo{author}{\bibfnamefont{S.}~\bibnamefont{{Karlsson}}},
  \bibinfo{author}{\bibfnamefont{P.}~\bibnamefont{{Strobel}}},
  \bibinfo{author}{\bibfnamefont{P.}~\bibnamefont{{Toulemonde}}},
  \bibinfo{author}{\bibfnamefont{M.-A.} \bibnamefont{{Measson}}},
  \bibinfo{author}{\bibfnamefont{M.}~\bibnamefont{{Cazayous}}},
  \bibinfo{author}{\bibfnamefont{A.}~\bibnamefont{{Sacuto}}},
  \bibinfo{author}{\bibfnamefont{S.}~\bibnamefont{{Kasahara}}},
  \bibnamefont{et~al.}, \bibinfo{journal}{ArXiv e-prints}
  (\bibinfo{year}{2016}), \eprint{1603.01492},
  \urlprefix\url{http://adsabs.harvard.edu/abs/2016arXiv160301492M}.

\bibitem[{\citenamefont{Terashima et~al.}(2016)\citenamefont{Terashima,
  Kikugawa, Kiswandhi, Graf, Choi, Brooks, Kasahara, Watashige, Matsuda,
  Shibauchi et~al.}}]{taichi2016}
\bibinfo{author}{\bibfnamefont{T.}~\bibnamefont{Terashima}},
  \bibinfo{author}{\bibfnamefont{N.}~\bibnamefont{Kikugawa}},
  \bibinfo{author}{\bibfnamefont{A.}~\bibnamefont{Kiswandhi}},
  \bibinfo{author}{\bibfnamefont{D.}~\bibnamefont{Graf}},
  \bibinfo{author}{\bibfnamefont{E.-S.} \bibnamefont{Choi}},
  \bibinfo{author}{\bibfnamefont{J.~S.} \bibnamefont{Brooks}},
  \bibinfo{author}{\bibfnamefont{S.}~\bibnamefont{Kasahara}},
  \bibinfo{author}{\bibfnamefont{T.}~\bibnamefont{Watashige}},
  \bibinfo{author}{\bibfnamefont{Y.}~\bibnamefont{Matsuda}},
  \bibinfo{author}{\bibfnamefont{T.}~\bibnamefont{Shibauchi}},
  \bibnamefont{et~al.}, \bibinfo{journal}{Phys. Rev. B}
  \textbf{\bibinfo{volume}{93}}, \bibinfo{pages}{094505}
  (\bibinfo{year}{2016}),
  \urlprefix\url{http://link.aps.org/doi/10.1103/PhysRevB.93.094505}.

\bibitem[{\citenamefont{{Kothapalli} et~al.}(2016)\citenamefont{{Kothapalli},
  {B{\"o}hmer}, {Jayasekara}, {Ueland}, {Das}, {Sapkota}, {Taufour}, {Xiao},
  {Alp}, {Bud'ko} et~al.}}]{goldman2016}
\bibinfo{author}{\bibfnamefont{K.}~\bibnamefont{{Kothapalli}}},
  \bibinfo{author}{\bibfnamefont{A.~E.} \bibnamefont{{B{\"o}hmer}}},
  \bibinfo{author}{\bibfnamefont{W.~T.} \bibnamefont{{Jayasekara}}},
  \bibinfo{author}{\bibfnamefont{B.~G.} \bibnamefont{{Ueland}}},
  \bibinfo{author}{\bibfnamefont{P.}~\bibnamefont{{Das}}},
  \bibinfo{author}{\bibfnamefont{A.}~\bibnamefont{{Sapkota}}},
  \bibinfo{author}{\bibfnamefont{V.}~\bibnamefont{{Taufour}}},
  \bibinfo{author}{\bibfnamefont{Y.}~\bibnamefont{{Xiao}}},
  \bibinfo{author}{\bibfnamefont{E.~E.} \bibnamefont{{Alp}}},
  \bibinfo{author}{\bibfnamefont{S.~L.} \bibnamefont{{Bud'ko}}},
  \bibnamefont{et~al.}, \bibinfo{journal}{ArXiv e-prints}
  (\bibinfo{year}{2016}), \eprint{1603.04135},
  \urlprefix\url{http://arxiv.org/abs/1603.04135}.

\bibitem[{\citenamefont{{Wang} et~al.}(2016)\citenamefont{{Wang}, {Sun}, {Cui},
  {Song}, {Li}, {Yu}, {Lei}, and {Yu}}}]{yu2016}
\bibinfo{author}{\bibfnamefont{P.}~\bibnamefont{{Wang}}},
  \bibinfo{author}{\bibfnamefont{S.}~\bibnamefont{{Sun}}},
  \bibinfo{author}{\bibfnamefont{Y.}~\bibnamefont{{Cui}}},
  \bibinfo{author}{\bibfnamefont{W.}~\bibnamefont{{Song}}},
  \bibinfo{author}{\bibfnamefont{T.}~\bibnamefont{{Li}}},
  \bibinfo{author}{\bibfnamefont{R.}~\bibnamefont{{Yu}}},
  \bibinfo{author}{\bibfnamefont{H.}~\bibnamefont{{Lei}}}, \bibnamefont{and}
  \bibinfo{author}{\bibfnamefont{W.}~\bibnamefont{{Yu}}},
  \bibinfo{journal}{ArXiv e-prints}  (\bibinfo{year}{2016}),
  \eprint{1603.04589},
  \urlprefix\url{http://adsabs.harvard.edu/abs/2016arXiv160304589W}.

\bibitem[{\citenamefont{{Fanfarillo} et~al.}(2016)\citenamefont{{Fanfarillo},
  {Mansart}, {Toulemonde}, {Cercellier}, {Le Fevre}, {Bertran}, {Valenzuela},
  {Benfatto}, and {Brouet}}}]{fanfarillo2016}
\bibinfo{author}{\bibfnamefont{L.}~\bibnamefont{{Fanfarillo}}},
  \bibinfo{author}{\bibfnamefont{J.}~\bibnamefont{{Mansart}}},
  \bibinfo{author}{\bibfnamefont{P.}~\bibnamefont{{Toulemonde}}},
  \bibinfo{author}{\bibfnamefont{H.}~\bibnamefont{{Cercellier}}},
  \bibinfo{author}{\bibfnamefont{P.}~\bibnamefont{{Le Fevre}}},
  \bibinfo{author}{\bibfnamefont{F.}~\bibnamefont{{Bertran}}},
  \bibinfo{author}{\bibfnamefont{B.}~\bibnamefont{{Valenzuela}}},
  \bibinfo{author}{\bibfnamefont{L.}~\bibnamefont{{Benfatto}}},
  \bibnamefont{and} \bibinfo{author}{\bibfnamefont{V.}~\bibnamefont{{Brouet}}},
  \bibinfo{journal}{ArXiv e-prints}  (\bibinfo{year}{2016}),
  \eprint{1605.02482}, \urlprefix\url{http://arxiv.org/abs/1605.02482}.

\bibitem[{\citenamefont{Chubukov et~al.}(2015)\citenamefont{Chubukov,
  Fernandes, and Schmalian}}]{chubukov2015}
\bibinfo{author}{\bibfnamefont{A.~V.} \bibnamefont{Chubukov}},
  \bibinfo{author}{\bibfnamefont{R.~M.} \bibnamefont{Fernandes}},
  \bibnamefont{and}
  \bibinfo{author}{\bibfnamefont{J.}~\bibnamefont{Schmalian}},
  \bibinfo{journal}{Phys. Rev. B} \textbf{\bibinfo{volume}{91}},
  \bibinfo{pages}{201105} (\bibinfo{year}{2015}),
  \urlprefix\url{http://link.aps.org/doi/10.1103/PhysRevB.91.201105}.

\bibitem[{\citenamefont{Glasbrenner et~al.}(2015)\citenamefont{Glasbrenner,
  Mazin, Jeschke, Hirschfeld, Fernandes, and Valent{\'\i}}}]{valenti2015}
\bibinfo{author}{\bibfnamefont{J.}~\bibnamefont{Glasbrenner}},
  \bibinfo{author}{\bibfnamefont{I.}~\bibnamefont{Mazin}},
  \bibinfo{author}{\bibfnamefont{H.~O.} \bibnamefont{Jeschke}},
  \bibinfo{author}{\bibfnamefont{P.}~\bibnamefont{Hirschfeld}},
  \bibinfo{author}{\bibfnamefont{R.}~\bibnamefont{Fernandes}},
  \bibnamefont{and}
  \bibinfo{author}{\bibfnamefont{R.}~\bibnamefont{Valent{\'\i}}},
  \bibinfo{journal}{Nature Physics} \textbf{\bibinfo{volume}{11}},
  \bibinfo{pages}{953} (\bibinfo{year}{2015}),
  \urlprefix\url{http://www.nature.com/nphys/journal/v11/n11/abs/nphys3434.htm%
l}.

\bibitem[{\citenamefont{{Wang} and {Wang}}(2015)}]{wangshuai2015}
\bibinfo{author}{\bibfnamefont{S.}~\bibnamefont{{Wang}}} \bibnamefont{and}
  \bibinfo{author}{\bibfnamefont{F.}~\bibnamefont{{Wang}}},
  \bibinfo{journal}{ArXiv e-prints}  (\bibinfo{year}{2015}),
  \eprint{1510.05476},
  \urlprefix\url{http://adsabs.harvard.edu/abs/2015arXiv151005476W}.

\bibitem[{\citenamefont{Wang et~al.}(2015)\citenamefont{Wang, Kivelson, and
  Lee}}]{wangfa2015}
\bibinfo{author}{\bibfnamefont{F.}~\bibnamefont{Wang}},
  \bibinfo{author}{\bibfnamefont{S.~A.} \bibnamefont{Kivelson}},
  \bibnamefont{and} \bibinfo{author}{\bibfnamefont{D.-H.} \bibnamefont{Lee}},
  \bibinfo{journal}{Nature Physics} p. \bibinfo{pages}{959}
  (\bibinfo{year}{2015}),
  \urlprefix\url{http://www.nature.com/nphys/journal/v11/n11/full/nphys3456.ht%
ml}.

\bibitem[{\citenamefont{Yu and Si}(2015)}]{yu2015}
\bibinfo{author}{\bibfnamefont{R.}~\bibnamefont{Yu}} \bibnamefont{and}
  \bibinfo{author}{\bibfnamefont{Q.}~\bibnamefont{Si}}, \bibinfo{journal}{Phys.
  Rev. Lett.} \textbf{\bibinfo{volume}{115}}, \bibinfo{pages}{116401}
  (\bibinfo{year}{2015}),
  \urlprefix\url{http://link.aps.org/doi/10.1103/PhysRevLett.115.116401}.

\bibitem[{\citenamefont{{Lai} et~al.}(2016)\citenamefont{{Lai}, {Hu}, {Yu}, and
  {Si}}}]{lai2016}
\bibinfo{author}{\bibfnamefont{H.-H.} \bibnamefont{{Lai}}},
  \bibinfo{author}{\bibfnamefont{W.-J.} \bibnamefont{{Hu}}},
  \bibinfo{author}{\bibfnamefont{R.}~\bibnamefont{{Yu}}}, \bibnamefont{and}
  \bibinfo{author}{\bibfnamefont{Q.}~\bibnamefont{{Si}}},
  \bibinfo{journal}{ArXiv e-prints}  (\bibinfo{year}{2016}),
  \eprint{1603.03027}, \urlprefix\url{http://arxiv.org/abs/1603.03027}.

\bibitem[{\citenamefont{Wang et~al.}(2016)\citenamefont{Wang, Hu, and
  Nevidomskyy}}]{wang2016}
\bibinfo{author}{\bibfnamefont{Z.}~\bibnamefont{Wang}},
  \bibinfo{author}{\bibfnamefont{W.-J.} \bibnamefont{Hu}}, \bibnamefont{and}
  \bibinfo{author}{\bibfnamefont{A.~H.} \bibnamefont{Nevidomskyy}},
  \bibinfo{journal}{Phys. Rev. Lett.} \textbf{\bibinfo{volume}{116}},
  \bibinfo{pages}{247203} (\bibinfo{year}{2016}),
  \urlprefix\url{http://link.aps.org/doi/10.1103/PhysRevLett.116.247203}.

\bibitem[{\citenamefont{{Luo} et~al.}(2016)\citenamefont{{Luo}, {Datta}, and
  {Yao}}}]{yao2016}
\bibinfo{author}{\bibfnamefont{C.}~\bibnamefont{{Luo}}},
  \bibinfo{author}{\bibfnamefont{T.}~\bibnamefont{{Datta}}}, \bibnamefont{and}
  \bibinfo{author}{\bibfnamefont{D.-X.} \bibnamefont{{Yao}}},
  \bibinfo{journal}{ArXiv e-prints}  (\bibinfo{year}{2016}),
  \eprint{1603.03273},
  \urlprefix\url{http://adsabs.harvard.edu/abs/2016arXiv160303273L}.

\bibitem[{\citenamefont{{Busemeyer} et~al.}(2016)\citenamefont{{Busemeyer},
  {Dagrada}, {Sorella}, {Casula}, and {Wagner}}}]{wagner2016}
\bibinfo{author}{\bibfnamefont{B.}~\bibnamefont{{Busemeyer}}},
  \bibinfo{author}{\bibfnamefont{M.}~\bibnamefont{{Dagrada}}},
  \bibinfo{author}{\bibfnamefont{S.}~\bibnamefont{{Sorella}}},
  \bibinfo{author}{\bibfnamefont{M.}~\bibnamefont{{Casula}}}, \bibnamefont{and}
  \bibinfo{author}{\bibfnamefont{L.~K.} \bibnamefont{{Wagner}}},
  \bibinfo{journal}{ArXiv e-prints}  (\bibinfo{year}{2016}),
  \eprint{1602.02054}, \urlprefix\url{https://arxiv.org/abs/1602.02054}.

\bibitem[{\citenamefont{Zhu et~al.}(2016)\citenamefont{Zhu, Cao, Xie, Hou,
  Chen, Xiang, and Gong}}]{zhu2016}
\bibinfo{author}{\bibfnamefont{H.-F.} \bibnamefont{Zhu}},
  \bibinfo{author}{\bibfnamefont{H.-Y.} \bibnamefont{Cao}},
  \bibinfo{author}{\bibfnamefont{Y.}~\bibnamefont{Xie}},
  \bibinfo{author}{\bibfnamefont{Y.-S.} \bibnamefont{Hou}},
  \bibinfo{author}{\bibfnamefont{S.}~\bibnamefont{Chen}},
  \bibinfo{author}{\bibfnamefont{H.}~\bibnamefont{Xiang}}, \bibnamefont{and}
  \bibinfo{author}{\bibfnamefont{X.-G.} \bibnamefont{Gong}},
  \bibinfo{journal}{Phys. Rev. B} \textbf{\bibinfo{volume}{93}},
  \bibinfo{pages}{024511} (\bibinfo{year}{2016}),
  \urlprefix\url{http://link.aps.org/doi/10.1103/PhysRevB.93.024511}.

\bibitem[{\citenamefont{Zhuo et~al.}(2016)\citenamefont{Zhuo, Qin, Dong, Li,
  and Liu}}]{zhuo2016}
\bibinfo{author}{\bibfnamefont{W.~Z.} \bibnamefont{Zhuo}},
  \bibinfo{author}{\bibfnamefont{M.~H.} \bibnamefont{Qin}},
  \bibinfo{author}{\bibfnamefont{S.}~\bibnamefont{Dong}},
  \bibinfo{author}{\bibfnamefont{X.~G.} \bibnamefont{Li}}, \bibnamefont{and}
  \bibinfo{author}{\bibfnamefont{J.-M.} \bibnamefont{Liu}},
  \bibinfo{journal}{Phys. Rev. B} \textbf{\bibinfo{volume}{93}},
  \bibinfo{pages}{094424} (\bibinfo{year}{2016}),
  \urlprefix\url{http://link.aps.org/doi/10.1103/PhysRevB.93.094424}.

\bibitem[{\citenamefont{Cao et~al.}(2015)\citenamefont{Cao, Chen, Xiang, and
  Gong}}]{cao2015}
\bibinfo{author}{\bibfnamefont{H.-Y.} \bibnamefont{Cao}},
  \bibinfo{author}{\bibfnamefont{S.}~\bibnamefont{Chen}},
  \bibinfo{author}{\bibfnamefont{H.}~\bibnamefont{Xiang}}, \bibnamefont{and}
  \bibinfo{author}{\bibfnamefont{X.-G.} \bibnamefont{Gong}},
  \bibinfo{journal}{Phys. Rev. B} \textbf{\bibinfo{volume}{91}},
  \bibinfo{pages}{020504} (\bibinfo{year}{2015}),
  \urlprefix\url{http://link.aps.org/doi/10.1103/PhysRevB.91.020504}.

\bibitem[{\citenamefont{Wysocki et~al.}(2011)\citenamefont{Wysocki,
  Belashchenko, and Antropov}}]{wysocki2011}
\bibinfo{author}{\bibfnamefont{A.~L.} \bibnamefont{Wysocki}},
  \bibinfo{author}{\bibfnamefont{K.~D.} \bibnamefont{Belashchenko}},
  \bibnamefont{and} \bibinfo{author}{\bibfnamefont{V.~P.}
  \bibnamefont{Antropov}}, \bibinfo{journal}{Nature Physics}
  \textbf{\bibinfo{volume}{7}}, \bibinfo{pages}{485} (\bibinfo{year}{2011}),
  \urlprefix\url{http://www.nature.com/nphys/journal/v7/n6/abs/nphys1933.html}.

\bibitem[{\citenamefont{Hu et~al.}(2012)\citenamefont{Hu, Xu, Liu, Hao, and
  Wang}}]{hu2012}
\bibinfo{author}{\bibfnamefont{J.}~\bibnamefont{Hu}},
  \bibinfo{author}{\bibfnamefont{B.}~\bibnamefont{Xu}},
  \bibinfo{author}{\bibfnamefont{W.}~\bibnamefont{Liu}},
  \bibinfo{author}{\bibfnamefont{N.-N.} \bibnamefont{Hao}}, \bibnamefont{and}
  \bibinfo{author}{\bibfnamefont{Y.}~\bibnamefont{Wang}},
  \bibinfo{journal}{Phys. Rev. B} \textbf{\bibinfo{volume}{85}},
  \bibinfo{pages}{144403} (\bibinfo{year}{2012}),
  \urlprefix\url{http://link.aps.org/doi/10.1103/PhysRevB.85.144403}.

\bibitem[{\citenamefont{Yu et~al.}(2012)\citenamefont{Yu, Wang, Goswami,
  Nevidomskyy, Si, and Abrahams}}]{yu2012}
\bibinfo{author}{\bibfnamefont{R.}~\bibnamefont{Yu}},
  \bibinfo{author}{\bibfnamefont{Z.}~\bibnamefont{Wang}},
  \bibinfo{author}{\bibfnamefont{P.}~\bibnamefont{Goswami}},
  \bibinfo{author}{\bibfnamefont{A.~H.} \bibnamefont{Nevidomskyy}},
  \bibinfo{author}{\bibfnamefont{Q.}~\bibnamefont{Si}}, \bibnamefont{and}
  \bibinfo{author}{\bibfnamefont{E.}~\bibnamefont{Abrahams}},
  \bibinfo{journal}{Phys. Rev. B} \textbf{\bibinfo{volume}{86}},
  \bibinfo{pages}{085148} (\bibinfo{year}{2012}),
  \urlprefix\url{http://link.aps.org/doi/10.1103/PhysRevB.86.085148}.

\bibitem[{\citenamefont{Glasbrenner et~al.}(2014)\citenamefont{Glasbrenner,
  Velev, and Mazin}}]{mazin2014}
\bibinfo{author}{\bibfnamefont{J.~K.} \bibnamefont{Glasbrenner}},
  \bibinfo{author}{\bibfnamefont{J.~P.} \bibnamefont{Velev}}, \bibnamefont{and}
  \bibinfo{author}{\bibfnamefont{I.~I.} \bibnamefont{Mazin}},
  \bibinfo{journal}{Phys. Rev. B} \textbf{\bibinfo{volume}{89}},
  \bibinfo{pages}{064509} (\bibinfo{year}{2014}),
  \urlprefix\url{http://link.aps.org/doi/10.1103/PhysRevB.89.064509}.

\bibitem[{\citenamefont{Stanek et~al.}(2011)\citenamefont{Stanek, Sushkov, and
  Uhrig}}]{stanek2011}
\bibinfo{author}{\bibfnamefont{D.}~\bibnamefont{Stanek}},
  \bibinfo{author}{\bibfnamefont{O.~P.} \bibnamefont{Sushkov}},
  \bibnamefont{and} \bibinfo{author}{\bibfnamefont{G.~S.} \bibnamefont{Uhrig}},
  \bibinfo{journal}{Phys. Rev. B} \textbf{\bibinfo{volume}{84}},
  \bibinfo{pages}{064505} (\bibinfo{year}{2011}),
  \urlprefix\url{http://link.aps.org/doi/10.1103/PhysRevB.84.064505}.

\bibitem[{\citenamefont{Bilbao~Ergueta and Nevidomskyy}(2015)}]{bilbao2015}
\bibinfo{author}{\bibfnamefont{P.}~\bibnamefont{Bilbao~Ergueta}}
  \bibnamefont{and} \bibinfo{author}{\bibfnamefont{A.~H.}
  \bibnamefont{Nevidomskyy}}, \bibinfo{journal}{Phys. Rev. B}
  \textbf{\bibinfo{volume}{92}}, \bibinfo{pages}{165102}
  (\bibinfo{year}{2015}),
  \urlprefix\url{http://link.aps.org/doi/10.1103/PhysRevB.92.165102}.

\bibitem[{\citenamefont{Jiang et~al.}(2009)\citenamefont{Jiang, Kr\"uger,
  Moore, Sheng, Zaanen, and Weng}}]{jiang2009}
\bibinfo{author}{\bibfnamefont{H.~C.} \bibnamefont{Jiang}},
  \bibinfo{author}{\bibfnamefont{F.}~\bibnamefont{Kr\"uger}},
  \bibinfo{author}{\bibfnamefont{J.~E.} \bibnamefont{Moore}},
  \bibinfo{author}{\bibfnamefont{D.~N.} \bibnamefont{Sheng}},
  \bibinfo{author}{\bibfnamefont{J.}~\bibnamefont{Zaanen}}, \bibnamefont{and}
  \bibinfo{author}{\bibfnamefont{Z.~Y.} \bibnamefont{Weng}},
  \bibinfo{journal}{Phys. Rev. B} \textbf{\bibinfo{volume}{79}},
  \bibinfo{pages}{174409} (\bibinfo{year}{2009}),
  \urlprefix\url{http://link.aps.org/doi/10.1103/PhysRevB.79.174409}.

\bibitem[{\citenamefont{White}(1992)}]{white1992}
\bibinfo{author}{\bibfnamefont{S.~R.} \bibnamefont{White}},
  \bibinfo{journal}{Phys. Rev. Lett.} \textbf{\bibinfo{volume}{69}},
  \bibinfo{pages}{2863} (\bibinfo{year}{1992}),
  \urlprefix\url{http://link.aps.org/doi/10.1103/PhysRevLett.69.2863}.

\bibitem[{sem()}]{semi}
\bibinfo{note}{The ground-state energy for the magnetic and quadrupolar states
  obtained from the site-factorized wavefunction calculations can be found in
  the Supplemental Material of Ref. 39.}

\bibitem[{\citenamefont{McCulloch and Gul{\'a}csi}(2002)}]{mcculloch2002}
\bibinfo{author}{\bibfnamefont{I.}~\bibnamefont{McCulloch}} \bibnamefont{and}
  \bibinfo{author}{\bibfnamefont{M.}~\bibnamefont{Gul{\'a}csi}},
  \bibinfo{journal}{Europhysics Letters} \textbf{\bibinfo{volume}{57}},
  \bibinfo{pages}{852} (\bibinfo{year}{2002}),
  \urlprefix\url{http://iopscience.iop.org/0295-5075/57/6/852}.

\bibitem[{\citenamefont{White and Chernyshev}(2007)}]{white2007}
\bibinfo{author}{\bibfnamefont{S.~R.} \bibnamefont{White}} \bibnamefont{and}
  \bibinfo{author}{\bibfnamefont{A.~L.} \bibnamefont{Chernyshev}},
  \bibinfo{journal}{Phys. Rev. Lett.} \textbf{\bibinfo{volume}{99}},
  \bibinfo{pages}{127004} (\bibinfo{year}{2007}),
  \urlprefix\url{http://link.aps.org/doi/10.1103/PhysRevLett.99.127004}.

\bibitem[{\citenamefont{Gong et~al.}(2013)\citenamefont{Gong, Sheng, Motrunich,
  and Fisher}}]{gong2013}
\bibinfo{author}{\bibfnamefont{S.-S.} \bibnamefont{Gong}},
  \bibinfo{author}{\bibfnamefont{D.~N.} \bibnamefont{Sheng}},
  \bibinfo{author}{\bibfnamefont{O.~I.} \bibnamefont{Motrunich}},
  \bibnamefont{and} \bibinfo{author}{\bibfnamefont{M.~P.~A.}
  \bibnamefont{Fisher}}, \bibinfo{journal}{Phys. Rev. B}
  \textbf{\bibinfo{volume}{88}}, \bibinfo{pages}{165138}
  (\bibinfo{year}{2013}),
  \urlprefix\url{http://link.aps.org/doi/10.1103/PhysRevB.88.165138}.

\bibitem[{\citenamefont{Gong et~al.}(2014)\citenamefont{Gong, Zhu, Sheng,
  Motrunich, and Fisher}}]{gong2014}
\bibinfo{author}{\bibfnamefont{S.-S.} \bibnamefont{Gong}},
  \bibinfo{author}{\bibfnamefont{W.}~\bibnamefont{Zhu}},
  \bibinfo{author}{\bibfnamefont{D.~N.} \bibnamefont{Sheng}},
  \bibinfo{author}{\bibfnamefont{O.~I.} \bibnamefont{Motrunich}},
  \bibnamefont{and} \bibinfo{author}{\bibfnamefont{M.~P.~A.}
  \bibnamefont{Fisher}}, \bibinfo{journal}{Phys. Rev. Lett.}
  \textbf{\bibinfo{volume}{113}}, \bibinfo{pages}{027201}
  (\bibinfo{year}{2014}),
  \urlprefix\url{http://link.aps.org/doi/10.1103/PhysRevLett.113.027201}.

\bibitem[{\citenamefont{Sandvik}(2012)}]{sandvik2012}
\bibinfo{author}{\bibfnamefont{A.~W.} \bibnamefont{Sandvik}},
  \bibinfo{journal}{Phys. Rev. B} \textbf{\bibinfo{volume}{85}},
  \bibinfo{pages}{134407} (\bibinfo{year}{2012}),
  \urlprefix\url{http://link.aps.org/doi/10.1103/PhysRevB.85.134407}.

\bibitem[{\citenamefont{Zhu et~al.}(2013)\citenamefont{Zhu, Huse, and
  White}}]{zhu2013}
\bibinfo{author}{\bibfnamefont{Z.}~\bibnamefont{Zhu}},
  \bibinfo{author}{\bibfnamefont{D.~A.} \bibnamefont{Huse}}, \bibnamefont{and}
  \bibinfo{author}{\bibfnamefont{S.~R.} \bibnamefont{White}},
  \bibinfo{journal}{Phys. Rev. Lett.} \textbf{\bibinfo{volume}{110}},
  \bibinfo{pages}{127205} (\bibinfo{year}{2013}),
  \urlprefix\url{http://link.aps.org/doi/10.1103/PhysRevLett.110.127205}.

\bibitem[{\citenamefont{Zhu and White}(2015)}]{zhu2015}
\bibinfo{author}{\bibfnamefont{Z.}~\bibnamefont{Zhu}} \bibnamefont{and}
  \bibinfo{author}{\bibfnamefont{S.~R.} \bibnamefont{White}},
  \bibinfo{journal}{Phys. Rev. B} \textbf{\bibinfo{volume}{92}},
  \bibinfo{pages}{041105} (\bibinfo{year}{2015}),
  \urlprefix\url{http://link.aps.org/doi/10.1103/PhysRevB.92.041105}.

\bibitem[{\citenamefont{Hu et~al.}(2015)\citenamefont{Hu, Gong, Zhu, and
  Sheng}}]{hu2015}
\bibinfo{author}{\bibfnamefont{W.-J.} \bibnamefont{Hu}},
  \bibinfo{author}{\bibfnamefont{S.-S.} \bibnamefont{Gong}},
  \bibinfo{author}{\bibfnamefont{W.}~\bibnamefont{Zhu}}, \bibnamefont{and}
  \bibinfo{author}{\bibfnamefont{D.~N.} \bibnamefont{Sheng}},
  \bibinfo{journal}{Phys. Rev. B} \textbf{\bibinfo{volume}{92}},
  \bibinfo{pages}{140403} (\bibinfo{year}{2015}),
  \urlprefix\url{http://link.aps.org/doi/10.1103/PhysRevB.92.140403}.

\bibitem[{\citenamefont{Saadatmand and McCulloch}(2016)}]{ian2016}
\bibinfo{author}{\bibfnamefont{S.~N.} \bibnamefont{Saadatmand}}
  \bibnamefont{and} \bibinfo{author}{\bibfnamefont{I.~P.}
  \bibnamefont{McCulloch}}, \bibinfo{journal}{Phys. Rev. B}
  \textbf{\bibinfo{volume}{94}}, \bibinfo{pages}{121111}
  (\bibinfo{year}{2016}),
  \urlprefix\url{http://link.aps.org/doi/10.1103/PhysRevB.94.121111}.

\bibitem[{\citenamefont{Yan et~al.}(2011)\citenamefont{Yan, Huse, and
  White}}]{yan2011}
\bibinfo{author}{\bibfnamefont{S.}~\bibnamefont{Yan}},
  \bibinfo{author}{\bibfnamefont{D.~A.} \bibnamefont{Huse}}, \bibnamefont{and}
  \bibinfo{author}{\bibfnamefont{S.~R.} \bibnamefont{White}},
  \bibinfo{journal}{Science} \textbf{\bibinfo{volume}{332}},
  \bibinfo{pages}{1173} (\bibinfo{year}{2011}),
  \urlprefix\url{http://science.sciencemag.org/content/332/6034/1173}.

\bibitem[{\citenamefont{Gong et~al.}(2015)\citenamefont{Gong, Zhu, and
  Sheng}}]{gong2015}
\bibinfo{author}{\bibfnamefont{S.-S.} \bibnamefont{Gong}},
  \bibinfo{author}{\bibfnamefont{W.}~\bibnamefont{Zhu}}, \bibnamefont{and}
  \bibinfo{author}{\bibfnamefont{D.~N.} \bibnamefont{Sheng}},
  \bibinfo{journal}{Phys. Rev. B} \textbf{\bibinfo{volume}{92}},
  \bibinfo{pages}{195110} (\bibinfo{year}{2015}),
  \urlprefix\url{http://link.aps.org/doi/10.1103/PhysRevB.92.195110}.

\bibitem[{\citenamefont{Towns et~al.}(2014)\citenamefont{Towns, Cockerill,
  Dahan, Foster, Gaither, Grimshaw, Hazlewood, Lathrop, Lifka, Peterson
  et~al.}}]{xsede}
\bibinfo{author}{\bibfnamefont{J.}~\bibnamefont{Towns}},
  \bibinfo{author}{\bibfnamefont{T.}~\bibnamefont{Cockerill}},
  \bibinfo{author}{\bibfnamefont{M.}~\bibnamefont{Dahan}},
  \bibinfo{author}{\bibfnamefont{I.}~\bibnamefont{Foster}},
  \bibinfo{author}{\bibfnamefont{K.}~\bibnamefont{Gaither}},
  \bibinfo{author}{\bibfnamefont{A.}~\bibnamefont{Grimshaw}},
  \bibinfo{author}{\bibfnamefont{V.}~\bibnamefont{Hazlewood}},
  \bibinfo{author}{\bibfnamefont{S.}~\bibnamefont{Lathrop}},
  \bibinfo{author}{\bibfnamefont{D.}~\bibnamefont{Lifka}},
  \bibinfo{author}{\bibfnamefont{G.~D.} \bibnamefont{Peterson}},
  \bibnamefont{et~al.}, \bibinfo{journal}{Computing in Science and Engineering}
  \textbf{\bibinfo{volume}{16}}, \bibinfo{pages}{62} (\bibinfo{year}{2014}),
  ISSN \bibinfo{issn}{1521-9615}.

\bibitem[{\citenamefont{Lacroix et~al.}(2011)\citenamefont{Lacroix, Mendels,
  and Mila}}]{Mila}
\bibinfo{author}{\bibfnamefont{C.}~\bibnamefont{Lacroix}},
  \bibinfo{author}{\bibfnamefont{P.}~\bibnamefont{Mendels}}, \bibnamefont{and}
  \bibinfo{author}{\bibfnamefont{F.}~\bibnamefont{Mila}},
  \emph{\bibinfo{title}{Introduction to Frustrated Magnetism}}
  (\bibinfo{publisher}{Springer Science \& Business Media},
  \bibinfo{year}{2011}).

\bibitem[{\citenamefont{Kittel}(1960)}]{Kittel}
\bibinfo{author}{\bibfnamefont{C.}~\bibnamefont{Kittel}},
  \bibinfo{journal}{Phys. Rev.} \textbf{\bibinfo{volume}{120}},
  \bibinfo{pages}{335} (\bibinfo{year}{1960}),
  \urlprefix\url{http://link.aps.org/doi/10.1103/PhysRev.120.335}.

\bibitem[{\citenamefont{Raghu et~al.}(2008)\citenamefont{Raghu, Qi, Liu,
  Scalapino, and Zhang}}]{Raghu2008}
\bibinfo{author}{\bibfnamefont{S.}~\bibnamefont{Raghu}},
  \bibinfo{author}{\bibfnamefont{X.-L.} \bibnamefont{Qi}},
  \bibinfo{author}{\bibfnamefont{C.-X.} \bibnamefont{Liu}},
  \bibinfo{author}{\bibfnamefont{D.~J.} \bibnamefont{Scalapino}},
  \bibnamefont{and} \bibinfo{author}{\bibfnamefont{S.-C.} \bibnamefont{Zhang}},
  \bibinfo{journal}{Phys. Rev. B} \textbf{\bibinfo{volume}{77}},
  \bibinfo{pages}{220503} (\bibinfo{year}{2008}),
  \urlprefix\url{http://link.aps.org/doi/10.1103/PhysRevB.77.220503}.

\bibitem[{\citenamefont{Fazekas}(1999)}]{Fazekas_Book}
\bibinfo{author}{\bibfnamefont{P.}~\bibnamefont{Fazekas}},
  \emph{\bibinfo{title}{Lecture Notes on Electron Correlation and Magnetism}}
  (\bibinfo{publisher}{World Scientific, Singapore}, \bibinfo{year}{1999}).

\end{thebibliography}

\end{document}